\documentclass[aps,prb,twocolumn,superscriptaddress,floatfix,letterpaper]{revtex4-1}
\usepackage[inline]{enumitem}
\usepackage{tikz}
\usepackage[export]{adjustbox}
\usepackage{float}
\usepackage[normalem]{ulem}
\usetikzlibrary{decorations.pathreplacing}

 \usepackage{pdftricks2}
 \usepackage{epsfig}
 \usepackage{pst-grad} % For gradients
 \usepackage{pst-plot} % For axes

\def\arraystretch{1.25} 

\newcommand{\calD}{\mathcal{D}}
\newcommand{\calM}{\mathcal{M}}

\newcommand{\tAw}[3]{{#1}_{e_{#2#3} }^{v_{#2}v_{#3}}}

\newcommand{\tC}[5]{{#1}_{v_{#2}v_{#3}v_{#4}v_{#5};
\phi_{#2#3#5} \phi_{#3#4#5}
}^{
e_{#2#3} e_{#2#4} e_{#2#5} e_{#3#4} e_{#3#5} e_{#4#5};
\phi_{#2#3#4} \phi_{#2#4#5}
}}

\newcommand{\nn}{\nonumber}

\renewcommand{\tilde}{\widetilde}
\renewcommand{\bar}{\overline}

\newcommand{\tlambda}{\tilde{\lambda}}

\renewcommand{\hat}[1]{\widehat{#1}}

\newcolumntype{L}{>{$}l<{$}}
\newcolumntype{R}{>{$}r<{$}}
\newcolumntype{C}{>{$}c<{$}}

\newcommand{\calA}{\mathcal{A}}
\newcommand{\calC}{\mathcal{C}}
\newcommand{\calG}{\mathcal{G}}
\newcommand{\calN}{\mathcal{N}}
\newcommand{\calH}{\mathcal{H}}

\newcommand{\calT}{\mathcal{T}}

\newcommand{\calZ}{\mathcal{Z}}

\newcommand{\ZZ}{\mathbb{Z}}

\newcommand{\fh}{\frac{1}{2}}

\newcommand{\fixme}[1]{{\bf {\color{red}[#1]}}}

\newcommand{\zb}{
\begin{tikzpicture}[scale=.5,baseline={([yshift=-.5ex]current bounding box.center)}]
\draw (0,0) rectangle (1.2,1.2);
\end{tikzpicture}\,
}
\newcommand{\zbh}{
\begin{tikzpicture}[scale=.5,baseline={([yshift=-.5ex]current bounding box.center)}]
\draw (0,0) rectangle (1.2,1.2);
\draw [thick] (0,.8)--(1.2,.8);
\end{tikzpicture}\,
}
\newcommand{\zbv}{
\begin{tikzpicture}[scale=.5,baseline={([yshift=-.5ex]current bounding box.center)}]
\draw (0,0) rectangle (1.2,1.2);
\draw [thick] (.8,0)--(.8,1.2);
\end{tikzpicture}\,
}
\newcommand{\zbx}{
\begin{tikzpicture}[scale=.5,baseline={([yshift=-.5ex]current bounding box.center)}]
\draw (0,0) rectangle (1.2,1.2);
\draw [thick] (0,.8)--(1.2,.8);
\draw [thick] (.8,0)--(.8,1.2);
\end{tikzpicture}\,
}

\begin{document}

\begin{titlepage}
\title{A unified view on symmetry, anomalous symmetry,\\
 and non-invertible gravitational anomaly
%Invertible and non-invertible anomaly from a global symmetry\\
%nd their constraint on low energy dynamics
}

\author{Wenjie Ji}
\affiliation{Department of Physics, Massachusetts Institute of
Technology, Cambridge, Massachusetts 02139, USA}
\affiliation{Department of Physics, University of California, Santa Barbara, California 93106, USA}

%\author{Shu-Heng Shao}
%\affiliation{School of Natural Sciences, Institute for Advanced Study, Princeton, New Jersey 08540, USA}

\author{Xiao-Gang Wen}
\affiliation{Department of Physics, Massachusetts Institute of
Technology, Cambridge, Massachusetts 02139, USA}

\begin{abstract} 

In this paper, using 1+1D models as examples, we study symmetries and anomalous
symmetries via multi-component partition functions obtained through symmetry
twists, and their transformations under the mapping class group of spacetime.
This point of view allows us to treat symmetries and anomalous symmetries as
non-invertible gravitational anomalies (which are also described by
multi-component partition functions, transforming covariantly under the mapping
group transformations).  This allows us to directly see how symmetry and
anomalous symmetry constrain the low energy dynamics of the systems, since the
low energy dynamics is directly encoded in the partition functions.  More
generally, symmetries, anomalous symmetries, non-invertible gravitational
anomalies, and their combinations, can all be viewed as constraints on low
energy dynamics.  In this paper, we demonstrate that they all can be viewed
uniformly and systematically as pure (non-invertible) gravitational
anomalies.

\end{abstract}


\maketitle

\end{titlepage}

{\small \setcounter{tocdepth}{1} \tableofcontents }

\section{Introduction}

A global symmetry in a quantum many-body system on a lattice can be on-site or
not on-site.\cite{CGW1107,W1313} If the global symmetry transformation $W$ is
on-site:
\begin{align}
 W = \bigotimes_i W_i,
\end{align}
where $W_i$ acts only on site-$i$, then the system can have a gapped symmetric
ground state with no topological order (\ie a product
state).\cite{W8987,CGW1038} If the global symmetry is not on-site (or more
precisely, the non-on-siteness cannot be removed by local unitary
transformations), then the system cannot have a symmetric gapped ground state
with no topological order.\cite{CLW1141} In this case, the system can be in a
gapless state, a symmetry breaking state, and/or topological ordered state. 

This result is obtained by noting that non-on-site global symmetry can be
realized at the boundary of a system with on-site global symmetry. The
non-on-siteness (the anomaly) at the boundary is directly related\cite{W1313}
to symmetry protected topological (SPT) order\cite{GW0931,CLW1141,CGL1314} or
symmetry enriched topological (SET) order in the bulk\cite{CGW1038}.  Now, we
can see why anomalous symmetry (\ie no-on-site symmetry) do not allow trivially
gapped ground state: according to the topological holographic principle that
the boundary uniquely determines the bulk,\cite{KW1458} a non-trivial SPT or
SET phase cannot have a symmetric boundary with no topological order, since a
symmetric boundary with no topological order will imply the bulk to be a
trivial SPT or SET state.

For a quantum field theory with global symmetry, it is hard to determine if the
symmetry is on-site or not since the short-distance cut-off may not be
specified.  In this case, we examine the obstruction of gauging the global
symmetry in the quantum theory, which is referred to as 't Hooft anomaly for
the global symmetry.\cite{H8035} The 't Hooft anomaly -- an IR property, and
non-on-siteness -- a UV property, of a symmetry is directly related: \emph{a
quantum field theory with 't Hooft anomaly does not have any lattice
regularization where the global symmetry $G$ is realized as an on-site
symmetry.}  

However,  a quantum field theory with 't Hooft anomaly can be realized as a
boundary of a lattice model with SPT order in one higher dimension, where the
global symmetry $G$ acts as an on-site symmetry in the bulk.  The SPT order in
one higher dimension corresponds to the non-on-siteness (\ie the 't Hooft
anomaly) of lattice global symmetry.  As a result, the 't Hooft anomaly for a
global symmetry $G$ is related to\cite{RML1204} and classified
by\cite{W1313,KT1430} the SPT order protected by $G$ in one higher dimension.  

From the above discussion, we see that the 't Hooft anomaly in a field theory
can be viewed as the obstruction to have a lattice regularization \emph{in the
same dimensions} where the global symmetry is on-site.  Even without symmetry,
a field theory can be anomalous, which is referred to as gravitational anomaly.
Similar to the 't Hooft anomaly discussed above, the gravitational anomaly is
an obstruction to have a lattice regularization in the same dimension.  More
precisely, gravitational anomaly is an obstruction for the field theory to have
a realization in the same dimensions where the total Hilbert space has the
tensor product decomposition into local Hilbert spaces
$\cV_i$\cite{KP0604,LW0605,W1313,YS13094596,KW1458}
\begin{align}
 \cV_\text{tot}=\bigotimes_i \cV_i.
\end{align}
However, a field theory with a gravitational anomaly can have a lattice
regularization in one higher dimension, and the field theory appears as an
effective boundary theory for this lattice model in one higher dimension.  To
reproduce 
%{\color{blue} [Does it actually mean ``cancel" here?] XG: I do mean reproduce} 
the gravitational anomaly on the boundary, the lattice model must be in a
topologically ordered phase.  Thus the gravitational anomaly simply corresponds
to the topological order in one higher dimension.\cite{W1313,KW1458}

Recently, it was pointed out that the gravitational anomaly defined this way is
more general than the usual gravitational anomaly defined via the obstruction
to define a path integral (\ie to have a diffeomorphism invariant path
integral).  In particular, the gravitational anomaly from the point of view of
lattice regularization (or tensor product decomposition) can be non-invertible
(\ie cannot be canceled by any other gravitational anomalies), 
%{\color{blue} in field theories describing invertible phases?} XG: cannot be
%canceled by even non-invertible gravitational anomalies)
\cite{KW1458,FV14095723,M14107442,JW190513279} while the gravitational anomaly
from  the point of view of diffeomorphism invariance is always invertible.  

What is non-invertible anomaly?  We first note that the 't Hooft anomaly and
global gravitational anomaly discussed in most previous literature focuses
mostly on invertible anomalies. The invertible anomaly is associated with
invertible topological orders in one higher dimensions.  The invertible
topological orders are only a small class of topological orders.  So it is
natural to generalize the notion of anomaly to include non-invertible anomaly,
which is associated with generic topological orders in one higher dimension.
This more general point of view on anomaly was introduced in
\Ref{W1313,KW1458,FV14095723,M14107442}, which generalize an anomaly in-flow
picture of \Ref{CH8527} that matches the anomaly to the topological term in one
higher dimensions.  Recent detailed and systematic discussions of
non-invertible gravitational anomalies were presented in
\Ref{JW190513279,JW191213492,KZ200308898,KZ200514178}.

In this paper, we will study non-invertible gravitational anomalies with a
global symmetry, in other words, the mixed case of 't Hooft anomaly and
non-invertible gravitational anomaly.  We will propose a very general unified
point of view for both 't Hooft anomaly and global gravitational anomaly.  Our
theory not only applies to 't Hooft anomaly and global gravitational anomaly, it
also applies to a new class of anomaly -- \emph{non-invertible anomaly} with
symmetry.  Including non-invertible anomaly is the key to obtain a unified
theory for 't Hooft anomaly and gravitational anomaly, as well as their
non-invertible generalization.  

A $d+1$D quantum field theory with a global symmetry that has 't Hooft anomaly
means the global symmetry cannot be gauged, \ie to be promoted to a gauge
symmetry.  However, if we consider the anomalous symmetry as the effective
boundary symmetry of an on-site symmetry in one higher dimension, then we can
``gauge'' the anomalous symmetry by gauging the on-site symmetry in the whole
$d+2$D system with a boundary.  This way, even global symmetry with 't Hooft
anomaly can be ``gauged''.  ``Gauging'' the anomalous (or anomaly-free) global
symmetry in a $d+1$D theory, after first coupling the theory to a theory in
$d+2$D,  produces another coupled theory, in which the boundary theory has a
non-invertible gravitational anomaly and the bulk theory describes a
topological order. 
%{\color{blue} [WJ: This does not generally hold when the boundary has an
%anomalous $1$-form symmetry and the bulk is a $1$-form SPT. For example, the
%$3+1$d $m=1$ $Z_2$ 1-form SPT, after gauging is still a 1-form SPT and not
%topologically ordered.] XG: We should discuss. I think it works} 
A systematic understanding of the non-invertible anomaly of a quantum field
theory as a boundary of the topological order in one higher dimension has been
developed in \Ref{JW190513279}.  

Comparing the boundary system before and after the gauging procedure described above allows us to identify a systematic treatment of
anomaly-free global symmetry $G$,  anomalous global symmetry $G$,  and the
(non-invertible) gravitational anomaly in the same footing.  Specifically,  if we restrict ourselves to consider the symmetric sub-Hilbert space of the (anomalous) symmetry
$\cV_\text{symm}$, symmetric sub-Hilbert space does not have a tensor product
decomposition \cite{JW191213492}
\begin{align}
 \cV_\text{symm} \neq \bigotimes_i \cV_i .
\end{align}
This constrained Hilbert space can be viewed as having a non-invertible
gravitational anomaly. This implies that we can treat anomaly-free and
anomalous global symmetry as a non-invertible gravitational anomaly.
\Ref{JW191213492,KZ200308898,KZ200514178} refer to non-invertible gravitational
anomaly as categorical symmetry to stress its connection to symmetry and
anomalous symmetry mentioned above.

For example, a lattice model with tensor product decomposed Hilbert space and
an anomaly-free or anomalous global symmetry can be viewed as a boundary of a
SPT state. We may then gauge the global symmetry in the bulk. The boundary
Hilbert space of the gauged bulk matches the symmetric sub-Hilbert space of the
original lattice model.  We see that restricting to the symmetric sub-Hilbert
space corresponds to a non-invertible anomaly.  The gauged bulk has a
topological order which matches the non-invertible anomaly on the boundary. 

We develop a unified theory that covers finite symmetry, 't Hooft anomalous
finite symmetry, and non-invertible gravitational anomaly on the equal footing,
in terms of multi-component partition functions, indexed by symmetry twists,
bulk anyons, as well as topological symmetry defects.  In this unified
description, We naturally see how symmetry, anomalous symmetry and gravitational
anomaly constrain the partition functions.  Physically speaking, the
(anomalous) symmetry and non-invertible gravitational anomaly are characterized
by the same data that characterize the topological order in one higher
dimension.  Such data includes the group cocycle of the (gauged) symmetry and
mapping class group representation.  We find that the multi-component partition
function of a symmetric or anomalous theory must satisfy some conditions
specified by the data that characterize the bulk topological order.  Those
conditions on partition functions allow us to constrain the low energy
properties of the anomalous system\cite{CLW1141,LS190404833} (\ie the low
energy properties of the boundary of the bulk topological orders). For example,
restricting the central charge and the quantum dimensions of operators in a
symmetry representation, given the (anomalous) symmetry.

In this paper, we will also consider a question: given a low energy effective
field theory with a global symmetry, how to compute the 't Hooft anomaly of the
symmetry?  If the symmetry is continuous, \emph{part} of its  't Hooft anomaly
can be computed via the correlation function of associated conserved
current.\cite{A6926,BJ6947} However, such an approach cannot detect all
possible 't Hooft anomaly of the symmetry. Also, the approach fails utterly
for discrete symmetry, since there is no associated conserved current.
\Ref{RZ1232} propose a way to compute 't Hooft anomaly for discrete $Z_2$
symmetry in 1+1D system via modular transformation properties.  A
generalization of this result to general Abelian symmetry and to higher
dimensions is presented in \Ref{TR171004730}.  In this paper, we will generalize
such an approach to include the mixed case of 't Hooft anomaly and
non-invertible gravitational anomaly, via the properties of the multicomponent
partition functions $Z^i(M^D,A)$ with symmetry twists $A$.  As pointed out by
\Ref{TR171004730}, for the invertible anomalies (described by single-component
partition function), the properties of the partition function $Z(M^D,A)$ is
directly related to the topological invariance\cite{W1447,HW1339,K1459,KW1458}
of the SPT order and the invertible topological order in one higher dimension.
Therefore, such an approach may allow us to compute all the 't Hooft anomaly of
any symmetry, as well as the global gravitational anomaly.

We have mentioned that 't Hooft anomaly can be probed via partition functions
with symmetry twists.  Since the symmetry twists can be represented as domain
walls in $D$-dimensional spacetime, the 't Hooft anomaly of the symmetry can
also be encoded in the fusion properties of the symmetry twists or domain
walls.  In other words, the fusion of the domain walls form a pointed  fusion
$(D-1)$-category,\cite{CY180204445,LS190404833} which in turn characterizes the
't Hooft anomaly.  In particular, a pointed fusion $(D-1)$-category is
characterized by a group cocycle $\om_{D+1} \in \cH^{D+1}(G;\R/\Z)$ where $G$
is the group that describes the pointed fusion of the domain wall.  We can also
derive the properties of the partition function from the pointed fusion
$(D-1)$-category of the symmetry twist domain walls, or equivalently, from the
group cocycle $\om_{D+1}$ that characterize the 't Hooft anomaly.  In this
paper, we derive explicitly the property of the 1+1D partition function of an
anomalous theory in terms of this group 3-cocycle $\om_3$.

%The organization of the paper is as follows. {\bf [$\cdots$]} In the second
%part of the paper, we will present some more general discussions.

%\subsection{Representations of mapping class group}

%Consider $1+1$ dimensional theories defined on a two dimensional surface
%$\Sigma$. Possible path integrals on the surface $Z(\Sigma)$ span a vector
%space $\calH(\Sigma)$.  A diffeomorphism $f: \Sigma\rightarrow \Sigma$
%corresponds to an automorphism $f_*: \calH(\Sigma)\rightarrow \calH(\Sigma)$,
%so that compositions of diffeomorphisms of $\Sigma$ corresponds to
%compositions of automorphisms on the vector space. In particular, large
%diffeomorphisms $f$, \ie equivalence classes of diffeomorphisms quotient by
%diffeomorphisms that are continuously connected to the identity, form the
%mapping class group $\MCG(\Sigma)$ of $\Sigma$. For example, when the two
%dimensional surface is a torus $\Sigma=\calT^2$, the mappling class group is
%$\MCG(\Sigma)=SL(2,\ZZ)$, also known as the modular group. The mapping class
%group acts as automorphisms of $\calH (\Sigma)$. Or rather, $\calH(\Sigma)$
%forms a (possibly projective) representation of $\MCG(\Sigma)$. 
 
%Consider conformal field theories on a two-dimensional surface $\Sigma$. It is
%a general belief that a CFT that describes the low energy physics of lattice
%models always has a partition function $Z(\Sigma)$ that is invariant under
%mapping class group transformations. That is, $Z (\Sigma)$ is a trivial
%one-dimensional representation of $\MCG(\Sigma)$. 

\section{1+1D theory with 't Hooft anomaly}

\subsection{Partition functions with symmetry twist}

For a system with a global discrete group $G$, it is important to understand
how the symmetry constrains the dynamics of the system. For example, for simple
discrete symmetries, we can determine whether the symmetry is anomalous by
considering the partition functions of the system under symmetry twists along
spatial and temporal directions. Consider a system on a spacetime torus twisted
by a symmetry defect $g\in G$ and acted by a symmetry transformation $h\in G$,
we denote its partition function as $Z_{g,h}$, with the condition that $gh=hg$,
since in a $g$-twisted sector, where local field satisfies the boundary
condition $\phi (L)=g\phi (0)$, the remaining global symmetry is $H=\{h |
hg=gh, h\in G\}$. In particular, the component with $(g,h)=(1,1)\in G\times G$
represents the untwisted/periodic boundary condition in both space and time
directions. We will denote $Z_{(g,h)}$ the partition function in a
symmetry-twisted basis. 

Under modular transformations, the twisted partition functions are in general
not invariant, but mapped into each other,
\begin{align}
\label{ZpropG}
 Z_{g',h'}(-1/\tau) &=  S_{(g',h'),(g,h)} Z_{g,h}(\tau),
\nonumber\\
 Z_{g',h'}(\tau+1) &=  T_{(g',h'),(g,h)} Z_{g,h}(\tau),
\nonumber\\
 Z_{g',h'}(\tau) &=  R_{(g',h'),(g,h)}(u) Z_{g,h}(\tau),
\nonumber\\
S_{(g',h'),(g,h)} &= \del_{(g',h'),(h^{-1},g)}s(g,h),
\nonumber\\
T_{(g',h'),(g,h)} &= \del_{(g',h'),(g,hg)}t(g,h),
\nonumber\\
R_{(g',h'),(g,h)}(u) &= \del_{(g',h'),(ugu^{-1},uhu^{-1})}r(u,g,h)
.
\end{align}
In the above, we also include the symmetry transformation that
does not change the shape of the torus.  The above  three transformations
generate a walk in the space of symmetry twist.  For some combinations of the
three transformations, the generated ``walk'' may form a loop.  In this case,
we obtain a property of the boundary partition function (\ie the partition
function for an anomalous system) \cite{HW1339}
\begin{align}
\label{Zloop}
Z_{g,h}(\tau)=\ee^{\ii \th} Z_{g,h}(\tau'). 
\end{align}
where the phase factor $\ee^{\ii \th}$ is a combination of the phase factors
$s(g,h)$, $t(g,h)$, and $r(u,g,h)$ in \eqn{ZpropG}.  The phase factor $\ee^{\ii
\th}$ is identified as the SPT invariant for the bulk in \Ref{HW1339}, and thus
can be used to identify the anomaly.

\subsection{Detecting anomaly via symmetry twists and spacetime transformations}

When the theory is anomaly-free for the symmetry $G$, the phases in (\ref{ZpropG}) is trivial,
\begin{align} 
s(h,g)=1, \quad t(h,g)=1 .
\end{align}
We may then gauge the symmetry and consider the partition function of the gauge theory. That is,
\begin{align}
Z^G=\frac{1}{|G|}\sum_{h,g\in G,\, [h,g]=0} Z_{h,g}.
\end{align}
Since the theory comes from gauging an anomalous free symmetry, it can have a lattice realization. It follows that the above partition function is invariant under the modular group.

When the theory is anomalous under $G$ (which is classified by group cohomological
classes: $\om_3 \in H^3[G, U(1)]$), the matrix elements $s(h,g)$ and $t(h,g)$
are a $U(1)$ phase. Later in this paper, we show how to determine the anomaly
$\om_3 \in H^3[G, U(1)]$ from the $S$ and $T$ matrix elements.  We will also
show how to determine $S$ and $T$ matrix elements from the anomaly $\om_3 \in
H^3[G, U(1)]$.

In the following, we consider the symmetry to be $\ZZ_2$ or $\ZZ_N$, and discuss how to detect the 't Hooft anomaly in systems whose low energy dynamics is described by rational conformal field theories. 

%\begin{align}
%\begin{tikzpicture}[scale=.5,baseline={([yshift=-.5ex]current bounding box.center)}]
%\draw (0,0) rectangle (2,2);
%\end{tikzpicture}
%\end{align}

\subsubsection{$\ZZ_2$ symmetry}

When $G=Z_2$, there is only one nontrivial element $-1\in Z_2$. There are four twisted partition functions, 
\begin{align}
&Z_{1,1},\quad Z_{1,-1},\quad Z_{-1,1}, \quad Z_{-1,-1}.
\end{align}
The first two describe the spectrum in the untwisted Hilbert space, and the last two describe that in the $\ZZ_2$-twisted Hilbert space.

Since $H^3(\ZZ_2, U(1))=\ZZ_2$, there are two types of low energy field theories under the $\ZZ_2$ symmetry, one is anomalous free, and the other is anomalous. As discussed above, the modular transformation properties of the above
4-component partition function are determined by the SPT invariant in one
higher dimension.  In the basis of twisted partition functions, the $S$ and $T$
matrix for anomaly-free theory is
\begin{align}
\label{Z2ST1}
S=\begin{pmatrix}
1 & 0 & 0 & 0 \\
0 & 0 & 1 & 0 \\
0 & 1 & 0 & 0 \\
0 & 0 & 0 & 1
\end{pmatrix},\quad 
T=\begin{pmatrix}
1 & 0 & 0 & 0 \\
0 & 1 & 0 & 0 \\
0 & 0 & 0 & 1 \\
0 & 0 & 1 & 0
\end{pmatrix}.
\end{align}
The $S$ and $T$ matrix for $ Z_2$ anomalous theory is
\begin{align}
\label{Z2ST2}
S=\begin{pmatrix}
1 & 0 & 0 & 0 \\
0 & 0 & 1 & 0 \\
0 & 1 & 0 & 0 \\
0 & 0 & 0 & -1
\end{pmatrix},\quad 
T=\begin{pmatrix}
1 & 0 & 0 & 0 \\
0 & 1 & 0 & 0 \\
0 & 0 & 0 & -1 \\
0 & 0 & 1 & 0
\end{pmatrix}.
\end{align}
See Appendix \ref{SPTinv} for a detailed calculation of the above $S,T$
matrices for the anomalous $Z_2$ symmetry.

Let us discuss the gapless phases described by $su(2)_k\otimes \overline{su(2)}_k$
conformal field theories (CFTs), as an example.  The $su(2)_k\otimes
\overline{su(2)}_k$ CFTs has $SU(2)_R\times SU(2)_L$ symmetry.  The $SU(2)_R$ and
$SU(2)_L$ symmetries both have perturbative anomalies.  Here we will
concentrate on the center of $SU(2)_L$ -- a $Z_2^L$ symmetry.
We would like to ask if such a $Z_2^L$ symmetry is anomalous.

The diagonal untwisted partition function for
$su(2)_k$ CFT is
\begin{align}
Z_{1,1}(\tau)=&\sum_{j=0,\frac{1}{2},\cdots,\frac{k}{2}} |\chi_j|^2.
\end{align}
With the $Z_2^L$ symmetry twist in the time direction, we have
\begin{align}
Z_{1,-1}=&\sum_{j=0,\frac{1}{2},\cdots,\frac{k}{2}} (-1)^{2j}|\chi_j|^2.
\label{zbh}
\end{align}
We have
\begin{align}
Z_{1,1}(\tau+1)= Z_{1,1}(\tau) , \ \ \ \ \ 
Z_{1,-1}(\tau+1)= Z_{1,-1}(\tau). 
\end{align}
Start from  $Z_{1,-1}$ and the modular transformation properties of the
$su(2)_k$ characters $\chi_j$, we can determine $Z_{-1,1}$ up to a phase factor
via
\begin{align}
Z_{-1,1}(-1/\tau)= \ee^{\ii \phi} Z_{1,-1}(\tau) .
\end{align}
Since the expansion of $Z_{-1,1}(\tau)$ in terms of the powers of $q=\ee^{\ii 2\pi
\tau}$ must have positive integer coefficients, this fixes the phase factor
$\ee^{\ii \phi}=1$.  Now, $Z_{-1,1}$ and $Z_{1,-1}$ are known. This allows us to find
that
\begin{align}
Z_{1,-1}(-1/\tau)= Z_{-1,1}(\tau).
\end{align}
  
%Then using 
%\begin{align}
%\zbx(\tau) = \ee^{\ii \th_T}\zbv(\tau+1)
%\end{align}
%we can determine $\zbx(\tau)$.  After obtaining $\zb,\zbh,\zbv,\zbx$, we can
%determine the $S,T$ matrices via \eqn{ZpropG}.
%However, the resulting $S,T$ contain an unknown phase factor
%$\ee^{\ii \th_T}$, which can be fixed by requiring

%We start with assuming $\zbh (\tau+1)= \zbh (\tau)$, since the global symmetry commute with $H$ and $P$. Now let us define 
%\begin{align}
%\zbx \,(\tau)= Z_2^L\, \zbv\, (\tau)
%\end{align}
%Generically, 

Next, we have, up to some phase factors (see \eqn{ZpropG}),
\begin{align}
&Z_{-1,1} (\tau+1)=\ee^{\ii \theta_{T,1}}\, Z_{-1,-1} (\tau),\nn\\
&Z_{-1,-1} (\tau+1)=\ee^{\ii \theta_{T,2}}\, Z_{-1,-1} (\tau),\nn\\
&Z_{-1,-1} \left(-\frac{1}{\tau}\right)=\ee^{\ii \theta_{S}}\, Z_{-1,-1} (\tau),
\end{align}
which implies that 
\begin{align}
S=\begin{pmatrix}
1 & 0 & 0 & 0 \\
0 & 0 & 1 & 0 \\
0 & 1 & 0 & 0 \\
0 & 0 & 0 & \ee^{\ii \theta_{S}}
\end{pmatrix}\,~~T=\begin{pmatrix}
1 & 0 & 0 & 0 \\
0 & 1 & 0 & 0 \\
0 & 0 & 0 & \ee^{\ii \theta_{T,1}} \\
0 & 0 & \ee^{\ii \theta_{T,2}} & 0
\end{pmatrix}.
\end{align}
Since, $S,T$ generate a representation of $SL(2,\Z)$, they satisfy
\begin{align}
 S^2=(ST)^3=C,\ \ \ C^2=1.
\end{align}
From $S^2=(ST)^3$, we obtain
\begin{align}
\begin{pmatrix}
1 & 0 & 0 & 0 \\
0 & 1 & 0 & 0 \\
0 & 0 & 1 & 0 \\
0 & 0 & 0 & \ee^{\ii 2\theta_{S}}
\end{pmatrix}=\begin{pmatrix}
1 & 0 & 0 & 0 \\
0 & \ee^{\ii \theta} & 0 & 0 \\
0 & 0 & \ee^{\ii \theta} & 0 \\
0 & 0 & 0 & \ee^{\ii \theta}
\end{pmatrix},
\end{align}
where $\theta= \theta_{T,1}+\theta_{T,2}+\theta_{S}$.  This gives us two
solutions, together with $T^2 $ given as follows,
\begin{enumerate}
\item $\theta_{S}=0\mod 2\pi,\ \theta_{T,1}=-\theta_{T,2}\mod 2\pi$, and $T^2$ is identity, which corresponds to anomaly-free $Z_2^L$;
\item $\theta_{S}=\pi\mod 2\pi,\ \theta_{T,1}=\pi-\theta_{T,2}\mod 2\pi$,
\begin{align}
T^2=\begin{pmatrix}
1 & 0 & 0 & 0 \\
0 & 1 & 0 & 0 \\
0 & 0 & -1 & 0 \\
0 & 0 & 0 & -1
\end{pmatrix} ,
\end{align}
which corresponds to anomalous $Z_2^L$.
\end{enumerate} 

In the above, we determined the possible modular transformation properties of
multi-component partition function with a $\Z_2$ symmetry.  Next, we can
construct multi-component partition functions of $\Z_2$ symmetric CFT's using
the symmetry twists, and examine their modular transformation properties.  We
find that
\begin{itemize}
\item $su(2)_1$ is $Z_2^L$ anomalous
\begin{align}
\label{Z2STZ}
Z_{1,1}=& |\chi^{su2_1}_0|^2+|\chi^{su2_1}_\fh|^2,\nn\\
Z_{1,-1}=&|\chi^{su2_1}_0|^2-|\chi^{su2_1}_\fh|^2,\nn\\
Z_{-1,1}=& \chi^{su2_1}_0\bar\chi^{su2_1}_{\fh}+\chi^{su2_1}_\fh\bar\chi^{su2_1}_0,\nn\\
Z_{-1,-1}=&-\ii \ee^{\ii \th_T}( \chi^{su2_1}_0\bar\chi^{su2_1}_{\fh}-\chi^{su2_1}_\fh\bar\chi^{su2_1}_0),
\end{align}
which transform as
\begin{align}
\label{Z2ST2th}
S=\begin{pmatrix}
1 & 0 & 0 & 0 \\
0 & 0 & 1 & 0 \\
0 & 1 & 0 & 0 \\
0 & 0 & 0 & -1
\end{pmatrix},\quad 
T=\begin{pmatrix}
1 & 0 & 0 & 0 \\
0 & 1 & 0 & 0 \\
0 & 0 & 0 & -\ee^{\ii\th_T} \\
0 & 0 & \ee^{-\ii\th_T} & 0
\end{pmatrix}.
\end{align}
\item $su(2)_2$ is $ Z_2^L$ anomaly-free
\begin{align}
\label{su2_2}
Z_{1,1}=& |\chi^{su2_2}_0|^2+|\chi^{su2_2}_\fh|^2+|\chi^{su2_2}_1|^2,\nn\\
Z_{1,-1}=&|\chi^{su2_2}_0|^2-|\chi^{su2_2}_\fh|^2+|\chi^{su2_2}_1|^2,\nn\\
Z_{-1,1}=& \chi^{su2_2}_0\bar\chi^{su2_2}_{1}+|\chi^{su2_2}_\fh|^2+ \chi^{su2_2}_1\bar\chi^{su2_2}_0,\nn\\
Z_{-1,-1}=&\ee^{\ii\th_T}(-\chi^{su2_2}_0\bar\chi^{su2_2}_{1}+|\chi^{su2_2}_\fh|^2- \chi^{su2_2}_1\bar\chi^{su2_2}_0),
\end{align}
which transform as
\begin{align}
\label{Z2ST1th}
S=\begin{pmatrix}
1 & 0 & 0 & 0 \\
0 & 0 & 1 & 0 \\
0 & 1 & 0 & 0 \\
0 & 0 & 0 & 1
\end{pmatrix},\quad 
T=\begin{pmatrix}
1 & 0 & 0 & 0 \\
0 & 1 & 0 & 0 \\
0 & 0 & 0 & \ee^{\ii\th_T} \\
0 & 0 & \ee^{-\ii\th_T} & 0
\end{pmatrix}.
\end{align}
\end{itemize}
In general $su(2)_k$ is $Z_2^L$ anomalous if $k$ is odd
and $Z_2^L$ anomaly-free if $k$ is even. 
%{\bf [ To elaborate: While in the CFT as an effective field theory, one enjoys
%a natural choice of $e^{\ii\theta_T}$, defined by $Z_{a_x, a_t}= TZ_{a_x, a_x
%a_t}$. ]}

In addition to the $Z_2^L$ anomaly, the above calculation also reveals a mixed
anomaly between $Z_2^L$ and $SU(2)$ (the diagonal part of $SU(2)_R\times
SU(2)_L$) for odd $k$.  This is because the $Z_2^L$ twist in the space
direction induces half-integer spin when $k$ is odd as one can see from the
partition function $Z_{-1,1}$. More precisely, there is spin-$\frac{k}{2}$ excitation in the $Z_2^L$ twisted sector of the $su(2)_k$ CFT. When $k$ is odd, this excitation carries half-integer spin.
 In contrast, for even $k$, there is no
such a mixed anomaly between $Z_2^L$ and $SU(2)$. 
The $su(2)_1$ CFT is the boundary of $\ZZ_2$-symmetry protected topological
phase. The lattice model is the CZX model first proposed in \Ref{CLW1141}. 
%{\bf [Show there is no $\ZZ_2$ symmetric perturbation we can add to get the
%edge]}
In Section \ref{spin12}, we discuss a physical realization of the $Z_2^L$
anomalous symmetry by a spin-$\frac12$ chain with translation symmetry.
Through this model, we can see how $Z_2^L$ anomalous symmetry protect
gaplessness if we do not break the $SU(2)$ spin rotation and translation
symmetry.

\subsubsection{$\ZZ_N$ symmetry}

We start with a CFT with $\Z_N=\langle a|a^N=1\rangle $ global symmetry. Under
symmetry twist, there are in total $N^2$ sectors $Z_{a^n, a^m}$. The $T$ matrix
in the symmetry twist basis is (see Appendix \ref{SPTinv}).
\begin{align}
T_{(a^{n'},a^{m'})(a^n,a^m)}= \delta_{n',n}\delta_{m',m+1}\,\omega_q (a^n, a^m, a^n),
\end{align}
where $\omega_q$ is the $3$-cocycle of $\ZZ_N$ of the class $q\in H^3(\ZZ_N, U(1))=\ZZ_N$. The $S$ matrix is
\begin{align}
S_{(a^{n'},a^{m'})(a^n,a^m)}=\delta_{n',m}\delta_{m',N-n}\, \omega_q (a^m, a^n, a^{N-n}).
\end{align}
When $\ZZ_N$ has 't Hooft anomaly, the $S$ and $T$ matrices are not purely permutation on different symmetry twisted sectors $Z_{a^n,a^m}$, but decorated with 
phase factors from $3$-cocycles.  

In general, the phase factors in the $S,T$ matrices in general depend on the
arbitrary choices of the phase factors of the basis vector. We find the
following matrix element does not depend on the choices of the phase factors of
the basis vector
\begin{align}
\langle a_x=a,a_t=1 | \calT^N |a_x=a,a_t=1\rangle=\ee^{\ii \theta_T}.
\end{align}
and is directly related to the cohomology class of cocycles. The 't Hooft anomaly of $ Z_N$ symmetry is classified by $H^3(\ZZ_N, U(1))=\ZZ_N$. A class $[q]\in H^3(\ZZ_N, U(1))$, is given by a $3$-cocycle. To be concrete, let us choose a normalized $3$-cocycle. For $a^l, a^m, a^n\in \ZZ_N$, they are
\begin{align}
\omega_q (a^l,a^m,a^n)= \ee^{\ii \frac{2\pi}{N^2}q l\left(m+n-\langle m+n\rangle_N\right)},
\end{align}
where $\langle x\rangle_N$ denote $x\mod N$. %The $S$ and $T$ matrices reduce to 
%\begin{align}
%S_{(n',m')(n,m)}=\delta_{n',m}\delta_{m',N-n}\, \ee^{\ii \frac{2\pi}{N}qm}
%\end{align}
We find that (shown in Appendix \ref{TNZN}),
\begin{align}
\ee^{\ii \theta_T}= \ee^{\ii \frac{2\pi}{N}q}
%\prod_{k=0}^{N-1}\omega_q (a,a^k,a)
\end{align}
%More explicitly, 
%\begin{align}
%T^N Z_{a^{N-1},1}= \ee^{-\ii \frac{2\pi}{N}q} Z_{a^{N-1},1}
%T^N Z_{a,1}= \ee^{\ii \frac{2\pi}{N}q} Z_{a,1}
%\end{align}
can measure the anomaly of $\ZZ_N$ symmetry.\cite{HW1339,Chang:2018iay}

Let us consider the example of  $\ZZ_3$ symmetry in $SU(3)_1$ CFT. In this CFT, except the vacuum $\hat{0}=[1,0,0]$, in terms of the Dynkin label, there is the fundamental representation $\hat{1}=[0,1,0]$ and anti-fundamental representation $\hat{2}=[0,0,1]$, under the $SU(3)$ level $1$ current algebra. In the basis of these three states, 
the generator $a$ of the $\ZZ_3$ symmetry, written in the basis of the three states is
\begin{align}
a=\begin{pmatrix}
1 & 0 & 0 \\ 0 & \omega & 0 \\ 0 & 0 & \omega^2
\end{pmatrix},
\end{align}
where $\omega=\ee^{\ii \frac{2\pi}{3}}$. 
There is an automorphism that is related to the action of the $ \ZZ_3$ symmetry by a $S$ transformation
\begin{align}
S A S^\dagger= a.
\end{align}
In $SU(3)_1$, the $S$ and $T$ matrix are 
\begin{align}
S=\frac{1}{\sqrt{3}}\begin{pmatrix}
1 & 1 & 1 \\ 1 & \omega & \omega^2 \\ 1 & \omega^2 & \omega
\end{pmatrix},\quad T=\begin{pmatrix}
1 & 0 & 0 \\ 0 & \omega & 0 \\ 0 & 0 & \omega
\end{pmatrix}.
\end{align}
It follows that the automorphism is given by
\begin{align}
A=\begin{pmatrix}
0 & 0 & 1 \\ 1 & 0 & 0 \\ 0 & 1 & 0
\end{pmatrix}.
\end{align}
Start with the untwisted partition function, (from now on we lighten the notation by dropping the hat of representation $\hat \lambda$,) 
\begin{align}
&Z_{1,1}=\sum_{\mu, \lambda} \bar\chi_{\mu} \calN[0,0]_{\mu\lambda}\chi_{\lambda},~~\calN[0,0]_{\mu\lambda}= \delta_{\mu,\lambda}.
\end{align}

The twisted partition funciton $Z_{n,m}$, with $m$ times the symmetry twist in the spatial direction twisted and $m$ times the symmetry action at fixed time slices, is obtained by the action of $A$ and $a$, 
\begin{align}
&Z_{n,m}=\sum_{\mu \lambda} \bar \chi_{\mu} \calN [ n,m]_{\mu,\lambda}\chi_\lambda,~~ \calN[n,m]=a^m A^n .
\end{align}
The actions of $S$ and $T$ are
\begin{align}
T^\dagger a^ mA^n T=&\ee^{\ii \theta[n,m]}a^{n+m}A^n ,\nn\\
S^\dagger a^n A^mS=&\ee^{\ii \theta_S[n,m]}A^n a^{-m}.
\end{align}
The anomaly is given by
\begin{align}
(T^\dagger)^3AT^3=\ee^{\ii \theta_T}A,~~e^{\ii\theta_T}=e^{\ii \sum_{m=0}^{2}\theta [1,m]},\label{TN}
\end{align}
where we have used $a^3=1$. Then in $SU(3)_1$, we find $e^{\ii \theta_T}=1$, the $\ZZ_3$ symmetry is non-anomalous.

% For any discrete finite unitary group $G$, the symmetry twisted partition functions are generated from $Z_{1,h}$, $h$ runs through a representative in each conjugacy class of $G$, and $Z_{h,1}=SZ_{1,h}$. Therefore, $T^{n_h}$ can detect the anomaly in each conjugacy class $[h]$, where $n_h$ is the order of $h$.

\iffalse
The anomaly can be measured by a local gauge transformation in the $g$-twisted sector. 

\begin{figure}
\includegraphics[scale=.7]{figure/Ftwistsec}
\caption{A local gauge transformation in the $g$-twisted sector. $g$ is a generator of the group $G$ and $g^n=1$. }
\end{figure}
\fi

As a second example, we can consider $su(N)_1$ CFT. It describes $N$-flavor of complex free fermions $\psi_i$ in one dimension. It can also describe the $SU(N)$ spin chain with anti-ferromagnetic Heisenberg interaction. The CFT has the symmetry $SU(N)_R\times SU(N)_L$. The subgroup $\ZZ_N^L\subset SU(N)_L$ acts on the fundamental representation of $SU(N)$ as follows,
\begin{align}
\ZZ_N^L: |j_L\rangle \rightarrow \ee^{\ii 2\pi\frac{m}{N}}|j_L\rangle , ~|j_R\rangle \rightarrow |j_R\rangle ,~~j_{L/R}=1,\cdots,N.
\end{align}
Through a similar analysis as described above, we can find the symmetry $\ZZ_N^L\subset SU(N)_L$ is anomalous when $N$ is even.

\section{Anomalies in spin-1/2 Heisenberg chain}
\label{spin12}

After determining the $Z_2^L$ anomaly and the $Z_2^L$-$SO(3)$ mixed anomaly in
the $su2_k\otimes \overline{su2}_k$ CFT, let us consider some lattice models
that can realize the $Z_2^L$ symmetry.  We first consider a spin-1/2
$J_1$-$J_2$ Heisenberg
model\cite{haldane1982spontaneous,okamoto1992fluid,eggert1996numerical} on a 1D
ring of $L$ sites:
\begin{align}
 H = \sum_i \v \si_i \cdot \v\si_{i+1} + 0.25\, \v \si_i \cdot \v\si_{i+2} .
\end{align}
Even without the second term, the model realizes a gapless phase described by a
CFT.  With the second term, the model realizes the same gapless phase described
by the same CFT. Nevertheless, the second term makes the lattice model closer to the CFT
fixed point, \ie the strength of marginally irrelevant operator $J_L\cdot J_R$
in the Hamiltonian is tuned to be small.

\begin{figure}[tb]
\begin{center}
\includegraphics[scale=0.35]{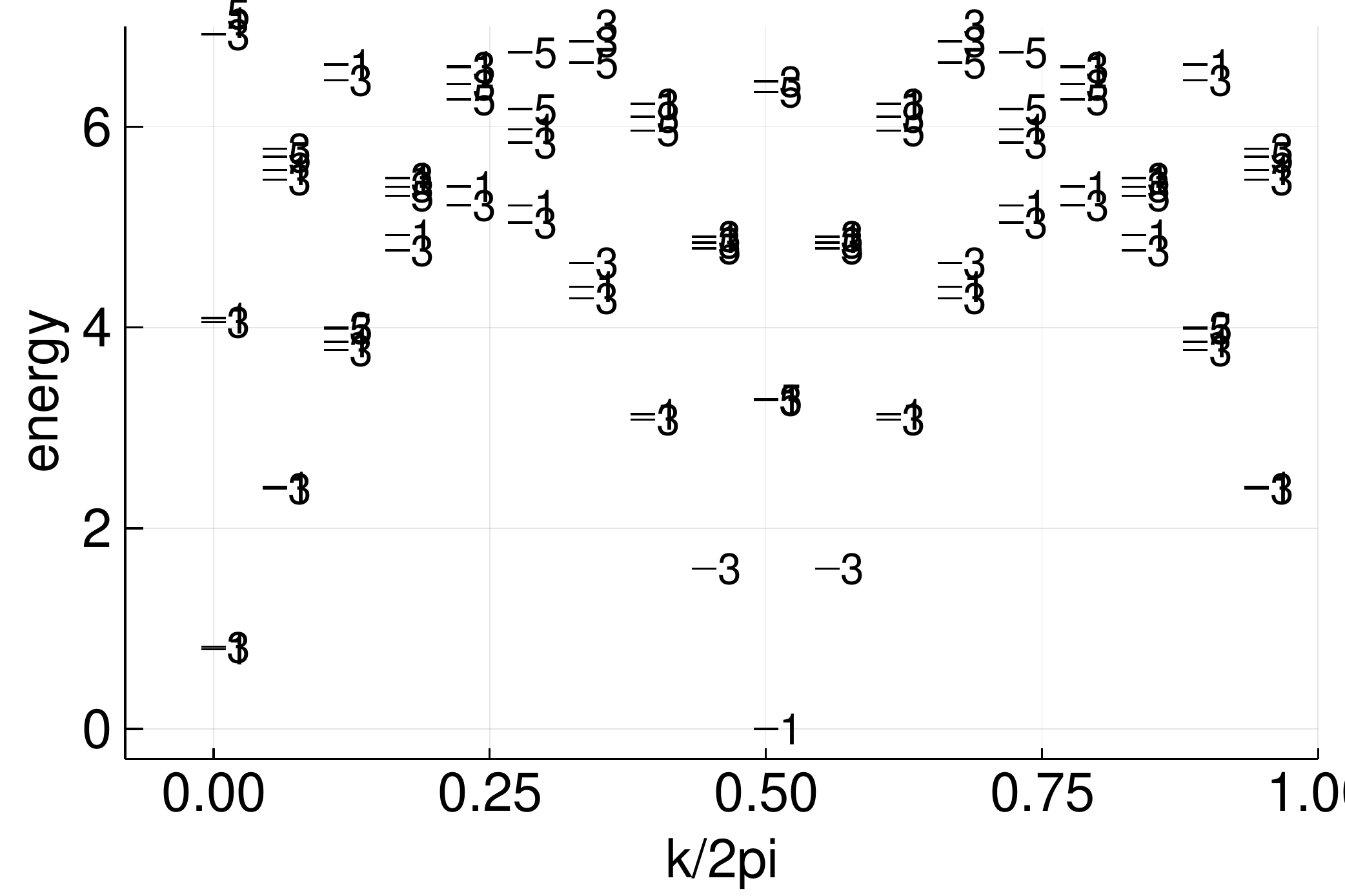} \end{center}
%Fig. 1
\caption{ The many-body energy spectrum of $J_1$-$J_2$ Heisenberg model on a
ring of 18 sites, where the ground state energy is shifted to 0.  The
horizontal axis is the crystal momentum $k/2\pi$.  The numbers by the bars
indicate the $SU(2)$ multiplets (the degeneracies).  The energy levels at
energy $\approx 3.2$ and $k = \pi$ contain $SU(2)$ multiplets 1, 3, and 5.  The
energy levels at energy $\approx 0.8$ and $k = 0$ contain $SU(2)$ multiplets 1
and 3 (which correspond to spin $\frac12 \otimes \frac12$).
}
\label{HeisCFT18}
\end{figure}

\begin{figure}[tb]
\begin{center}
\includegraphics[scale=0.35]{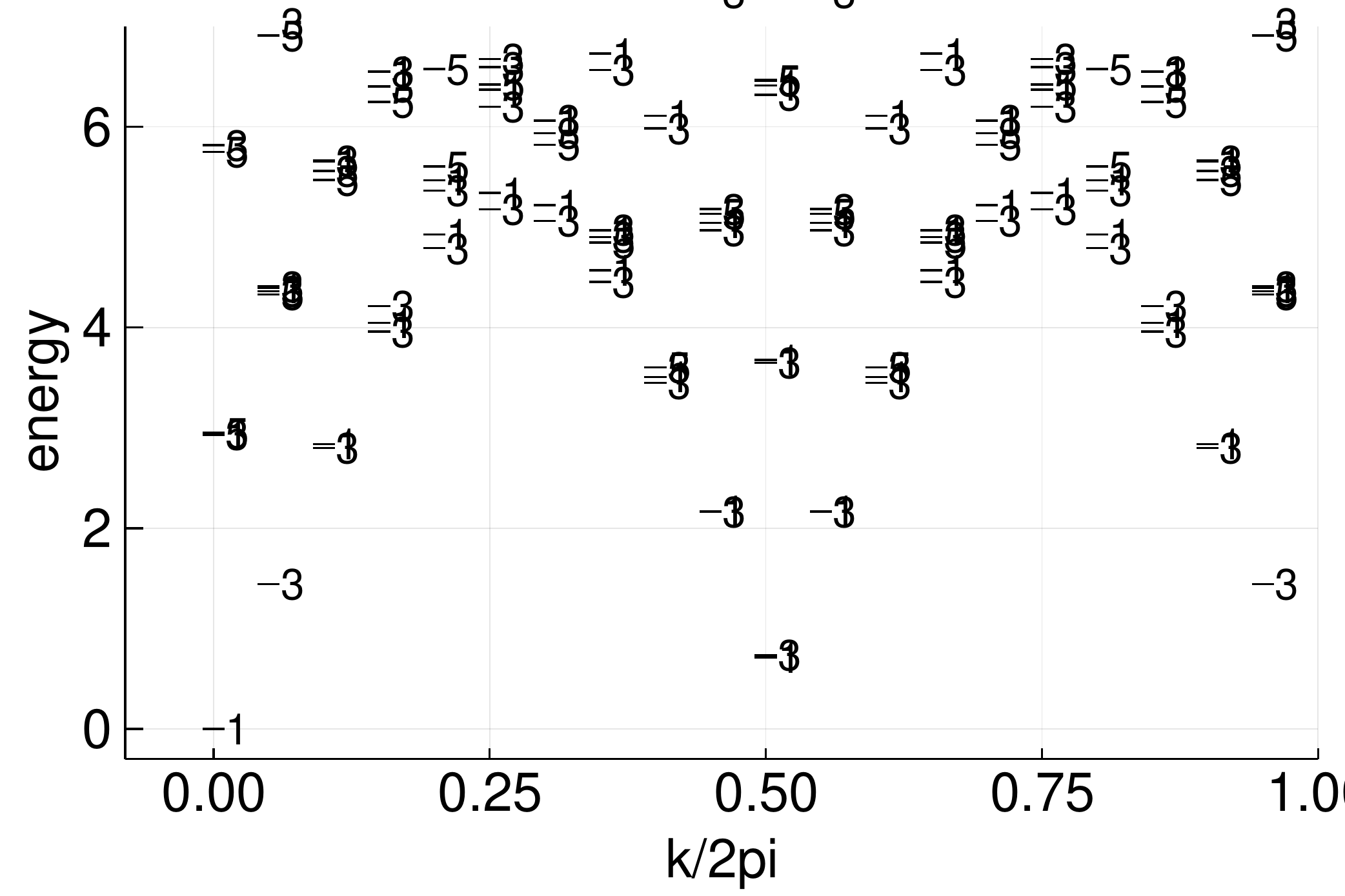} \end{center}
%Fig. 1
\caption{ The many-body energy spectrum of $J_1$-$J_2$ Heisenberg model on a
ring of 20 sites, where the ground state energy is shifted to 0.  The horizontal axis is the crystal momentum $k/2\pi$.  The
numbers by the bars indicate the $SU(2)$ multiplets (the degeneracies).  The
energy levels at energy $\approx 3.0$ and $k = 0$ contain $SU(2)$ multiplets 1,
3, and 5.  The energy levels at energy $\approx 0.8$ and $k = \pi$ contain
$SU(2)$ multiplets 1 and 3 (which correspond to spin $\frac12 \otimes \frac12$).
}
\label{HeisCFT20}
\end{figure}

On a ring of 20 sites, the spectrum of $H$ is given by Fig. \ref{HeisCFT20}.
The low energy spectrum contains a sector $|\chi^{su2_1}_0|^2$ at momentum
$k\sim 0$, and a sector $|\chi^{su2_1}_\frac12|^2$ at momentum $k\sim \pi$.
This implies that the operator $V_\frac12 \bar V_\frac12$ (that generates the
sector $|\chi^{su2_1}_\frac12|^2$ from the ground state in the sector
$|\chi^{su2_1}_0|^2$) carries a momentum $\pi$.  On a ring of 18 sites, the
spectrum of $H$ is given by Fig. \ref{HeisCFT18}.  In this case, the sector
$|\chi^{su2_1}_0|^2$ is at momentum $k\sim \pi$, and a sector
$|\chi^{su2_1}_\frac12|^2$ at momentum $k\sim 0$.  Let us define a pseudo
translation by one lattice spacing as 
\begin{align}
\t T = \ee^{\ii \pi N/2 } T, 
\end{align}
where $T$ is the usual translation by one lattice spacing and $N$ is the number
of sites.  We see that the pseudo momentum is always $\t k \sim 0$ for  the
sector $|\chi^{su2_1}_0|^2$, and $\t k \sim \pi$ for the sector
$|\chi^{su2_1}_{\frac12}|^2$.

From $Z_{1,-1}=|\chi^{su2_1}_0|^2-|\chi^{su2_1}_\fh|^2$, we also see that the
operator $V_\frac12 \bar V_\frac12$ carries a $Z_2^L$ charge $-1$, since it
contains $\bar V_\frac12$.  This allows us to identify $Z_2^L$ with the pseudo
translation by one lattice spacing.  As a result, the $Z_2^L$ symmetry twist is realized in the system on a ring with an odd number of sites.

\begin{figure}[tb]
\begin{center}
\includegraphics[scale=0.35]{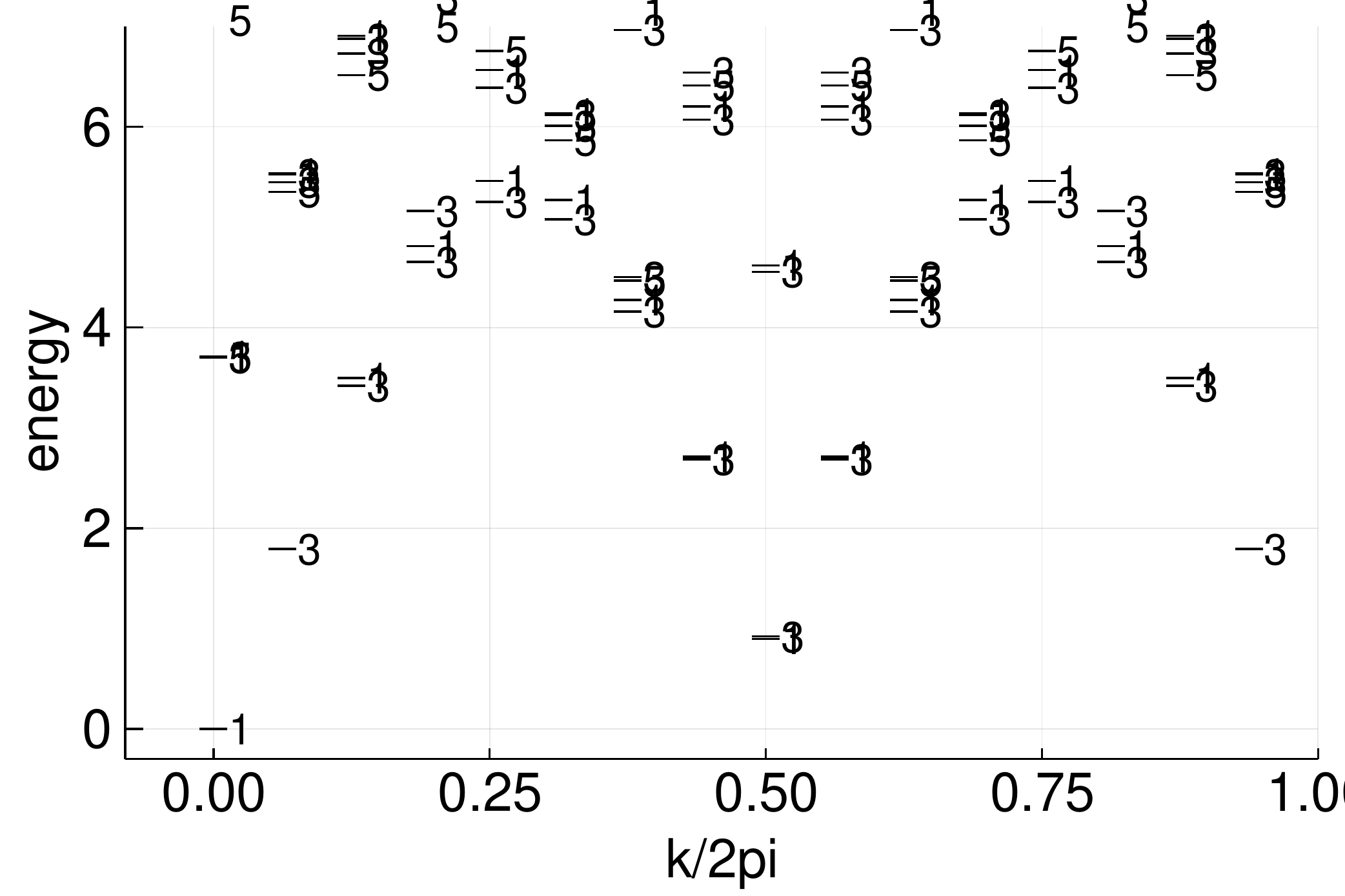} \end{center}
%Fig. 1
\caption{ The many-body energy spectrum of $J_1$-$J_2$ Heisenberg model on a
ring of 16 sites, where the ground state energy is shifted to 0.  The
horizontal axis is the crystal momentum $k/2\pi$.  The numbers by the bars
indicate the $SU(2)$ multiplets (the degeneracies).  The energy levels at
energy $\approx 3.2$ and $k = \pi$ contain $SU(2)$ multiplets 1, 3, and 5.  The
energy levels at energy $\approx 0.8$ and $k = 0$ contain $SU(2)$ multiplets 1
and 3 (which correspond to spin $\frac12 \otimes \frac12$).
}
\label{HeisCFT16}
\end{figure}

\begin{figure}[tb]
\begin{center}
\includegraphics[scale=0.35]{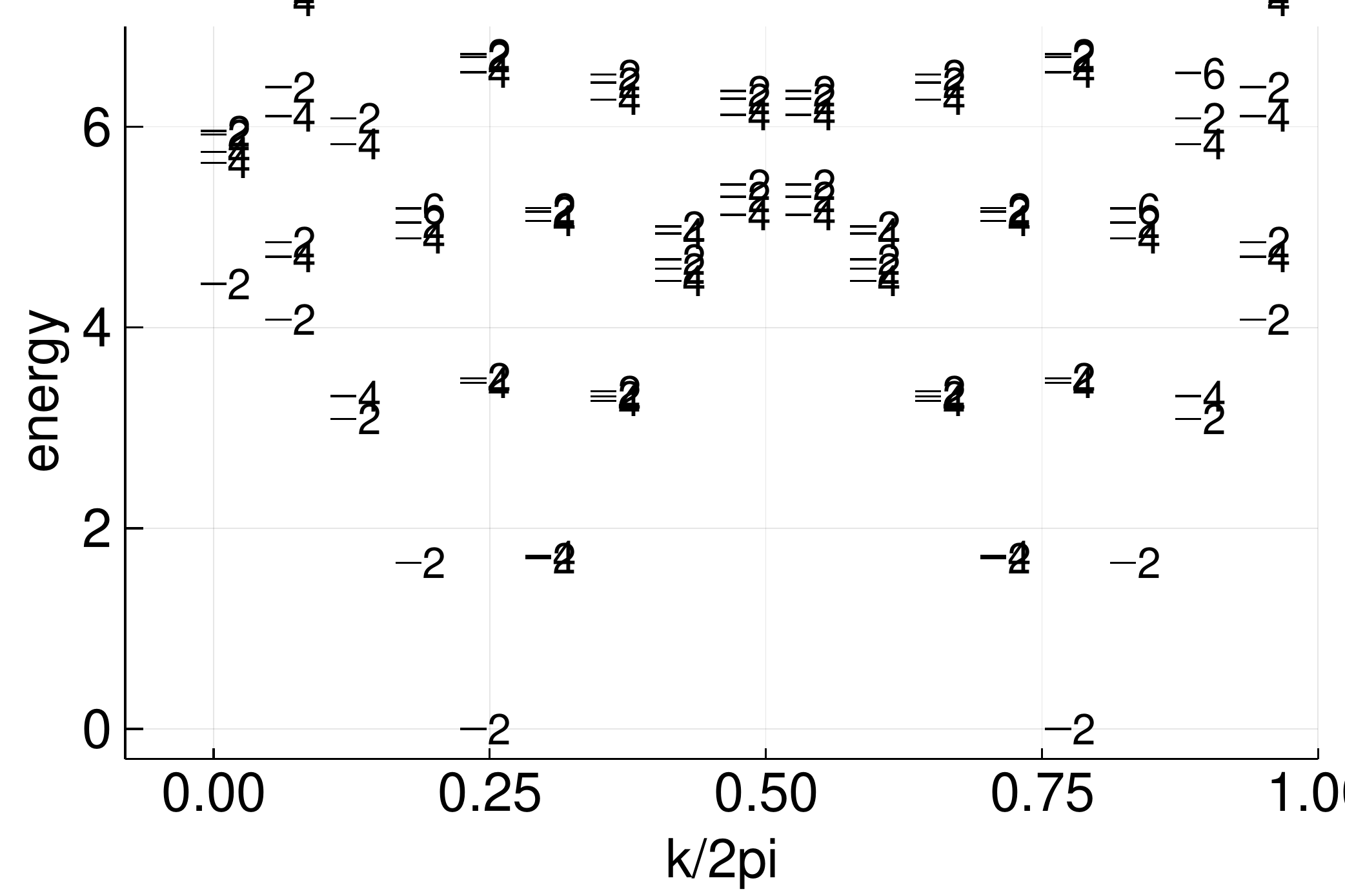} \end{center}
%Fig. 1
\caption{ The many-body energy spectrum of $J_1$-$J_2$ Heisenberg model on a
ring of 17 
sites, where the ground state energy is shifted to 0.  The horizontal axis is the crystal momentum $k/2\pi$.  The
numbers by the bars indicate the $SU(2)$ multiplets (the degeneracies).  
}
\label{HeisCFT17}
\end{figure}

\begin{figure}[tb]
\begin{center}
\includegraphics[scale=0.35]{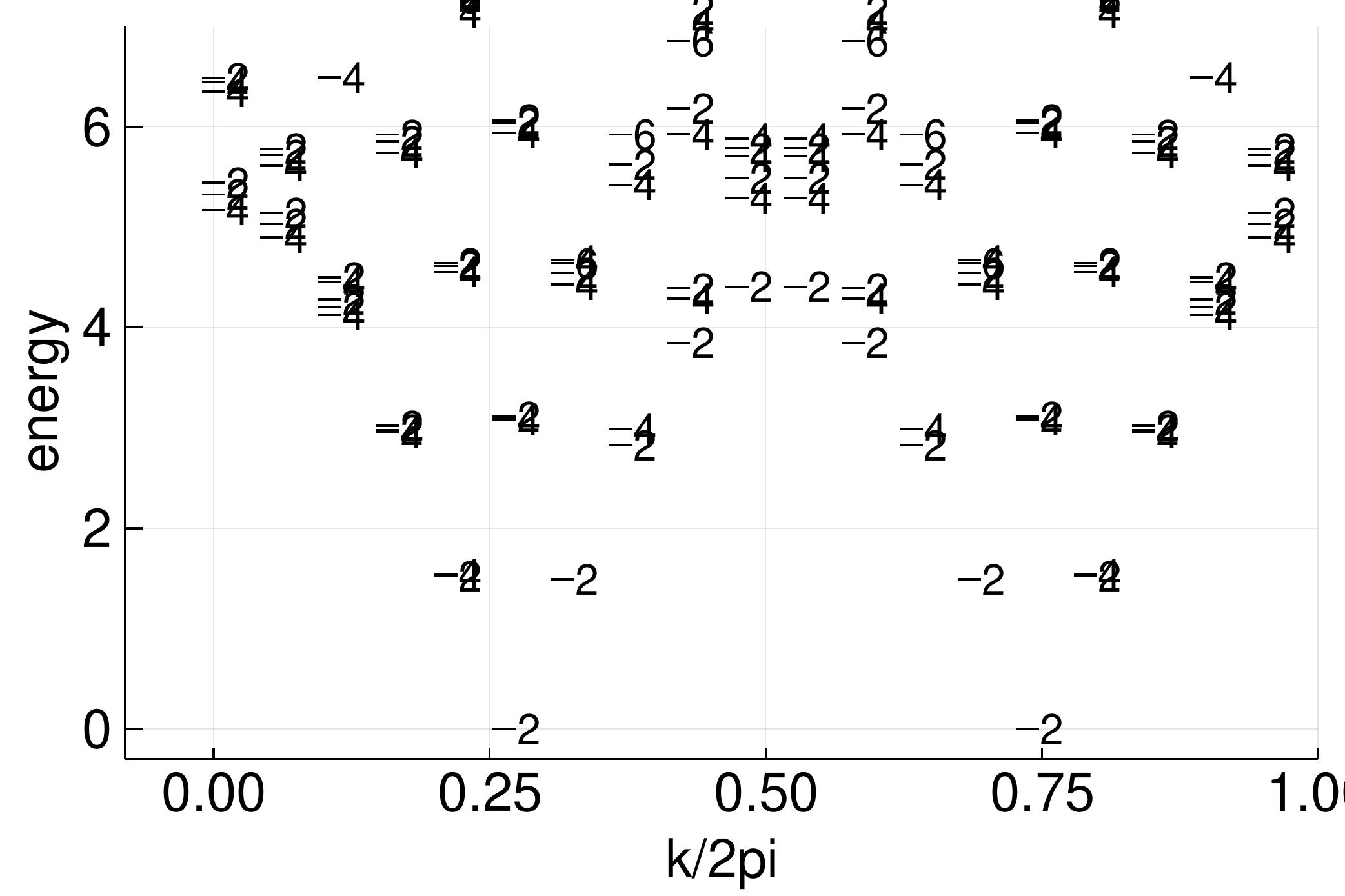} \end{center}
%Fig. 1
\caption{ The many-body energy spectrum of $J_1$-$J_2$ Heisenberg model on a
ring of 19 sites, where the ground state energy is shifted to 0.  The horizontal axis is the crystal momentum $k/2\pi$.  The
numbers by the bars indicate the $SU(2)$ multiplets (the degeneracies).  
}
\label{HeisCFT19}
\end{figure}

\begin{figure}[tb]
\begin{center}
\includegraphics[scale=0.35]{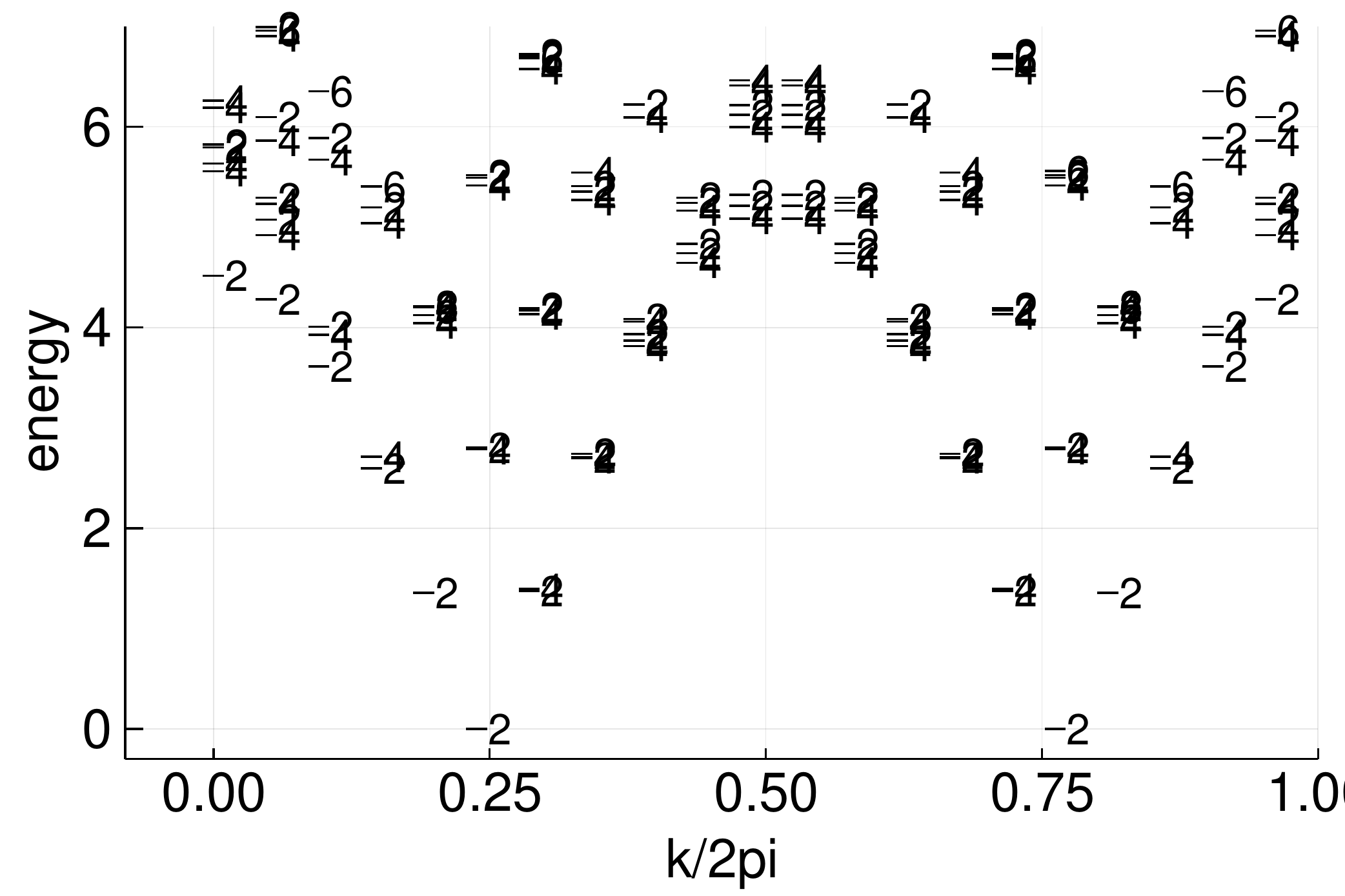} \end{center}
%Fig. 1
\caption{ The many-body energy spectrum of $J_1$-$J_2$ Heisenberg model on a
ring of 21 sites.  The horizontal axis is the crystal momentum $k/2\pi$.  The
numbers by the bars indicate the $SU(2)$ multiplets (the degeneracies).  
}
\label{HeisCFT21}
\end{figure}

On a ring of 21 sites, the spectrum of $H$ is given by Fig. \ref{HeisCFT21}. The lowest excitations in each tower are a spin-doublet. They are described by $V_0\bar V_{\frac{1}{2}}$ and $V_\frac{1}{2}\bar V_{0}$. We see that the  $Z_2^L$ symmetry twist gives rise to half-integer spins (which
is not surprising, since the ring now contains an odd number of spin-$1/2$'s).  Thus the
spin-1/2 Heisenberg chain has the mixed anomaly between $Z_2^L$ (translation)
and $SO(3)$ (spin rotation).  We also see that the low energy sectors with the
$Z_2^L$ symmetry twist carry pseudo momenta $\t k \sim 0,\pi$.  This
corresponds to $\th_T=\pi/2$ in \eqn{Z2STZ}.  Our numerical calculations (see
Fig.  \ref{HeisCFT16} -- \ref{HeisCFT21}) indicate that the partition functions
of spin-1/2 chain is given by \eqn{Z2STZ}, which transform as \eqn{Z2ST2th}.
Thus the  $Z_2^L$ symmetry has a 't Hooft anomaly.

However, we would like to stress that $Z_2^L$ is only a low energy emergent symmetry.
The actual symmetry is the $\Z$ translation symmetry.  For spin-1/2 Heisenberg
chain, the pseudo translation by 2 lattice spacing becomes the identity action at low
energies, meaning that $\Z$ translation reduces to the emergent low energy $Z_2^L$
symmetry.  This realization is important since $\Z_2^L$ has 't Hooft anomaly, while the $\ZZ$ symmetry where $\Z_2^L$ is embedded is not anomalous.  This is in accord with the fact that $\cH^3(Z_2^L,\R/\Z)=\Z_2$ while $\cH^3(\Z,\R/\Z)=0$. Therefore, the 't Hooft
anomaly for $Z_2^L$ is only an emergent anomaly at low energies for the
Heisenberg chain.  The spin-1/2 chain with translation symmetry actually does not
have the $Z_2^L$ 't Hooft anomaly.  The emergent anomaly has been studied in
\Ref{MT170707686}.  The emergent $Z_2^L$ 't Hooft anomaly implies that, near
the gapless state of the spin-1/2 Heisenberg chain, even the perturbations that
break the $SO(3)$ spin rotation symmetry cannot put the system in a gaped phase
if the translation symmetry is not broken.\cite{MT170707686}

But the mixed anomaly between $Z_2^L$ and $SO(3)$ remains to be a mixed anomaly
after we promote $Z_2^L$ to $\Z$ translation symmetry.  (Note that the spin-1/2
chain has a $SO(3)$ symmetry since on each site, we only have half-integer
spins, rather than both half-integer and integer spins.  If there were both
half-integer and integer spins on each site, the symmetry would be $SU(2)$.  If
there were only integer spins on each site, the symmetry would also be
$SO(3)$.) Thus, the spin-1/2 Heisenberg chain has an exact mixed anomaly
between $Z_2^L$ and $SO(3)$, or more precisely, an exact mixed anomaly between
$\Z$ translation and $SO(3)$ spin rotation. See \Ref{YO180506885} for a more
complete discussion.  As a result, the symmetric gapless phase of spin-1/2
Heisenberg chain must be described by $su2_k\otimes \overline{su2}_k$ with $k=$
odd, if the $\Z$ translation and $SO(3)$ spin rotation symmetries are not
broken.

Since the spin-1 chain can be obtained from the stacking of two spin-1/2
chains, and since the stacking cancels the exact mixed anomaly between $\Z$ and
$SO(3)$, The spin-1 chain does not have the mixed anomaly.  Its symmetric
gapless phase must be described by $su2_k\otimes \overline{su2}_k$ with $k=$
even.  This is the derivation of the symmetry protected gaplessness in
\Ref{FO150307292} via the mixed anomaly of $\Z$ and $SO(3)$.\cite{YO180506885}

In the above 1D lattice model, the $Z_2^L$ symmetry in the CFT is realized as
an emergent symmetry (\ie as a quotient ground of lattice translation $\Z$).
Do we have a lattice model where the $Z_2^L$ symmetry is realized as the exact
lattice on-site symmetry?  The answer is yes. But since $Z_2^L$ is anomalous,
the lattice realization must be in one higher dimension.  In fact, we can start
with a 2+1D $Z_2$ SPT state, the   $su2_1\otimes \overline{su2}_1$ CFT can be
realized at the boundary of the 2+1D $Z_2$ SPT state, with $Z_2^L$ symmetry in
the CFT is realized as the $Z_2$ symmetry of the 2+1D $Z_2$ SPT state.

\section{The non-invertible gravitational anomaly of the $1+1$D boundary of $2+1D$ topological order }

In the above, we have discussed 1+1D 't Hooft anomalies and their reflections
in modular transformation properties of the anomalous partition functions with
symmetry twist. We can view the non-on-site global symmetry as the boundary
effective symmetry of an on-site global symmetry in a 2+1D SPT state.  We can
then gauge the 2+1D on-site global symmetry to obtain 1+1D theories with a
non-invertible gravitational anomaly, which is characterized by 2+1D
topological order (\ie the gauged 2+1D SPT order).  The boundary of the topological order is described also in terms of a vector of partition functions yet in the basis of quasiparticles that are present in the bulk. 
The phase on the boundary of the topological order always needs to satisfy the following matching conditions,\cite{JW190513279}
\begin{align}
\label{ZST}
 Z^a(\tau+1)=T_{ab}Z^b(\tau),\ \ \
 Z^a(-1/\tau)=S_{ab}Z^b(\tau) .
\end{align}
In this section, we will show, through examples, how the theory with a global symmetry is related to the theory with non-invertible gravitational
anomaly, obtained via gauging on-site global symmetry in one-higher-dimensional
bulk. We will see that the boundary theory before gauging is described by a vector of partition functions indexed by symmetry twists, the boundary theory after gauging is described by a vector partition functions indexed by bulk anyons. Nevertheless, both partition functions are built from the same set of data - characters of conformal field theories and $3$-cocycles in $H^3(G, U(1))$, the two vectors are related by a matrix transformation. 

\subsection{$\ZZ_2$ topological order}

\begin{figure}
\includegraphics[scale=.7]{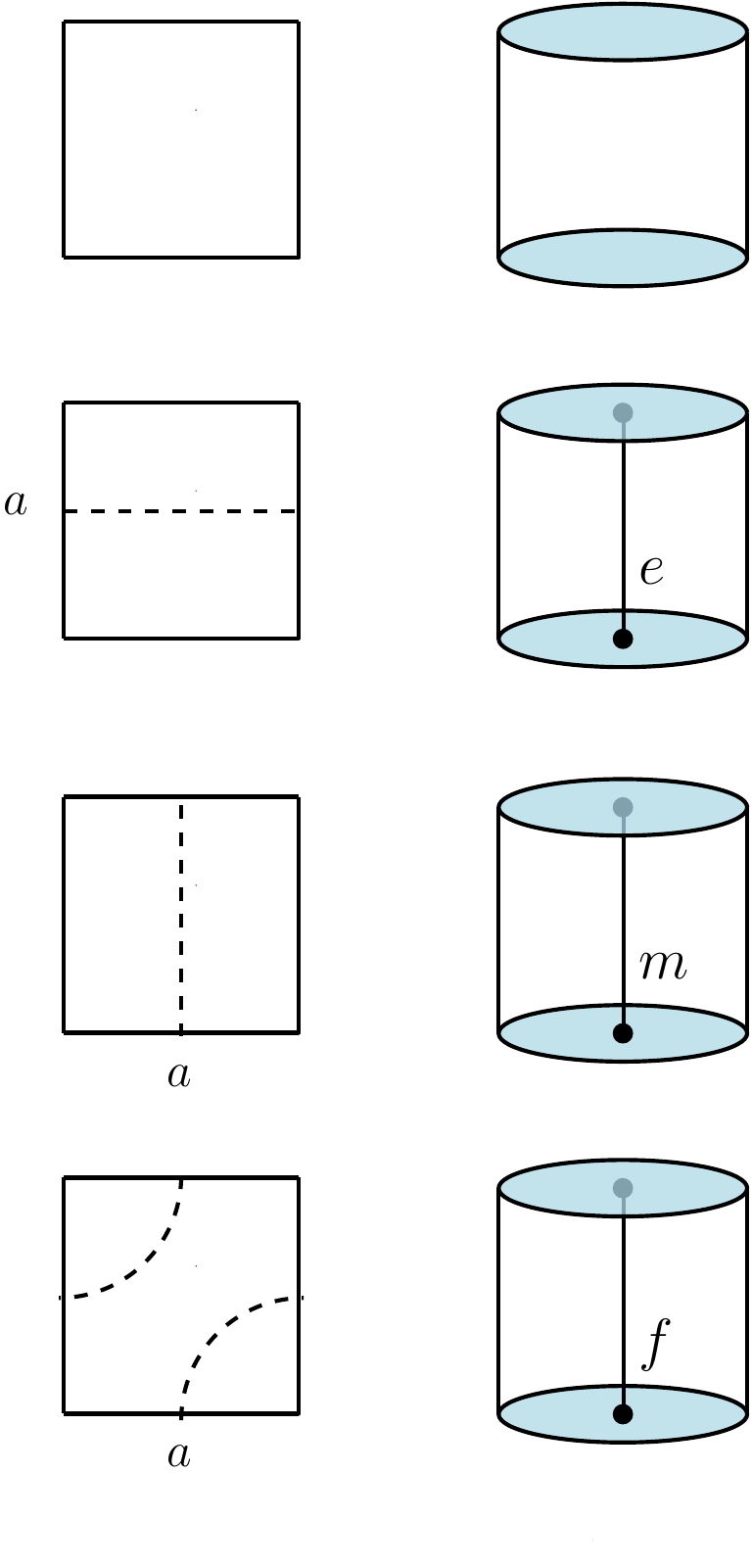}
\caption{Transformation between (a) boundaries of trivial $\ZZ_2$ SPT (2+1d Ising model) and (b) that of $\ZZ_2$ topological order.  Transformation from (a) to (b) corresponds to gauging the $\ZZ_2$ global symmetry in the bulk. }
\label{fig:Z2tfm}
\end{figure}

The 1+1D Ising model in a transverse field at the critical point has an anomaly-free $Z_2$ global
symmetry, which is described by the following 4 partition functions with
different $Z_2$ symmetry twist:
\begin{align}
\label{IsingZ1}
 Z_{1,1}(\tau) &= |\chi^\text{Is}_0|^2+|\chi^\text{Is}_\frac{1}{2}|^2 +|\chi^\text{Is}_\frac{1}{16}|^2, \nonumber
  \\ 
 Z_{1,-1}(\tau) &= |\chi^\text{Is}_0|^2+|\chi^\text{Is}_\frac{1}{2}|^2 -|\chi^\text{Is}_\frac{1}{16}|^2,
 \nonumber \\ 
Z_{-1,1}(\tau) &= 
|\chi^\text{Is}_\frac{1}{16}|^2
+\chi^\text{Is}_0 \bar \chi^\text{Is}_\frac{1}{2} + \chi^\text{Is}_\frac12 \bar \chi^\text{Is}_0,
 \nonumber \\ 
Z_{-1,-1}(\tau) &= |\chi^\text{Is}_\frac{1}{16}|^2 -
\chi^\text{Is}_0 \bar \chi^\text{Is}_\frac{1}{2} - \chi^\text{Is}_\frac12 \bar \chi^\text{Is}_0.
\end{align}
Under modular transformation, they transform as eqn. (\ref{Z2ST1}).
On the other hand, we may gauge the $2+1D$ system with a boundary, the bulk becomes $Z_2$ topological order which has four types of quasiparticles $\one, e,m$ and $f$. Correspondingly, the boundary partition function is in the quasiparticle basis. This partition function, is related to the one before gauging, via a basis transformation,
\begin{align}
\label{eqn:Z2tfn}
\begin{pmatrix}
Z^{\one} \\ Z^e \\ Z^m \\ Z^{f}
\end{pmatrix}
=\frac{1}{2}\begin{pmatrix}
1 & 1 &  0&0  \\
1 & -1 & 0&0 \\
 0& 0& 1 & 1 \\
 0& 0& 1 & -1 
\end{pmatrix}\begin{pmatrix}
Z_{1,1} \\ Z_{1,-1} \\ Z_{-1,1} \\ Z_{-1,-1}
\end{pmatrix}.
\end{align}
The correspondence of two vectors of partition functions, is illustrated in Fig.\ref{fig:Z2tfm}.

The partition functions in the quasiparticle basis\cite{JW190513279} are given
by
\begin{align}
\label{IsingZ2}
\begin{split}
 Z^{\one} &= |\chi^\text{Is}_0|^2+|\chi^\text{Is}_\frac{1}{2}|^2,
  \\ 
Z^{e} &= 
|\chi^\text{Is}_\frac{1}{16}|^2,
 \\ 
Z^{m} &= 
|\chi^\text{Is}_\frac{1}{16}|^2,
 \\ 
Z^{f} &= 
\chi^\text{Is}_0 \bar \chi^\text{Is}_\frac{1}{2} + \chi^\text{Is}_\frac12 \bar \chi^\text{Is}_0.
\end{split}
\end{align}
Under modular transformation, they transform as \eqn{ZST} and 
\begin{align}
\label{Z2sSTmat}
  T&=
\begin{pmatrix}
    1&0&0&0\\
    0&1&0&0\\
    0&0&1&0\\
    0&0&0&-1
  \end{pmatrix} ,
&
  S&=\frac12 \begin{pmatrix}
    1&1&1&1\\
    1&1&-1&-1\\
    1&-1&1&-1\\
    1&-1&-1&1
  \end{pmatrix}.
\end{align}
Note that $T$ is always diagonal in the quasiparticle basis.

According to \Ref{JW190513279}, $Z_\one$, $Z_e$, $Z_m$, and $Z_f$ are the 4
partition functions that describe a $c=\bar c=\frac 12$ CFT with a
non-invertible anomaly corresponding to a gapless boundary theory of the 2+1D $Z_2$-topological order.
Therefore, changing to the quasiparticle basis (\ie from \eqn{IsingZ1} to
\eqn{IsingZ2}) corresponds to gauging $Z_2$ global symmetry in the bulk.  This
turns a CFT with $Z_2$ global symmetry into a CFT with no symmetry but
with a non-invertible gravitational anomaly.

A second example is $su2_2\otimes \overline{su2}_2$ CFT
with $Z_2$ anomaly-free symmetry.
Its 4 partition functions with $Z_2$ symmetry twist are given by \eqn{su2_2}.
In the quansiparticle basis, we obtain the following 4 partition functions:
\begin{align}
\begin{split}
Z^{\one}=&\, |\chi^{su2_2}_0|^2+|\chi^{su2_2}_1|^2,\\
Z^e=&\, |\chi^{su2_2}_\fh|^2, \\
Z^m=&\, |\chi^{su2_2}_\fh|^2,\\
Z^f=&\, \chi^{su2_2}_0\bar\chi^{su2_2}_{1} + \chi^{su2_2}_1\bar\chi^{su2_2}_0.
\end{split}
\end{align}
The above 4 partition functions describe a $c=\bar c=\frac 32$ CFT with
non-invertible gravitational anomaly of the 2+1D $Z_2$ topological order.

\subsection{Double-semion topological order}

The $su2_1\otimes \overline{su2}_1$ CFT can have a $Z_2$ global
symmetry. Yet the symmetry is anomalous. The CFT under the $Z_2$ symmetry twists has the partition functions given in \eqn{Z2STZ}.
Under modular transformation, they transform as eqn. (\ref{Z2ST2}).

The CFT can appear on the boundary of the $\ZZ_2$ SPT. 
We can then gauge the $Z_2$ symmetry (in the 2+1D bulk). The bulk phase becomes the double-semion topological order, where the boundary system is described by the CFT with
non-invertible gravitational anomaly.  Explicitly, we obtain the boundary partition functions in the quasi particle basis, via a matrix transformation, 
\begin{align}
\label{eqn:DStfm}
\begin{pmatrix}
Z^{\one} \\ Z^b \\ Z^s \\ Z^{s^*}
\end{pmatrix}
=\frac{1}{2}\begin{pmatrix}
1 & 1 &  0&0  \\
1 & -1 & 0&0 \\
 0& 0& 1 & -\ii \\
 0& 0& 1 & \ii 
\end{pmatrix}\begin{pmatrix}
Z_{1,1} \\ Z_{1,-1} \\ Z_{-1,1} \\ Z_{-1,-1}
\end{pmatrix}.
\end{align}

\begin{figure}
\includegraphics[scale=.7]{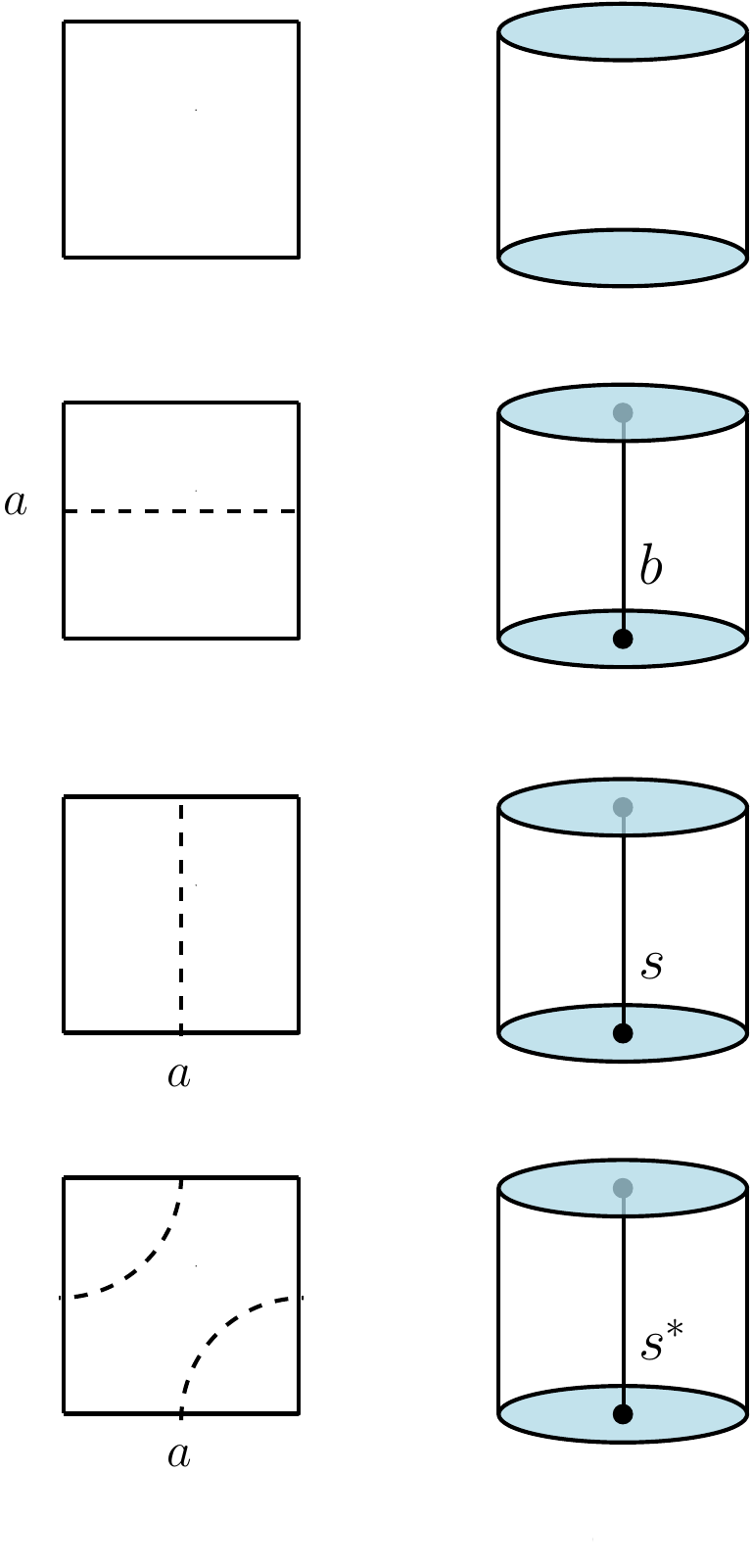}
\caption{Transformation between (a) boundaries of $\ZZ_2$ SPT and (b) that of double semion topological order. Transformation from (a) to (b) corresponds to gauging the $\ZZ_2$ global symmetry in the bulk.  }
\label{fig:anomalousZ2tfm}
\end{figure}
The partition functions in the quasiparticle basis\cite{JW190513279} are given
by
\begin{align}
\label{Zds1}
Z^{\one}=& |\chi^{su2_1}_0|^2,
\nn\\
Z^{s} =&\chi^{su2_1}_\fh\bar\chi^{su2_1}_0,
\nn\\
Z^{s^*}=& \chi^{su2_1}_0\bar\chi^{su2_1}_{\fh},
\nonumber\\
Z^{b} =&|\chi^{su2_1}_\fh|^2.
\end{align}
Under modular transformation, they transform as \eqn{ZST} and 
\begin{align}
\label{STDS}
  T&=
\begin{pmatrix}
    1&0&0&0\\
    0&\ii&0&0\\
    0&0&-\ii&0\\
    0&0&0&1
  \end{pmatrix} ,
&
  S&=\frac12 \begin{pmatrix}
    1&1&1&1\\
    1&-1&1&-1\\
    1&1&-1&-1\\
    1&-1&-1&1
  \end{pmatrix}.
\end{align}
As pointed out in \Ref{JW190513279},
$Z^\one$, $Z^s$, $Z^{s^*}$, and $Z^b$ are the 4 partition functions that
describe a $c=\bar c= 1$ CFT with a non-invertible anomaly corresponding to the
2+1D double-semion topological order.

\begin{table*}[!ht]
\setlength\extrarowheight{4pt}
\setlength{\tabcolsep}{6pt}
\centering
\begin{tabular}{|c|| c|c|c|c|c|c|c|c|}
\hline
$\otimes$ & $\bm 1$ &  $a^1$  & $a^2$ &  $b$  &  $b^1$  &  $b^2$  & $c$  & $c^1$    \\
\hline
\hline
$ \bm 1 $ & $\bm 1$ & $a^1$ & $a^2$   & $b$  & $b^1$  &  $b^2$          & $c$  & $c^1$    \\
$a^1$ & $a^1$ & $\bm 1$ & $a^2$	  & $b$  &$b^1$  & $b^2$      &  $c^1$  & $c$  \\
$a^2$ & $a^2$ & $a^2$   & $\bm 1\oplus a^1\oplus a^2$      & $b^1\oplus b^2$    & $b\oplus b^2$ & $b\oplus b^1$      & $c\oplus c^1$  & $c\oplus c^1$\\
$b$  & $b$  & $b$ & $b^1\oplus b^2$     & $\bm 1\oplus a^1\oplus b$ & $b^2\oplus a^2$  & $b^1\oplus a^2$     & $c\oplus c^1$   & $c\oplus c^1$  \\
$b^1$  & $b^1$   & $b^1$ & $b\oplus b^2$      & $b^2\oplus a^2$ & $\bm 1\oplus a^1\oplus b^1$  & $b\oplus a^2$      & $c\oplus c^1$ & $c\oplus c^1$ \\
$b^2$  & $b^2$  & $b^2$  & $b\oplus b^1$    & $b^1\oplus a^2$ & $b\oplus a^2$  & $\bm 1\oplus a^1\oplus b^2$      & $c\oplus c^1$  & $c\oplus c^1$ \\
$c$ & $c$ & $c^1$ & $c\oplus c^1$     & $c\oplus c^1$   & $c\oplus c^1$  & $c\oplus c^1$     & $\bm 1\oplus a^2\oplus b\oplus b^1\oplus b^2$ & $a^1 \oplus a^2\oplus b\oplus b^1\oplus b^2$  \\
$c^1$ & $c^1$ & $c$  & $c\oplus c^1$     & $c\oplus c^1$ & $c\oplus c^1$  & $c\oplus c^1$     & $a^1 \oplus a^2\oplus b\oplus b^1\oplus b^2$  & $\bm 1\oplus a^2\oplus b\oplus b^1\oplus b^2$ \\
\hline
\end{tabular}
\caption{Fusion rules $N^{ab}_c$ of 2+1D $S_3$ topological order. Here $b$
and $c$ correspond to pure flux excitations, $a^1$ and $a^2$ pure charge
excitations, $\bm 1$ the vacuum sector while  $b^1$, $b^2$, and $c^1$ are
charge-flux composites. 
}\label{S3FusionRules}
\end{table*}

\subsection{$S_3$-topological order}

For a CFT with $S_3$ symmetry, its partition function with symmetry twists
$Z_{g,h}$, $g,h\in S_3$, $gh=hg$, transform in the way under modular
transformations as in \ref{ZpropG}. What CFTs would have the non-anomalous $S_3$ symmetry? And how does the CFT transformation under a symmetry twist? It is not straightforward to find out. In this paper, we give one way to solve it. That is to first think of the CFT on the boundary of a $2+1D$ system with a global $S_3$ symmetry. Then we gauge the whole system, and find the CFT on the boundary of the $S_3$ gauge theory. The CFT with a global $S_3$ symmetry is related to this one by a basis transformation. 

\begin{table}\label{Table:S3anyon}
\setlength\extrarowheight{4pt}
\begin{tabular}{|c  c  c|}
\hline
symbol & (flux, charge) & quantum dimension \\
\hline
${\bf 1}$ & $([1],\lambda^1_1)$ & $1$ \\
$a^1$ & $([1],\lambda^1_-)$ & $1$ \\
$a^2$ & $([1],\lambda^1_2)$ & $2$ \\
\hline 
$b$ & $([s],\lambda^s_1)$ & $2$ \\
$b^1$ & $([s],\lambda^s_+)$ & $2$ \\
$b^2$ & $([s],\lambda^s_-)$ & $2$ \\
\hline
$c$ & $[r],\lambda^r_1)$ & $3$ \\
$c^1$ & $([r],\lambda^r_-)$ & $3$ \\
\hline
\end{tabular}
\caption{Anyons in $S_3$ topological order.}
\end{table}

To begin with, the  $S_3$ gauge theory describes the $\calD (S_3)$ topological order, which has 8 types of topological excitations labeled by $(\one, a^1,
a^2,b,b^1,b^2,c,c^1)$, as listed in Table \ref{Table:S3anyon}.
This topological order is characterized by the following modular matrices,
\begin{align*}
 & \text{Diag}(T)=(1,1,1,1,\ee^{\ii\frac{2\pi}{3}},\ee^{-\ii\frac{2\pi}{3}},1,-1),\\
  &S=\frac{1}{6}
  \begin{pmatrix}
    1&1&2&2&2&2&3&3\\
    1&1&2&2&2&2&-3&-3\\
    2&2&4&-2&-2&-2&0&0\\
    2&2&-2&4&-2&-2&0&0\\
    2&2&-2&-2&-2&4&0&0\\
    2&2&-2&-2&4&-2&0&0\\
    3&-3&0&0&0&0&3&-3\\
    3&-3&0&0&0&0&-3&3
  \end{pmatrix}.
\end{align*}
The fusion rules of the topological excitations are given in Table \ref{S3FusionRules}.

By matching the boundary with the bulk topological order, we find the 1+1D CFT with a gravitational anomaly given by the 2+1D $S_3$-topological order
has 8 partition functions.  If the CFT has central charges $c=\bar c= \frac
45$, the 8 partition functions are given by
\begin{align}
 Z^{\one} &=  |\chi^{m6}_{0}|^2 +  |\chi^{m6}_{3}|^2 +  |\chi^{m6}_{\frac{2}{5}}|^2 +  |\chi^{m6}_{\frac{7}{5}}|^2 
 \nonumber \\ 
Z^{a^1} &=  \chi^{m6}_{0} \bar\chi^{m6}_{3} +  \chi^{m6}_{3} \bar\chi^{m6}_{0} +  \chi^{m6}_{\frac{2}{5}} \bar\chi^{m6}_{\frac{7}{5}} +  \chi^{m6}_{\frac{7}{5}} \bar\chi^{m6}_{\frac{2}{5}} 
 \nonumber \\ 
Z^{a^2} &=  |\chi^{m6}_{\frac{2}{3}}|^2 +  |\chi^{m6}_{\frac{1}{15}}|^2 
 \nonumber \\ 
Z^{b} &=  |\chi^{m6}_{\frac{2}{3}}|^2 +  |\chi^{m6}_{\frac{1}{15}}|^2 
\\ 
Z^{b^1} &=  \chi^{m6}_{0} \bar\chi^{m6}_{\frac{2}{3}} +  \chi^{m6}_{3} \bar\chi^{m6}_{\frac{2}{3}} +  \chi^{m6}_{\frac{2}{5}} \bar\chi^{m6}_{\frac{1}{15}} +  \chi^{m6}_{\frac{7}{5}} \bar\chi^{m6}_{\frac{1}{15}} 
 \nonumber \\ 
Z^{b^2} &=  \chi^{m6}_{\frac{2}{3}} \bar\chi^{m6}_{0} +  \chi^{m6}_{\frac{2}{3}} \bar\chi^{m6}_{3} +  \chi^{m6}_{\frac{1}{15}} \bar\chi^{m6}_{\frac{2}{5}} +  \chi^{m6}_{\frac{1}{15}} \bar\chi^{m6}_{\frac{7}{5}} 
 \nonumber \\ 
Z^{c} &=  |\chi^{m6}_{\frac{1}{8}}|^2 +  |\chi^{m6}_{\frac{13}{8}}|^2 +  |\chi^{m6}_{\frac{1}{40}}|^2 +  |\chi^{m6}_{\frac{21}{40}}|^2 
 \nonumber \\ 
Z^{c^1} &=  \chi^{m6}_{\frac{1}{8}} \bar\chi^{m6}_{\frac{13}{8}} +  \chi^{m6}_{\frac{13}{8}} \bar\chi^{m6}_{\frac{1}{8}} +  \chi^{m6}_{\frac{1}{40}} \bar\chi^{m6}_{\frac{21}{40}} +  \chi^{m6}_{\frac{21}{40}} \bar\chi^{m6}_{\frac{1}{40}} .
\nonumber 
\end{align}
where $\chi^{m6}_{h}$ are characters of $\calM(6,5)$ minimal model.  In
other words, the 2+1D bosonic topological order described by $S_3$ gauge theory
may have a gapless boundary described by $c=\bar c=\frac 45$ CFT.  Such a
gapless boundary has only one relevant operator with scaling dimension
$2\frac25=\frac45$.  However, in the above, we only considered the modular
covariant partition functions on torus.  To be sure that $S_3$ gauge theory can
have a gapless boundary described by the minimal model $\calM (6,5)$, we also
need to check the covariance of the partition functions under mapping class
group actions for higher genus surfaces.

Since the $S_3$ gauge theory comes from gauging a non-anomalous $S_3$ global symmetry, we learn that the minimal model $\calM(6,5)$ can have a $S_3$ global symmetry. In the next section, we discuss starting with this CFT in the quasiparticle basis, how to find the CFT twisted by the $S_3$ symmetry.

\section{Correspondence between $G$-'t Hooft anomaly and non-invertible anomaly}

One advantage of the basis transformation is to identify the invertible anomaly through the non-invertible anomaly. In particular, consider the topological order described by $G$ gauge theory twisted by $\omega\in H^3(G, U(1))$. 

\begin{figure}
\includegraphics[scale=.85]{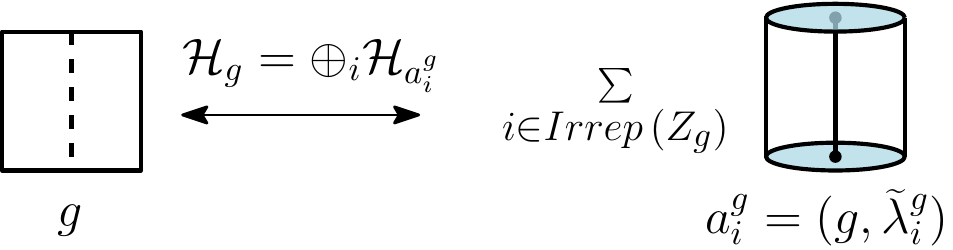}
\caption{Given the same CFT, as the boundary of invertible topological order with symmetry $G$ (left) and non-invertible topological order with gauge group $G$ (right) , twisted by the same cocycle $\omega\in H^3(G,U(1))$, the Hilbert space of $g$-twisted sector (left) is the same as the direct sum of that of the flux-$g$ anyon sectors $a_i^g=(g,\widetilde{\lambda}_i^g)$(right).  }
\end{figure}

The topological orders we consider are described by a discrete finite gauge group $G$. Particularly, we focus on the topological orders described by (1) untwisted topological field theories (TQFT) (2) TQFT twisted by $3$-cocycle $\omega\in H^3(G, U(1))$. We show that there is a basis transformation from the quasiparticle basis labeled by anyons in a $G$ gauge theory to, what we call, symmetry-twisting basis, labeled by $(a,b)\in G\times G$, where $a$ is a representative group element of each conjugacy class of $G$, and $b\in Z(a)$, the centralizer group of $a$ . The modular matrices $S, T$ in the symmetry-twisting basis, becomes \emph{permutation matrices}, up to phases determined from the slant product of $3$-cocycle $\omega \in H^3(G, U(1))$. More explicitly, start with the components $\{Z_{(1,g)}\}$, where $g$ is a representative group element of each conjugacy class of $G$. The actions of $S, T$ on the components give the whole vector of partition functions. 

We find there is a unitary matrix that relates the partition functions in the symmetry twisted basis and that in the quasiparticle basis. The former describes the boundary of a topological phase protected by $G$ symmetry (SPT), the latter describes the boundary of the topological order that is related to the SPT by gauging the global symmetry. 

To describe the unitary matrix, let us first discuss the dimension of the partition functions in the symmetry twisted basis. The independent components are  given by conjugate classes of commuting pair. A CFT with global symmetry $G$, in general, the CFT can be twisted in spatial direction by $g$ and time direction by $h$ for any $g,h\in G$ that commute. The total number of symmetry twisted sectors are then
$\sum_{[g]} \left| Z(g)\right|$, where $[g]$ is to sum over conjugacy classes of $G$. For example, when $G=S_3$, the number of twisted sectors is $6+3+2=11$. However, not all of them are independent. In fact, if $(g',h')\sim (g,h)$, where the equivalence $~$ means there is an element $x\in G$, such that $(g',h')= (xgx^{-1}, xhx^{-1})$, then the twisted sector $Z_{g',h'}$ and $Z_{g,h}$ are related by gauge transformation. In particular, if the symmetry $G$ is anomaly free, $Z_{g',h'}=Z_{g,h}$. Therefore, each independent component $Z_{(g,h)}$ given by a representative $g$ in each conjugacy class of $G$, and a representative of $h$ in each conjugacy class of the remaining global symmetry group $Z(g)$. The number of independent ones is $\calN=\sum_{[g]} k(Z(g))$, where $k(Z(g))$ is the class number (the number of conjugacy classes) of the centralizer group $Z(g)$. For example, for $S_3$, $\calN=8$. 

Now recall that a quasiparticle in the $\calD^{(\omega)} (G)$ topological order is given by $(a,\mu^a)$, where $a$ is the magnetic flux, given by a representative in a conjugacy class of $G$, and $\mu^a$ is a representation of the remaining gauge group $Z(a)$. For certain twisted quantum doubles, that is, those twisted by the $3$-cocycle that can be written in terms of $2$-cocycles, $\mu^a$ is the projective representation. 

The transformation between the quasiparticle basis and symmetry twisting basis in the bulk is\cite{hu2013twisted}
\begin{align}
Z^{(a,\mu^a)}=\frac{1}{\sqrt{|G|}}\sum_{a'\in [a], g\in Z_{a'}} \tilde{\chi}_\mu^{a'}(g)^*Z_{(a',g)}, \label{qptosym}
\end{align}
where $\tilde{\chi}_\mu^g (h)=\Tr\, \tilde{\mu}^g (h)$ is the projective character. That is, the transformation is given by the projective character table. When the symmetry is untwisted, this is the character table. 

Since the $S$ matrix in anyon basis is 
\begin{align}
\langle a,\mu^a | S |b,\lambda^b \rangle =&\frac{1}{|G|}\sum_{\substack{a'\in [a], b'\in [b], \\ a'b'=b'a'}}\tilde \chi_\mu^{a'}(b')^*\tilde \chi_\lambda^{b'}(a')^*\,,\nn\\
\langle a,\mu^a | T |b,\lambda^b \rangle =&\delta_{a,b}\delta_{\mu^a,\lambda^b} \frac{\tilde \chi^{a}(a)}{\dim \mu}.  \label{Sqp}
\end{align}
We can obtain that, the $S$ matrix in the symmetry twisting basis, when the symmetry is anomaly free, is a permutation matrix as follows, 
\begin{align}
\langle (a, g)| S |(b,h)\rangle = \delta_{a,h}\delta_{g,b^{-1}}.
\end{align}
And when the symmetry is anomalous, the $S$ matrix is the permutation matrix modified by the $2$-cocycles, which is the slant product applied to $\omega\in H^3(G, U(1))$. 
\begin{align}
\langle (a, g)| S |(b,h)\rangle = \delta_{a,h}\delta_{g,b^{-1}}\beta_h (b,b^{-1}),\label{Ssym}
\end{align}
where
\begin{align}
\beta_a (b,c)=\omega (a,b,c)\omega (b,b^{-1}ab,c)^{-1}\omega (b,c, (bc)^{-1}abc)\,.
\end{align}
We describe the proof that the modular matrices are related as above in Appendix \ref{basistrans}, as well as give the transformation when the topological order is $D_4$ quantum double, $Q_8$ quantum double and $\ZZ_3$ twisted quantum double. 

Since each sector in the bulk has a boundary described by one component of partition function, we conclude that the boundary partition functions transform the same way accordingly to (\ref{qptosym}).

The simplest examples are when $G=\ZZ_2$. If the symmetry is non-anomalous, the transformation (\ref{qptosym}) is explicitly shown in (\ref{eqn:Z2tfn}), relating the CFT twisted by the anomaly-free $\ZZ_2$ symmetry and the CFT with the non-invertible anomaly matched by $\ZZ_2$ topological order. 
If the symmetry is anomalous, the transformation (\ref{qptosym}) is shown in (\ref{eqn:DStfm}), relating the anomalous $\ZZ_2$ symmetry and the non-invertible anomaly matched by the double semion topological order. The above correspondence between the $\ZZ_2$ symmetry twists and non-invertible anomalies are illustrated in Fig. \ref{fig:Z2tfm} and Fig. \ref{fig:anomalousZ2tfm}. 

Let us apply the correspondence given by (\ref{qptosym}) to the example of $G=S_3$. The correspondence answers us how the $S_3$ symmetry acts on the minimal model $\calM (6,5)$. There are three conjugacy classes of $S_3$, and we choose the representative of each class as $1, s, r$. The gauge-inequivalent symmetry twisted sectors are
\begin{align}
&Z_{(1,1)}\,,~ Z_{(1,s)}\,,~Z_{(1,r)}\,,
\nn\\
&Z_{(s,1)}\,,~Z_{(s,s)}\,,~Z_{(s,s^2)}\,,\nn\\
&Z_{(r,1)}\,,~Z_{(r,r)}\,.
\end{align}

The symmetry-twisted basis is related to the quasi-particle basis indexed by anyons in $S_3$ gauge theory via the following transformation,
\begin{align}
\begin{pmatrix}
Z_{(1,1)}\\ Z_{(1,s)} \\ Z_{(1,r)} 
\end{pmatrix}=\begin{pmatrix}
1 & 1 & 2 \\ 1 & 1 & -1 \\ 1 & -1 & 0
\end{pmatrix}\begin{pmatrix}
Z^{\bf 1} \\ Z^{a^1} \\ Z^{a^2}
\end{pmatrix},
\end{align}
\begin{align}
\begin{pmatrix}
Z_{(s,1)} \\ Z_{(s,s)} \\ Z_{(s,s^2)}
\end{pmatrix}=\begin{pmatrix}
1 & 1 & 1 \\ 1 & \omega & \bar \omega \\ 1 & \bar \omega & \omega
\end{pmatrix}\begin{pmatrix}
Z^b \\ Z^{b^1} \\ Z^{b^2}
\end{pmatrix},
\end{align}
\begin{align}
\begin{pmatrix}
Z_{(r,1)} \\ Z_{(r,r)} 
\end{pmatrix}=\begin{pmatrix}
1 & 1 \\ 1 & -1 
\end{pmatrix}\begin{pmatrix}
Z^c \\ Z^{c^1}
\end{pmatrix},
\end{align}
where $\omega=\ee^{\ii \frac{2\pi}{3}}$. And the matrices appearing above are the character tables of $S_3$, $Z_3$ and $Z_2$ group, respectively.

From the above, we obtain that the $\calM(6,5)$ CFT twisted under the non-anomalous $S_3$ symmetry is, 
\begin{align}
 Z_{(1,1)} =&  |\chi^{m6}_{0}+\chi^{m6}_{3}|^2 +  |\chi^{m6}_{\frac{2}{5}}+\chi^{m6}_{\frac{7}{5}}|^2 +2|\chi^{m6}_{\frac{2}{3}}|^2 + 2 |\chi^{m6}_{\frac{1}{15}}|^2 ,
 \nonumber \\ 
 Z_{(1,s)}= &|\chi^{m6}_{0}+\chi^{m6}_{3}|^2 +  |\chi^{m6}_{\frac{2}{5}}+\chi^{m6}_{\frac{7}{5}}|^2 -|\chi^{m6}_{\frac{2}{3}}|^2 - |\chi^{m6}_{\frac{1}{15}}|^2, \nonumber\\
  Z_{(1,r)} =&  |\chi^{m6}_{0}-\chi^{m6}_{3}|^2 +  |\chi^{m6}_{\frac{2}{5}}-\chi^{m6}_{\frac{7}{5}}|^2 \nonumber \\
Z_{(s,1)}=& |\chi^{m6}_{\frac{2}{3}}|^2 +  |\chi^{m6}_{\frac{1}{15}}|^2 \nonumber \\
&+ (\chi^{m6}_{0} +  \chi^{m6}_{3} )\bar\chi^{m6}_{\frac{2}{3}} +  (\chi^{m6}_{\frac{2}{5}} +  \chi^{m6}_{\frac{7}{5}}) \bar\chi^{m6}_{\frac{1}{15}}, \nonumber \\
Z_{(s,s)}=& |\chi^{m6}_{\frac{2}{3}}|^2 +  |\chi^{m6}_{\frac{1}{15}}|^2 \nonumber \\
&+ \omega(\chi^{m6}_{0} +  \chi^{m6}_{3} )\bar\chi^{m6}_{\frac{2}{3}} +  \bar\omega(\chi^{m6}_{\frac{2}{5}} +  \chi^{m6}_{\frac{7}{5}}) \bar\chi^{m6}_{\frac{1}{15}}, \nonumber \\
Z_{(s,s^2)}=& |\chi^{m6}_{\frac{2}{3}}|^2 +  |\chi^{m6}_{\frac{1}{15}}|^2 \nonumber \\
&+\bar \omega(\chi^{m6}_{0} +  \chi^{m6}_{3} )\bar\chi^{m6}_{\frac{2}{3}} +  \omega(\chi^{m6}_{\frac{2}{5}} +  \chi^{m6}_{\frac{7}{5}}) \bar\chi^{m6}_{\frac{1}{15}} ,\nonumber \\
Z_{(r,1)}=&|\chi^{m6}_{\frac{1}{8}}+\chi^{m6}_{\frac{13}{8}}|^2 +  |\chi^{m6}_{\frac{1}{40}}+\chi^{m6}_{\frac{21}{40}}|^2 ,
 \nonumber\\
 Z_{(r,r)}=&|\chi^{m6}_{\frac{1}{8}}-\chi^{m6}_{\frac{13}{8}}|^2 +  |\chi^{m6}_{\frac{1}{40}}-\chi^{m6}_{\frac{21}{40}}|^2 .
\end{align}

\section{1+1D gravitational anomaly with symmetry}

CFTs with global gravitational anomaly can also have a global symmetry with
group $G$. This is the case when the bulk is a topologically ordered phase
enriched by a global symmetry $G$ (SET). We show through examples, that to
obtain the CFT on the boundary of a SET, one way is to gauge the global
symmetry and obtain the gauged topological order. In the gauged topological
order, there is a set of anyons that are self-bosons. If these anyons
condense, the condensed phase is the same as the topological order before
gauging. It is straightforward to obtain the CFT on the boundary of the gauged
topological order first. The input CFTs describing the boundary of the SET and that of the gauged topological
order are the same. The two partition functions are related by a basis
transformation.

Specifically, we study the gapless boundary of $Z_2$ symmetry enriched
topological order phases and their relations to the gapless boundary of
topologically ordered phases after gauging the $Z_2$ symmetry. The $Z_2$
symmetries we consider include a symmetry that does not permute anyons, the
charge conjugation symmetry, and electric-magnetic duality symmetry that
permute anyons. In all cases, we find (1) a gapless boundary of the SET phase in which for any sector, either with an anyon or a symmetry defect in the bulk has gapless excitations on the boundary. The vector of partition functions transforms under the modular $S$ and $T$ matrix of the $G$-crossed MTC;
(2) a correspondence between the gapless boundary of the SET and the gapless state with certain
anyon condensed on the boundary of the topological order after gauging. An
empirical characterization of a gapless state with certain anyon condensed is
summarized in (\ref{eqn:condensedbdy}). Similar formal relations have been
established relating chiral degrees of freedom in rational conformal field
theory and Abelian topological field theories (TFT) in one higher
dimensions.\cite{witten1999ads} To our knowledge, the relations between
non-chiral rational conformal field theory and TFT have not been studied
previously.

We further find it is always possible to relate this boundary state to a
gapless boundary state of the SET before gauging. We also find, the gapless
boundaries with no anyon condensation of a SET and the topological order after
gauging do not necessarily admit a field theory built from the characters of
the same conformal field theory, as in the example of $Z_3^{(1)}\boxtimes
\bar{SU(2)}_4$ topological order described in subsubsection.
\ref{subsubsec:EMinZ3}.

Let us begin with the description of the boundary of the SET phase in terms of a
vector of partition functions. One algebraic way to describe a SET phase is the
$G$-crossed unitary fusion category.\cite{barkeshli2019symmetry} That is first to generalize the unitary
braided fusion category $\calC_0$ describing the topological phase to a
$G$-graded fusion category, $\calC_{G}=\oplus_{g\in G}\calC_{g}$. The element
$a_g$ in $\calC_{g}$ with $g\neq 0$ labels the $g$-symmetry defect that appears
in the SET phase as an external topological defect. The next is to find the
compatible braiding of the topological defects and topological quasiparticle,
and promote $\calC_G$ to a ``G-crossed'' braided tensor category
$\calC_G^\times$. Now it is natural to expect that the vector of boundary
partition functions is also labeled by elements in $\calC_g$. More precisely,
each component of the boundary CFT is labeled by $Z^{a_g}_{g, h}$, the
subscript $g\in G$ means the bulk is in the $g$-twisted sector, the superscript
$a_g\in \calC_g$ means the presence of a topological defect $a_g$ in the bulk,
and the subscript $h\in Z(g)$ means a global symmetry action on the system.

\subsection{$\ZZ_2$ symmetry in the $\ZZ_2$ topological order} 

We begin with
the simplest example, and consider a $c=\bar c=\frac12$ CFT with simplest
non-invertible gravitational anomaly and symmetry.  The gravitational anomaly
is matched by a 2+1D $Z_2$ bosonic topologically ordered bulk, in which there are
4 types of excitations $\one,e,m,f$.  \ie the 1+1D CFT appears on the boundary
of the 2+1D $Z_2$ bosonic topological order.  The symmetry is $Z_2$ symmetry,
\ie the 2+1D $Z_2$ bosonic topological order and its boundary can also have a
$Z_2$ symmetry.  However, there are many 2+1D topological states that have $Z_2$
bosonic topological order with a $Z_2$ symmetry: those are topological states
with $Z_2$ symmetry, which become the $Z_2$ bosonic topological order after we
break the symmetry.  In \Ref{BBC1440,LW160205946}, 2+1D topological orders with
symmetry (including $Z_2$ symmetry) are full classified.  In particular, there
are in total $6$ types of $Z_2$  topological orders with $Z_2$ symmetry
\cite{mesaros2013classification,lu2016classification}.  Those $Z_2$ symmetric
$Z_2$ topological orders are clearly different types of SET phases, as they
become different types of topological orders after gauging the $Z_2$ symmetry.
In four types, the symmetry is not fractionalized-- there are no anyons
carrying projective representations. In this subsection, we focus on these
types and describe how the gapless boundary partition functions can be
determined and related when the bulk is the SET phase or the topological
ordered phase after the global symmetry is gauged. 

\subsubsection{The on-site $\ZZ_2$ symmetry in $\ZZ_2$ topological order} 

First, we
consider an on-site $\ZZ_2$ symmetry on the $Z_2$ topological order. After
gauging the $\ZZ_2$ symmetry, the topological phase becomes the $Z_2\times Z_2$
topological order (\ie the topological order described by $Z_2\times Z_2$ gauge
theory), or the quantum double model $\calD(Z_2\times Z_2)$.  This topological
order is also the same as the stack of two $Z_2$ topological orders, which we
denote as $\calD(Z_2)\boxtimes \calD(Z_2)$. Reversely,we can start with the stack of
two $Z_2$ topological orders and turn on certain interactions that let the
anyon $mm'$, the bound state of the magnetic flux in two layers, to condense. This way, we
obtain the $Z_2$ bosonic topological order with the $Z_2$ symmetry.  

How are the gapless boundary state of the SET and that of the TO after gauging
related? We give one scenario. One of the gapless boundary of $\calD(Z_2)\boxtimes
\calD(Z_2)$ topological order is described by an anomalous CFT has 16 partition
functions $Z^{i}_{g,h}$ where $i=\one,e,m,f$ and $g,h=\pm 1$. 

In the quasiparticle basis,  the following 16 partition functions $Z^{i,i'}$,
$i'=\one',e',m',f'$ describes a boundary where $mm'$ quasiparticle is
condensed,
\begin{align}
 Z^{\one;\one'} &= |\chi^\text{Is}_0|^2 + |\chi^\text{Is}_\frac{1}{2}|^2  ,
 \nonumber \\ 
Z^{\one;e'} &= 0,
 \nonumber \\ 
Z^{\one;m'} &= |\chi^\text{Is}_\frac{1}{16}|^2  ,
 \nonumber \\ 
Z^{\one;f'} &= 0,
\end{align}
\begin{align}
Z^{e;\one'} &= 0,
 \nonumber \\ 
Z^{e;e'} &= |\chi^\text{Is}_\frac{1}{16}|^2 ,
 \nonumber \\ 
Z^{e;m'} &= 0,
 \nonumber \\ 
Z^{e;f'} &= \chi^\text{Is}_0 \bar\chi^\text{Is}_\frac{1}{2} + \chi^\text{Is}_\frac{1}{2} \bar\chi^\text{Is}_0 ,
\end{align}
\begin{align}
Z^{m;\one'} &= |\chi^\text{Is}_\frac{1}{16}|^2 ,
 \nonumber \\ 
Z^{m;e'} &= 0,
 \nonumber \\ 
Z^{m;m'} &= |\chi^\text{Is}_0|^2 + |\chi^\text{Is}_\frac{1}{2}|^2 ,
 \nonumber \\ 
Z^{m;f'} &= 0,
\end{align}
\begin{align}
Z^{f;\one'} &= 0,
 \nonumber \\ 
Z^{f;e'} &= \chi^\text{Is}_0 \bar\chi^\text{Is}_\frac{1}{2} + \chi^\text{Is}_\frac{1}{2} \bar\chi^\text{Is}_0 ,
 \nonumber \\ 
Z^{f;m'} &= 0,
 \nonumber \\ 
Z^{f;f'} &= |\chi^\text{Is}_\frac{1}{16}|^2 .
\end{align}
Indeed, this solution describes the gapless boundary phase that  $mm'$ is condensed on the boundary. In this phase, the condensed anyon appears on the boundary does not cost extra energy, the boundary excitations in this sector are thus the same as those in the vacuum sector, those anyons that braid non-trivially with $mm'$ become gapped excitations when appearing on the boundary, while those anyons that braid trivially with $mm'$ remain gapless excitations on the boundary.  We can observe these phenomena from the above partition functions, as they satisfy the following properties. Say $a_0=mm'$ is condensed, 
\begin{align}
\label{eqn:condensedbdy}
\begin{split}
&Z^{\text{vacuum}}=Z^{a_0}, \\
&Z^{a}\begin{cases}
=0, & M_{aa_0}\neq 1, \\
\neq 0, & M_{aa_0} = 1,
\end{cases}
\end{split}
\end{align}
where $M_{ab}=\frac{S_{ab}^*S_{00}}{S_{0a}S_{0b}}$ means the braiding phase between $a$ and $b$ anyon.

To change to the symmetry twist basis, we use
\begin{align}
Z^{\one'}=&\frac{1}{2}(Z_{1,1}+Z_{1,-1}),\; Z^{e'}=\frac{1}{2}(Z_{1,1}-Z_{1,-1}),
\\
Z^{m'}=&\frac{1}{2}(Z_{-1,1}+Z_{-1,-1}),\; Z^{f'}=\frac{1}{2}(Z_{-1,1}-Z_{-1,1}),
\nonumber 
\end{align}
the same transformation for non-anomalous $Z_2$ symmetry as in (\ref{eqn:Z2tfn}).

We find that the boundary vector of partition functions on the $Z_2$ topological order enriched by the on-site $Z_2$ symmetry is the following. When the bulk is in the untwisted superselection sector, the components are
\begin{align}
 Z^{\one}_{1,1} &= |\chi^\text{Is}_0|^2 + |\chi^\text{Is}_\frac{1}{2}|^2,
 \nonumber \\ 
Z^{\one}_{1,-1} &= |\chi^\text{Is}_0|^2 + |\chi^\text{Is}_\frac{1}{2}|^2,
 \nonumber \\ 
 Z^{e}_{1,1} &= |\chi^\text{Is}_\frac{1}{16}|^2,
 \nonumber \\ 
Z^{e}_{1,-1} &= -|\chi^\text{Is}_\frac{1}{16}|^2 ,
\end{align}
\begin{align}
Z^{m}_{1,1} &= |\chi^\text{Is}_\frac{1}{16}|^2 ,
 \nonumber \\ 
Z^{m}_{1,-1} &= |\chi^\text{Is}_\frac{1}{16}|^2,
 \nonumber \\ 
Z^{f}_{1,1} &= \chi^\text{Is}_0 \bar\chi^\text{Is}_\frac{1}{2} + \chi^\text{Is}_\frac{1}{2} \bar\chi^\text{Is}_0,
 \nonumber \\ 
Z^{f}_{1,-1} &= - \chi^\text{Is}_0 \bar\chi^\text{Is}_\frac{1}{2} - \chi^\text{Is}_\frac{1}{2} \bar\chi^\text{Is}_0 ,
\end{align}
As expected, the $e$ and $f$ quasiparticle carry the on-site $\ZZ_2$ symmetry charges. 
When the bulk is in the $\ZZ_2$-twisted defect sector, the components are 
\begin{align}
Z^{\one}_{-1,1} &= |\chi^\text{Is}_\frac{1}{16}|^2  ,
 \nonumber \\ 
Z^{\one}_{-1,-1} &= |\chi^\text{Is}_\frac{1}{16}|^2,\nonumber\\
Z^{e}_{-1,1} &= \chi^\text{Is}_0 \bar\chi^\text{Is}_\frac{1}{2} + \chi^\text{Is}_\frac{1}{2} \bar\chi^\text{Is}_0,
 \nonumber \\ 
Z^{e}_{-1,-1} &= -\chi^\text{Is}_0 \bar\chi^\text{Is}_\frac{1}{2} - \chi^\text{Is}_\frac{1}{2} \bar\chi^\text{Is}_0 ,
\end{align}
\begin{align}
Z^{m}_{-1,1} &= |\chi^\text{Is}_0|^2 + |\chi^\text{Is}_\frac{1}{2}|^2 ,
 \nonumber \\ 
Z^{m}_{-1,-1} &= |\chi^\text{Is}_0|^2 + |\chi^\text{Is}_\frac{1}{2}|^2,\nonumber \\
Z^{f}_{-1,1} &= |\chi^\text{Is}_\frac{1}{16}|^2,
 \nonumber \\ 
Z^{f}_{-1,-1} &= -|\chi^\text{Is}_\frac{1}{16}|^2 .
\end{align}
\iffalse
\begin{align}
 Z^{\one}_{1,1} &= |\chi^\text{Is}_0|^2 + |\chi^\text{Is}_\frac{1}{2}|^2  ,
 \nonumber \\ 
Z^{\one}_{1,-1} &= |\chi^\text{Is}_0|^2 + |\chi^\text{Is}_\frac{1}{2}|^2,
 \nonumber \\ 
Z^{\one}_{-1,1} &= |\chi^\text{Is}_\frac{1}{16}|^2  ,
 \nonumber \\ 
Z^{\one}_{-1,-1} &= |\chi^\text{Is}_\frac{1}{16}|^2,
\end{align}
%
\begin{align}
Z^{e}_{1,1} &= |\chi^\text{Is}_\frac{1}{16}|^2,
 \nonumber \\ 
Z^{e}_{1,-1} &= -|\chi^\text{Is}_\frac{1}{16}|^2 ,
 \nonumber \\ 
Z^{e}_{-1,1} &= \chi^\text{Is}_0 \bar\chi^\text{Is}_\frac{1}{2} + \chi^\text{Is}_\frac{1}{2} \bar\chi^\text{Is}_0,
 \nonumber \\ 
Z^{e}_{-1,-1} &= -\chi^\text{Is}_0 \bar\chi^\text{Is}_\frac{1}{2} - \chi^\text{Is}_\frac{1}{2} \bar\chi^\text{Is}_0 ,
\end{align}
%
\begin{align}
Z^{m}_{1,1} &= |\chi^\text{Is}_\frac{1}{16}|^2 ,
 \nonumber \\ 
Z^{m}_{1,-1} &= |\chi^\text{Is}_\frac{1}{16}|^2,
 \nonumber \\ 
Z^{m}_{-1,1} &= |\chi^\text{Is}_0|^2 + |\chi^\text{Is}_\frac{1}{2}|^2 ,
 \nonumber \\ 
Z^{m}_{-1,-1} &= |\chi^\text{Is}_0|^2 + |\chi^\text{Is}_\frac{1}{2}|^2,
\end{align}
%
\begin{align}
Z^{f}_{1,1} &= \chi^\text{Is}_0 \bar\chi^\text{Is}_\frac{1}{2} + \chi^\text{Is}_\frac{1}{2} \bar\chi^\text{Is}_0,
 \nonumber \\ 
Z^{f}_{1,-1} &= - \chi^\text{Is}_0 \bar\chi^\text{Is}_\frac{1}{2} - \chi^\text{Is}_\frac{1}{2} \bar\chi^\text{Is}_0 ,
 \nonumber \\ 
Z^{f}_{-1,1} &= |\chi^\text{Is}_\frac{1}{16}|^2,
 \nonumber \\ 
Z^{f}_{-1,-1} &= -|\chi^\text{Is}_\frac{1}{16}|^2 .
\end{align}
\fi
Here we have used the knowledge that in the $Z_2^\times$ braided tensor category describing the bulk SET, the $Z_2$-twisted sector $\calC_{-1}$ also contains four types of abelian defects $a$ with $Z_2\times Z_2$ fusion and trivial $F$ symbols. And to simplify notations, we also label them as $\one, e,m,f$, as appeared in the components $Z_{-1, h}^{a}$. 
The above partition functions describe a gapless boundary with non-invertible
gravitational anomaly (with symmetry). We can see that in our gauge choice, the $e$ and $f$ topological defects also carry the $Z_2$ charges. Indeed, this assignment of local charges is consistent with fusion rules. Alternatively, one can also attach a trivial $Z_2$ SPT (the Ising model) to the bulk, and combine the charged anyons with the charged spins in the SPT. After this redefinition, all anyons are charged trivially.

We also observe that
\begin{align}
Z^{a_{-1}}_{(-1,1)}=Z^{a_{-1}\times m}_{(1,1)},
\end{align}
where $m$ is a topological defect in $\calC_{-1}$, and we have used the fusion rule in $\calC_{G}$ that $a_{g}\times b_{g'}=(ab)_{gg'}$.

\iffalse
\subsubsection{The second $Z_2$ symmetric topological order}
The second example of $Z_2$ topological order enriched by an on-site $Z_2$ symmetry  is similar to the previous example, except to first stack it with a $Z_2$ SPT. After gauging this $Z_2$ symmetry, the SET becomes the tensor product of $Z_2$ topological order and the double semion 
 topological order, $\calD(\ZZ_2)\boxtimes \calD^{[\omega]}(\ZZ_2)$,  where $\calD^{[\omega]}(\ZZ_2)$ corresponds to the double semion topological
order-- a twisted $\ZZ_2$ topological order. The
boundary of $\calD(\ZZ_2)\boxtimes \calD^{[\omega]}(\ZZ_2)$ topological order will
have a different non-invertible gravitational anomaly (with symmetry).
\fi

\def\arraystretch{1.25} \setlength\tabcolsep{3pt}
\begin{table*}[t] 
\caption{The two topological orders coming from gauging an on-site
$Z_2$ symmetry that is not fractionalized, in the $\ZZ_2$ topological order. The quantum dimensions and topological spins of quasiparticles in the two 2+1D bosonic topological orders with 9 types of topological excitations. ``$+$'' and ``-'' represents the $Z_2$ charge of the corresponding quasiparticle before the symmetry is gauged.``$\times$'' means the quasiparticle corresponds to a symmetry defect before gauging. 
} 
\label{mextZ2b} 
\setlength\extrarowheight{4pt}
\setlength{\tabcolsep}{6pt}
\centering
\begin{tabular}{ |c|c|c|c|  } 
\hline 
  label &$d_1,\cdots, d_9$ & $s_1,s_2,\cdots, s_9$ & $N^B_{c}$\\
\hline
$\text{Ising} \boxtimes\overline{\text{Ising}}$ & \tiny $1\times 4, 2,\red{\sqrt2\times 4}$ &\tiny $0, 0, \frac{1}{2}, \frac{1}{2}, 0, \red{\frac{15}{16}, \frac{1}{16}, \frac{7}{16}, \frac{9}{16}}$ & $9^{ B}_{ 0}$ \\
& $Z_2$-charge/defect: & $+,-,+,-,\pm,\red{\times,\times,\times,\times}$& \\
\hline
$SU(2)_2\boxtimes \bar{SU(2)}_2$ & \tiny $1\times 4, 2,\red{\sqrt2\times 4}$ &\tiny $0, 0, \frac{1}{2}, \frac{1}{2}, 0, \red{\frac{3}{16}, \frac{13}{16}, \frac{11}{16}, \frac{5}{16}}$ & $9^{ B}_{ 0}$ \\
& $Z_2$-charge: & $+,-,+,-,\pm,\red{\times,\times,\times,\times}$&  \\
\hline
\end{tabular} 
\end{table*}

\subsubsection{The electric-magnetic duality symmetry in $\ZZ_2$ topological order }\label{subsubsec:EMinZ2TO}

The  $Z_2$ symmetry on the 2+1D $Z_2$ topological order with $Z_2$ symmetry can also act as an exchange symmetry on the $e$ and $m$
topological excitations of the $Z_2$ topological order.  The topological orders enriched by this symmetry are studied in \Ref{LW160205946} and described in Table \ref{mextZ2b}.  After gauging the $Z_2$ symmetry (\ie after
the modular extension), they become $\text{Ising} \boxtimes \bar{\text{Ising}}$
topological order and $SU(2)_2\boxtimes \bar{SU(2)}_2$ topological
order (also known as the $\text{Ising}^{(3)}\boxtimes \bar{\text{Ising}^{(3)}}$ topological order).\cite{LW160205946} Here the ``bar'' the time reversal conjugate of the topological order. The $SU(2)_2$ topological order can be obtained
from the Ising topological order if we let the fermion $\psi$
in the Ising topological order to form a filling fraction
$\nu=1$ integer quantum Hall state.  The $\text{Ising} \boxtimes \bar{\text{Ising}}$ and $SU(2)_2\boxtimes \bar{SU(2)}_2$ topological order
both have 9 types of topological excitations, whose quantum dimensions $d_i$
and topological spins $s_i$ are given in the Table \ref{mextZ2b}.  The
excitations labeled by $i=6,7,8,9$ carry the $\pi$-flux of the gauged $Z_2$
symmetry.  We also listed the $Z_2$-charge of the topological excitation
\emph{before the $Z_2$-gauging}.

Here, we focus on the $e-m$ exchange symmetry, which after gauging the symmetry, we obtain the $\text{Ising}\boxtimes \bar{\text{Ising}}$ topological order. If we choose to first glue a $Z_2$ SPT onto the $Z_2$ topological order and perform the $e-m$ exchange symmetry, we will obtain the $3^B_{3/2}\boxtimes 3^B_{-3/2}$ \cite{barkeshli2019symmetry}(or rather $SU(2)_2\boxtimes \bar{SU(2)}_2 = \text{Ising}^{(3)}\boxtimes \bar{\text{Ising}}^{(3)}$).
\iffalse
The first two excitations labeled by $i=1,2$ are trivial excitations $\one$ in
the 2+1D $Z_2$ topological order, which can carry $Z_2$-charge $+$ or $-$. The
next  two excitations labeled by $i=3,4$ are the emergent fermions $f$ in the
2+1D $Z_2$ topological order, which again can carry $Z_2$-charge $+$ or $-$.
The fifth excitation with quantum dimension 2 corresponds to $e\oplus m$.
Since the $Z_2$ action exchanges $e$ and $m$, the potential trap that traps a
$e$ particle will also trap a $m$ particle.  So $e$ and $m$ always appear
together as an excitations with quantum dimension $d=2$.  Such an excitation is
labeled by $i=5$ or by $e\oplus m$.  Certainly, such an excitations carry
$Z_2$-charge $\pm$.  Those $Z_2$ charges are also listed in Table
\ref{mextZ2b}.  The above discussion can be summarized as the following
correspondence between the label $i=1,2,3,4,5$ and the label $\one, e, m,f$ for
the excitations of the $Z_2$ topological order
\begin{align}
\label{i1emf}
 1 &\sim \one_+, & 2 &\sim \one_-, 
\nonumber\\
 3 &\sim f_+, & 4 &\sim f_-, & 5 &\sim e\oplus m_\pm, 
\end{align}
where subscripts indicate the $Z_2$-charge.\fi

The $\text{Ising} \boxtimes \bar{\text{Ising}} $ topological order has a $c=\bar
c=\frac12$ gapless boundary with gravitational anomaly.  Such an anomalous CFT
is described by the following 9-component partition functions in the quasi-particle basis,
\begin{align}
\label{eqn:Isbdy}
\begin{split}
\begin{pmatrix}
Z^{\bf 1 \bar{\one}} \\ Z^{\psi} \\ Z^{\sigma}  \\ Z^{\bar \psi} \\ Z^{\psi\bar{\psi}} \\ Z^{\sigma \bar{ \psi}} \\ Z^{\bar{\sigma}} \\ Z^{\psi \bar{\sigma}} \\ Z^{\sigma\bar{\sigma}} \\
\end{pmatrix}=\begin{pmatrix}
|\chi_0^\text{Is}|^2 \\ 
\chi_{\frac{1}{2}}^\text{Is}\bar \chi_0^\text{Is}\\ 
\chi_{\frac{1}{16}}^\text{Is}\bar\chi_0^\text{Is} \\ 
\chi_0^\text{Is}\bar \chi_{\frac{1}{2}}^\text{Is} \\
\chi_\frac{1}{2}^\text{Is}\bar \chi_{\frac{1}{2}}^\text{Is}\\
\chi_\frac{1}{16}^\text{Is}\bar \chi_{\frac{1}{2}}^\text{Is}\\
\chi_0^\text{Is}\bar \chi_{\frac{1}{16}}^\text{Is}\\
\chi_\frac{1}{2}^\text{Is}\bar \chi_\frac{1}{16}^\text{Is}\\
\chi_\frac{1}{16}^\text{Is}\bar \chi_\frac{1}{16}^\text{Is}\\
\end{pmatrix}
\end{split}.
\end{align}
 From the expression $Z^{\one}=
\chi^\text{Is}_0(\tau) \bar\chi^\text{Is}_0(\bar\tau)$, we see that such a
$c=\bar c=\frac12$ gapless boundary of the third $Z_2$ topological order (with
the $e,m$ exchange symmetry) has no relevant operators and is stable.

The anyons in this bulk topological order comes from anyons in the $Z_2$ symmetry enriched $Z_2$ topological order which either carry the $Z_2$-charges or are $Z_2$ symmetry twist defect, as listed in Table \ref{mextZ2b}. From this correspondence, we can
obtain the partition functions in the symmetry twist basis:  
\iffalse
\begin{align}
Z^{\one}_{ 1,1} &= |\chi^\text{Is}_0|^2 +|\chi^\text{Is}_\frac12|^2 , 
\nonumber\\
Z^{\one}_{ 1,-1} &= |\chi^\text{Is}_0|^2 -|\chi^\text{Is}_\frac12|^2 , 
\nonumber\\
Z^{f}_{ 1,1} &= \chi^\text{Is}_0 \bar\chi^\text{Is}_\frac12+\chi^\text{Is}_\frac12 \bar\chi^\text{Is}_0, 
\nonumber\\
Z^{f}_{ 1,-1} &= \chi^\text{Is}_0 \bar\chi^\text{Is}_\frac12-\chi^\text{Is}_\frac12 \bar\chi^\text{Is}_0, 
\nonumber\\
Z^{e\oplus m}_{ 1,1} &= 2|\chi^\text{Is}_{\frac1{16}}|^2, 
\nonumber\\
Z^{e\oplus m}_{ 1,-1} &= 0, 
\nonumber\\
Z^{v}_{ -1,1} &=  (\chi^\text{Is}_{\frac1{16}} \bar\chi^\text{Is}_0 +\chi^\text{Is}_{\frac1{16}} \bar\chi^\text{Is}_\frac12) ,
\nonumber\\
Z^{\bar v}_{ -1,1} &=  (\chi^\text{Is}_0 \bar\chi^\text{Is}_{\frac1{16}} + \chi^\text{Is}_\frac12 \bar\chi^\text{Is}_{\frac1{16}} ) ,
\nonumber\\
Z^{v}_{ -1,-1} &=  \ee^{\ii \th}(\chi^\text{Is}_{\frac1{16}} \bar\chi^\text{Is}_0 -\chi^\text{Is}_{\frac1{16}} \bar\chi^\text{Is}_\frac12) ,
\nonumber\\
Z^{\bar v}_{ -1,-1} &=  (\chi^\text{Is}_0 \bar\chi^\text{Is}_{\frac1{16}} - \chi^\text{Is}_\frac12 \bar\chi^\text{Is}_{\frac1{16}} ).
\end{align}
\fi

\begin{align}
\begin{split}
Z^{\one}_{ 1,1} &= |\chi_0|^2 +|\chi_\frac12|^2 , \\
Z^{f}_{ 1,1} &= \chi_0 \bar\chi_\frac12+\chi_\frac12 \bar\chi_0, \\
Z^{e}_{ 1,1} &=|\chi_{\frac1{16}}|^2, \\
Z^{m}_{ 1,1} &=|\chi_{\frac1{16}}|^2, \\[1em]
Z^{\one}_{ 1,-1} &= |\chi_0|^2 -|\chi_\frac12|^2 , \\
Z^{f}_{ 1,-1} &= -\chi_0 \bar\chi_\frac12+\chi_\frac12 \bar\chi_0, \\[1em]
Z^{v_+}_{-1,1} &= \chi_{\frac1{16}}( \bar\chi_0 +\bar\chi_\frac12) ,\\
Z^{v_-}_{-1,1} &= (\chi_0 + \chi_\frac12) \bar\chi_{\frac1{16}} ,\\[1em]
Z^{v_+}_{-1,-1}  &= \eta\chi_{\frac1{16}}( \bar\chi_0 -\bar\chi_\frac12) ,\\
Z^{v_-}_{-1,-1}  &=  \eta^*  (\chi_0 - \chi_\frac12) \bar\chi _{\frac1{16}} .
\end{split}
\end{align}
where $\eta$ is a phase factor $|\eta|=1$. The $Z^{v_+}_{-1,1}$ and $Z^{v_-}_{-1,1}$ are the two kinds of twisting defects $v_{\pm}$ that exchange $e$ and $m$ particles. The fusion rules of them with particles in the $Z_2$ topological order are as follows,
\begin{align}
\begin{split}
 & e\otimes v_\pm = m\otimes v_\pm=v_\mp\,,\\
 & f\otimes v_\pm =v_\pm \,,\\
 & v_\pm \otimes v_\pm = 1\oplus f \,,\\
 & v_\pm \otimes v_\mp = e\oplus m\,.
 \end{split}
\end{align}

The modular matrices, which are unitary,\citep{barkeshli2019symmetry} are as follows 
\begin{align}
\label{eqn:EMcrossedST}
&\scriptsize{
S=\begin{pmatrix}
 S^{\calD (\ZZ_2)}& & & & & \\
 & & &\frac{1}{\sqrt{2}} &\frac{1}{\sqrt{2}} & & \\
 & & &-\frac{1}{\sqrt{2}} &\frac{1}{\sqrt{2}} & & \\
 & \frac{1}{\sqrt{2}} &-\frac{1}{\sqrt{2}} & & & & \\
 & \frac{1}{\sqrt{2}} &\frac{1}{\sqrt{2}} & & & & \\
 & && & & 0& \eta^2\\
& && & & \eta^{-2} & 0\\
\end{pmatrix}},\\
&\scriptsize{
T=\begin{pmatrix}
T^{\calD (\ZZ_2)} & & & & & \\
 & 1& 0& & & & \\
 & 0& -1& & & & \\
 & &  & & & \gamma^{-1}\eta^{-1}&0 \\
 & & & & & 0& \gamma\eta\\
 & & & \gamma^{-1}\eta & 0& & \\
& && 0& \gamma \eta^{-1} &  & \\
\end{pmatrix}},
\end{align}
where $S^{\calD (\ZZ_2)}$ and $T^{\calD (\ZZ_2)}$ are the modular matrix for $\ZZ_2$ topological order as in (\ref{Z2sSTmat}). 

The partition functions under $\ZZ_2$ twist is related to the vector of partition functions (\ref{eqn:Isbdy}) describing the fully gapless boundary of $\ZZ_2$ gauged via a basis transformation, 
\begin{align}
\begin{split}
&\begin{pmatrix}
Z^\one_{1,1} \\ Z^\one_{1,-1}
\end{pmatrix}=\begin{pmatrix}
1 & 1 \\ 1 & -1
\end{pmatrix}\begin{pmatrix}
Z^\one \\Z^{\psi\bar\psi}
\end{pmatrix},\\
&\begin{pmatrix}
Z^f_{1,1} \\ Z^f_{1,-1}
\end{pmatrix}=\begin{pmatrix}
1 & 1 \\ 1 & -1
\end{pmatrix}\begin{pmatrix}
Z^\psi \\Z^{\bar\psi}
\end{pmatrix},\\
&~~~~~Z^{[e]}_{1,1}~=~Z^{\sigma\bar\sigma},\\
&\begin{pmatrix}
Z^{v_+}_{-1,1} \\ Z^{v_+}_{-1,-1}
\end{pmatrix}=\begin{pmatrix}
1 & 1 \\ \eta & -\eta
\end{pmatrix}\begin{pmatrix}
Z^\sigma \\Z^{\sigma\bar\psi}
\end{pmatrix},\\
&\begin{pmatrix}
Z^{v_-}_{-1,1} \\ Z^{v_-}_{-1,-1}
\end{pmatrix}=\begin{pmatrix}
1 & 1 \\ \eta^* & -\eta^*
\end{pmatrix}\begin{pmatrix}
Z^{\bar \sigma} \\Z^{\psi \bar \sigma}
\end{pmatrix}.
\end{split}
\end{align}

Let us explain the correspondence of the SET boundary and the boundary of topological order after gauging described by the above transformation.  First, without symmetry twist,
we have multi-component partition function due to the non-invertible anomaly
that corresponds to the 2+1D $\ZZ_2$ topological order. Thus we may have four
partition functions $Z^\one_{1,1}$, $Z^e_{1,1}$, $Z^m_{1,1}$, and $Z^f_{1,1}$.
Next, $e$ and $m$ together form the single orbit under the $Z_2$ symmetry, and will become a single
quasiparticle $[e]=\{e,m\}$ once the symmetry is gauged.  So we actually have 3 partition functions
$Z^\one_{1,1}$, $Z^{[e]}_{1,1}$, and $Z^f_{1,1}$, in the absence of symmetry twists in the bulk.

The partition function $Z^\one_{1,1}$ contain trivial excitaion $\one$ which
can have $\ZZ_2$ charge $\pm$, \ie
%. From \eqn{i1emf}, we see that $Z^\one_{1,1}$
%contain excitations 
$i=1 \sim \one$ excitation and $i=2 \sim \psi\bar \psi$ excitation.  As a
result $Z^{\one}_{1,1} = |\chi^\text{Is}_0|^2 +|\chi^\text{Is}_\frac12|^2$.
Since $i=1 \sim \one$ charge $Z_2$ carries charge $+$, and $i=2 \sim
\psi\bar\psi$ carries charge $Z_2$ charge $-$, we see that $Z^{\one}_{1,-1}
= |\chi^\text{Is}_0|^2 -|\chi^\text{Is}_\frac12|^2$.  Similarly, we can obtain
$Z^f_{1,1}$ and $Z^f_{1,-1}$.

The $\ZZ_2$-orbit $[e]$ corresponds to $i= 5 \sim \sigma\bar\sigma$. Note that $\{e,m\}$ are exchanged by the $\ZZ_2$ symmetry, this subspace can be reduced to a trivial representation and a sign representation of $\ZZ_2$. Correspondingly, $Z^{[e]}_{ 1,1}$ contains two equal contributions with $\ZZ_2$ charge $\pm$. That is, $Z^{[e]}_{ 1,1} = |\chi^\text{Is}_{\frac1{16}}|^2$, and
$Z^{[e]}_{ 1,-1}=0$.

Next, consider the partition functions with a $\ZZ_2$ symmetry twist in the
space direction $Z^{v_+}_{-1,\pm 1}$.  This appears when in the bulk there are topological excitations labeled by $i=6,7,8,9$ which carry a gauge flux of the gauged $\ZZ_2$ symmetry. Thus, the components of partition functions are labeled by these topological excitations.  Let us consider a partition function $Z^{v_+}_{-1,\pm 1}$ that contains
$i= 7 \sim \sigma $: $Z^{v_+}_{-1,\pm 1} = \chi^\text{Is}_{\frac1{16}}
\bar\chi^\text{Is}_0+ \cdots$.  It should also contain the fusion of $i= 7  \sim \si$ with $\one_+$ and $\one_- \sim
\psi\bar\psi$.
From the Ising fusion rule
\begin{align}
 \psi \otimes \psi &=\one, \ \ \ \ \ \si \otimes \si =\one \oplus \psi,  \ \ \ \
 \psi \otimes \si = \si,
\nonumber\\
 \bar\psi \otimes \bar\psi &=\bar\one, \ \ \ \ \ \bar\si \otimes \bar\si =\bar\one \oplus \bar\psi, \ \ \ \ \
 \bar\psi \otimes \bar\si = \bar\si,
\end{align}
we see that $Z^{v_+}_{-1,\pm 1}$ should also contain $i= 9 \sim
(\frac1{16},\frac12) \sim \si\bar\psi$.  Therefore, we have $Z^{v_+}_{-1,1} =
\chi^\text{Is}_{\frac1{16}} \bar\chi^\text{Is}_0 +\chi^\text{Is}_{\frac1{16}}
\bar\chi^\text{Is}_\frac12$.  To obtain $Z^{v_+}_{-1,-1}$, we note that $\one_-
 \sim \psi\bar\psi$ is the $\ZZ_2$ gauge charge, as it has $\pi$ mutual statistics with the gauge fluxes $\sigma,\bar{\sigma}$. Thus the
two terms, $\chi^\text{Is}_{\frac1{16}} \bar\chi^\text{Is}_0$ and
$\chi^\text{Is}_{\frac1{16}} \bar\chi^\text{Is}_\frac12$, carry different $\ZZ_2$ charge.  However,  we do not know the net $\ZZ_2$ charge of
$\chi^\text{Is}_{\frac1{16}} \bar\chi^\text{Is}_0$ which can be fractional.
Therefore $Z^{v_+}_{-1,-1} = \ee^{\ii \th}(\chi^\text{Is}_{\frac1{16}}
\bar\chi^\text{Is}_0 -\chi^\text{Is}_{\frac1{16}} \bar\chi^\text{Is}_\frac12)$.
Using a similar method, we obtain $Z^{v_-}_{-1,\pm 1}$.  But in this case, we know that $\chi^\text{Is}_0
\bar\chi^\text{Is}_{\frac1{16}}$ carries $\ZZ_2$ charge $+$, since the $\ZZ_2$
charge is carried only by the sector $\one,\psi,\si$, not by the sector
$\bar\one,\bar\psi,\bar\si$.

A noticeable aspect of the unitary modular $S$ matrix (\ref{eqn:EMcrossedST}) describing the SET phase is that each block $S_{(g,h)}$ has dimensions greater than $1$. It means that on the boundary, under the $S$ transformation, a state in the untwisted sector, does not become a state in a single sector of twisted Hilbert space, but a linear superposition of several twisted sectors. For example, $Z_{1,-1}^{\one}(-1/\tau) = \frac{1}{\sqrt{2}}\left(Z_{-1,1}^{v_+}(\tau)+Z_{-1,1}^{v_-}(\tau)\right)$. This is a generic feature of a SET phase with defect types greater than $1$, or rather when there are invariant anyon types under the symmetry. We will study a further example with this feature in subsubsection \ref{subsubsec:EMinZ3}.

Finally, note that if we start with $\text{Ising}\boxtimes \bar{\text{Ising}}$ and let the anyon $\psi\bar\psi$ anyon condensed, the phase becomes the $\ZZ_2$ topological order with the $e-m$ exchange symmetry. This suggests there exists a boundary state of $\text{Ising}\boxtimes \bar{\text{Ising}}$ that $\psi\bar{\psi}$ is condensed, yet with gapless excitations on the boundary when the bulk anyon braids trivially with $\psi\bar{\psi}$. Indeed, we find this boundary as described in the following,
\begin{align}
 Z^{\one} &= |\chi^\text{Is}_0|^2 + |\chi^\text{Is}_\frac{1}{2}|^2,
 \nonumber \\ 
Z^{\sigma } &= 0,
 \nonumber \\ 
Z^{\psi } &= \chi^\text{Is}_0 \bar\chi^\text{Is}_\frac{1}{2} + \chi^\text{Is}_\frac{1}{2} \bar\chi^\text{Is}_0,
 \nonumber \\ 
Z^{\bar \sigma} &= 0,
 \nonumber \\ 
Z^{\sigma \bar \sigma} &= |\chi^\text{Is}_\frac{1}{16}|^2,
 \nonumber \\ 
Z^{\psi \bar \sigma} &= 0,
 \nonumber \\ 
Z^{\bar \psi} &= \chi^\text{Is}_0 \bar\chi^\text{Is}_\frac{1}{2} + \chi^\text{Is}_\frac{1}{2} \bar\chi^\text{Is}_0 ,
 \nonumber \\ 
Z^{\sigma \bar \psi} &= 0,
 \nonumber \\ 
Z^{\psi \bar \psi} &= |\chi^\text{Is}_0|^2 + |\chi^\text{Is}_\frac{1}{2}|^2. 
\end{align}
This boundary satisfies the conditions (\ref{eqn:condensedbdy}) to describe a boundary phase with anyon condensed.

\subsection{$\ZZ_2$ symmetry in $\ZZ_3$ topological order}

\subsubsection{$\ZZ_2$ charge conjugation symmetry in $\ZZ_3$ topological order}
We consider the bulk to be the $\ZZ_3$ topological order.  The anyons in the $\ZZ_3$ topological orders are $e^am^b, a,b=0,1,2$. The $\ZZ_3$ topological order can be enriched with a global $Z_2$ charge conjugation symmetry, $e^am^b\rightarrow e^{-a}m^{-b}$. When we gauge the $\ZZ_2$ symmetry in the bulk, we obtain the $S_3$ topological order, also known as the quantum double $\calD (S_3)$. 
Now consider the boundary vector of partition functions. We can expect that it is the same as that on the boundary of $\calD(S_3)$ topological order. 

The question is what is the transformation between the two vectors of partition functions: one is in the quasiparticle-twisted-by-symmetry basis, the other in the quasiparticle basis.

To begin with, on the boundary of $\ZZ_3$ topological order, the vector of partition functions is composed of characters of the three-state Potts conformal field theory. 
\begin{align}
\label{eqn:Z3bdy}
 Z^{\bf 1} &=  |\chi^{m6}_0 + \chi^{m6}_3|^2 +  |\chi^{m6}_\frac{2}{5} +  \chi^{m6}_\frac{7}{5}|^2,  
 \nonumber \\ 
Z^e &=  |\chi^{m6}_\frac{2}{3}|^2 +  |\chi^{m6}_\frac{1}{15}|^2,  
 \nonumber \\ 
Z^{e^2} &=  |\chi^{m6}_\frac{2}{3}|^2 +  |\chi^{m6}_\frac{1}{15}|^2,  
 \nonumber \\ 
Z^m &=  |\chi^{m6}_\frac{2}{3}|^2 +  |\chi^{m6}_\frac{1}{15}|^2,  
 \nonumber \\ 
Z^{me} &=  \chi^{m6}_0 \bar\chi^{m6}_\frac{2}{3} +  \chi^{m6}_3 \bar\chi^{m6}_\frac{2}{3} +  \chi^{m6}_\frac{2}{5} \bar\chi^{m6}_\frac{1}{15} +  \chi^{m6}_\frac{7}{5} \bar\chi^{m6}_\frac{1}{15} , 
 \nonumber \\ 
Z^{me^2} &=  \chi^{m6}_\frac{2}{3} \bar\chi^{m6}_0 +  \chi^{m6}_\frac{2}{3} \bar\chi^{m6}_3 +  \chi^{m6}_\frac{1}{15} \bar\chi^{m6}_\frac{2}{5} +  \chi^{m6}_\frac{1}{15} \bar\chi^{m6}_\frac{7}{5},  
 \nonumber \\ 
Z^{m^2} &=  |\chi^{m6}_\frac{2}{3}|^2 +  |\chi^{m6}_\frac{1}{15}|^2, 
 \nonumber \\ 
Z^{m^2e} &=  \chi^{m6}_\frac{2}{3} \bar\chi^{m6}_0 +  \chi^{m6}_\frac{2}{3} \bar\chi^{m6}_3 +  \chi^{m6}_\frac{1}{15} \bar\chi^{m6}_\frac{2}{5} +  \chi^{m6}_\frac{1}{15} \bar\chi^{m6}_\frac{7}{5},  
 \nonumber \\ 
Z^{m^2e^2} &=  \chi^{m6}_0 \bar\chi^{m6}_\frac{2}{3} +  \chi^{m6}_3 \bar\chi^{m6}_\frac{2}{3} +  \chi^{m6}_\frac{2}{5} \bar\chi^{m6}_\frac{1}{15} +  \chi^{m6}_\frac{7}{5} \bar\chi^{m6}_\frac{1}{15} .
\end{align}

The topological order after gauging the $\ZZ_2$ symmetry has the following anyons. The symmetry defect becomes a fluxeon, let us call it $c$ anyon. The other fluxeons come from the anyons before gauging. The anyons invariant can carry either trivial or sign representation of the $\ZZ_2$, and therefore split. The anyons related to each other under $\ZZ_2$ symmetry form a $\ZZ_2$ orbit, a single anyon in the gauged model. 

Let us derive these anyons, as well as the boundary vector of partition functions. First, consider the vacuum sector. It is invariant under the symmetry. Therefore certain excitations on the boundary of the vacuum sector carry the sign representation under the on-site $\ZZ_2$ action, 
\begin{align}
Z^{{\bf 1}}_{1,-1}= |\chi^{m6}_0 -\chi^{m6}_3|^2 +  |\chi^{m6}_\frac{2}{5} - \chi^{m6}_\frac{7}{5}|^2.
\end{align}

Combining this sector with $Z^{{\bf 1}}=Z^{{\bf 1}}_{1,1}$, we can separate the two anyonic sectors in the topological order after gauging. One anyon ${\tilde{{\bf 1}}}$, carrying trivial representation of the on-site $\ZZ_2$ symmetry, and the other   $a^1$, carrying the sign representation. That is through the linear combination 
\begin{align}
\begin{pmatrix}
Z^{\bf 1}_{1,1} \\ Z^{\bf 1}_{1,-1} 
\end{pmatrix}=\begin{pmatrix}
1 & 1  \\ 
1 & -1  
\end{pmatrix}\begin{pmatrix}
Z^{\tilde{\bf 1}} \\ Z^{a^1} 
\end{pmatrix}.
\end{align}
The resulting components of partition functions are
\begin{align}
Z^{\tilde{{\bf 1}}} = &|\chi^{m6}_0|^2  +  |\chi^{m6}_3|^2 +  |\chi^{m6}_\frac{2}{5}|^2  +  |\chi^{m6}_\frac{7}{5}|^2,\nonumber\\
Z^{a^1} = & \chi^{m6}_0 \bar\chi^{m6}_3 + \chi^{m6}_3 \bar\chi^{m6}_0 + \chi^{m6}_\frac{2}{5} \bar\chi^{m6}_\frac{7}{5} +  \chi^{m6}_\frac{7}{5} \bar\chi^{m6}_\frac{2}{5}.
\end{align}

We can obtain the twisted sector $Z^{\bf 1}_{-1,1}$ by the $S$ transformation of the sector $Z^{\bf 1}_{1,-1}$, 
\begin{align}
Z^{{\bf 1}}_{-1,1}=|\chi^{m6}_{\frac 1 8}+\chi^{m6}_{\frac{13}{8}}|^2+|\chi^{m6}_{\frac{1}{40}}+\chi^{m6}_{\frac{21}{40}}|^2.
\end{align}
It is the $Z_2$ twisted sector. Under the $\ZZ_2$ symmetry, now acted as an on-site symmetry, it becomes
\begin{align}
Z^{{\bf 1}}_{-1,-1}=|\chi^{m6}_{\frac 1 8}-\chi^{m6}_{\frac{13}{8}}|^2+|\chi^{m6}_{\frac{1}{40}}-\chi^{m6}_{\frac{21}{40}}|^2.
\end{align}
$Z^{{\bf 1}}_{-1,1}$ contains both the trivial representation as well as the sign representation of the on-site $\ZZ_2$ symmetry. They correspond to two defect anyon sectors $c$ and $c^1$. The partition function of the two sectors are related by the following,
\begin{align}
\begin{pmatrix}
 Z^{\bf 1}_{-1,1} \\ Z^{\bf 1}_{-1,-1}
\end{pmatrix}=\begin{pmatrix}
 1 & 1 \\
 1 & -1
\end{pmatrix}\begin{pmatrix}
 Z^{c} \\ Z^{c^1}
\end{pmatrix}.
\end{align} 
The gapless excitations on their boundary are as follows,
\begin{align}
Z^{c}=&|\chi^{m6}_{\frac 1 8}|^2+|\chi^{m6}_{\frac{13}{8}}|^2+|\chi^{m6}_{\frac{1}{40}}|^2+|\chi^{m6}_{\frac{21}{40}}|^2,\nonumber\\
Z^{c^1}=&\chi^{m6}_{\frac 1 8}\bar \chi^{m6}_{\frac{13}{8}}+\chi^{m6}_{\frac{13}{8}}\bar \chi^{m6}_{\frac{1}{8}}+\chi^{m6}_{\frac{1}{40}}\bar \chi^{m6}_{\frac{21}{40}}+\chi^{m6}_{\frac{21}{40}}\bar \chi^{m6}_{\frac{1}{40}}.
\end{align}

Let us summarize. Before gauging, the exchange symmetry acts on anyons in two ways, exchanging anyons, and as the on-site $\ZZ_2$ symmetry on those anyons invariant under exchanging. In the gauged model, the symmetry defects become dynamical, added to the family of anyons.

Now consider anyons get permuted under the symmetry. After the symmetry is gauged, they get combined and form $\ZZ_2$-orbits. The boundary local excitations of those anyons, before gauging are the same. Those local excitations contribute to one component of the vector of partition functions, labeled by a representative anyon in the $\ZZ_2$-orbit. In the current case, we have four $\ZZ_2$-orbits. They become four types of anyons that have quantum dimension $2$ in the gauged model, 
\begin{align}
[e]&=\{e, e^2\}\sim a^2,\nonumber \\
[m]&=\{m,m^2\}\sim b,~~\nonumber \\
[me]&=\{ me,m^2e^2 \}\sim b^1,\nonumber \\
[m^2e]&=\{m^2e,me^2\}\sim b^2.
\end{align}
Among them, $[e]\sim a^2$ carries the representation of the $\ZZ_3$ gauge group, the other three carry the $\ZZ_3$ flux. 

The boundary partition functions are the following,
\begin{align}
Z^{[e]} &=  |\chi^{m6}_\frac{2}{3}|^2 +  |\chi^{m6}_\frac{1}{15}|^2,  
 \nonumber \\ 
 Z^{[m]} &=  |\chi^{m6}_\frac{2}{3}|^2 +  |\chi^{m6}_\frac{1}{15}|^2,  
 \nonumber \\ 
 Z^{[me]} &=  \chi^{m6}_0 \bar\chi^{m6}_\frac{2}{3} +  \chi^{m6}_3 \bar\chi^{m6}_\frac{2}{3} +  \chi^{m6}_\frac{2}{5} \bar\chi^{m6}_\frac{1}{15} +  \chi^{m6}_\frac{7}{5} \bar\chi^{m6}_\frac{1}{15},  
 \nonumber \\ 
Z^{[m^2e]} &=  \chi^{m6}_\frac{2}{3} \bar\chi^{m6}_0 +  \chi^{m6}_\frac{2}{3} \bar\chi^{m6}_3 +  \chi^{m6}_\frac{1}{15} \bar\chi^{m6}_\frac{2}{5} +  \chi^{m6}_\frac{1}{15} \bar\chi^{m6}_\frac{7}{5} . 
\end{align}

In summary, we obtain the full partition functions on the boundary. 
\begin{align}
 Z^{\one} &=  |\chi^{m6}_{0}|^2 +  |\chi^{m6}_{3}|^2 +  |\chi^{m6}_{\frac{2}{5}}|^2 +  |\chi^{m6}_{\frac{7}{5}}|^2, 
 \nonumber \\ 
Z^{a^1} &=  \chi^{m6}_{0} \bar\chi^{m6}_{3} +  \chi^{m6}_{3} \bar\chi^{m6}_{0} +  \chi^{m6}_{\frac{2}{5}} \bar\chi^{m6}_{\frac{7}{5}} +  \chi^{m6}_{\frac{7}{5}} \bar\chi^{m6}_{\frac{2}{5}} ,
 \nonumber \\ 
Z^{a^2} &=  |\chi^{m6}_{\frac{2}{3}}|^2 +  |\chi^{m6}_{\frac{1}{15}}|^2 ,
 \nonumber \\ 
Z^{b} &=  |\chi^{m6}_{\frac{2}{3}}|^2 +  |\chi^{m6}_{\frac{1}{15}}|^2 ,\nonumber
\\ 
Z^{b^1} &=  \chi^{m6}_{0} \bar\chi^{m6}_{\frac{2}{3}} +  \chi^{m6}_{3} \bar\chi^{m6}_{\frac{2}{3}} +  \chi^{m6}_{\frac{2}{5}} \bar\chi^{m6}_{\frac{1}{15}} +  \chi^{m6}_{\frac{7}{5}} \bar\chi^{m6}_{\frac{1}{15}} ,
 \nonumber \\ 
Z^{b^2} &=  \chi^{m6}_{\frac{2}{3}} \bar\chi^{m6}_{0} +  \chi^{m6}_{\frac{2}{3}} \bar\chi^{m6}_{3} +  \chi^{m6}_{\frac{1}{15}} \bar\chi^{m6}_{\frac{2}{5}} +  \chi^{m6}_{\frac{1}{15}} \bar\chi^{m6}_{\frac{7}{5}} ,
 \nonumber \\ 
Z^{c} &=  |\chi^{m6}_{\frac{1}{8}}|^2 +  |\chi^{m6}_{\frac{13}{8}}|^2 +  |\chi^{m6}_{\frac{1}{40}}|^2 +  |\chi^{m6}_{\frac{21}{40}}|^2 ,
 \nonumber \\ 
Z^{c^1} &=  \chi^{m6}_{\frac{1}{8}} \bar\chi^{m6}_{\frac{13}{8}} +  \chi^{m6}_{\frac{13}{8}} \bar\chi^{m6}_{\frac{1}{8}} +  \chi^{m6}_{\frac{1}{40}} \bar\chi^{m6}_{\frac{21}{40}} +  \chi^{m6}_{\frac{21}{40}} \bar\chi^{m6}_{\frac{1}{40}} .
\end{align} 
This boundary is exactly the same as that of a gapless boundary of the $\calD(S_3)$ topological order, where no anyon is condensed on the boundary. 
%\begin{align}
%\calS: Z^{\one}-Z^{a^1}\rightarrow Z^c +Z^{c^1}.
%\end{align}

Reversely, in the $S_3$ topological order, the Abelian anyon $a^1$ is a self-boson. In modern terminology, this worldline of this anyon generates a $\ZZ_2$ one-form symmetry that is anomaly-free. Turning on an appropriate interaction, we can let $a^1$ condensed. This leads to confinement of the $c$ and $c^1$ anyons, and the split of $a^1, b, b^1, b^2$. This condensed phase is the phase with $\ZZ_3$ topological order, with a gapless boundary described by (\ref{eqn:Z3bdy}).

We close this subsection, by giving another gapless boundary of $S_3$ topological order, where the $a^1$ anyon is condensed on the boundary, 
\begin{align}
 Z^{\one} &=  |\chi^{m6}_{0}+\chi^{m6}_{3}|^2 +  |\chi^{m6}_{\frac{2}{5}}+\chi^{m6}_{\frac{7}{5}}|^2 , 
 \nonumber \\ 
Z^{a^1} &= |\chi^{m6}_{0}+\chi^{m6}_{3}|^2 +  |\chi^{m6}_{\frac{2}{5}}+\chi^{m6}_{\frac{7}{5}}|^2,
 \nonumber \\ 
Z^{a^2} &=  |\chi^{m6}_{\frac{2}{3}}|^2 +  |\chi^{m6}_{\frac{1}{15}}|^2 , 
 \nonumber \\ 
Z^{b} &=  |\chi^{m6}_{\frac{2}{3}}|^2 +  |\chi^{m6}_{\frac{1}{15}}|^2 , 
 \nonumber \\ 
Z^{ b^1} &=  \chi^{m6}_{0} \bar\chi^{m6}_{\frac{2}{3}} +  \chi^{m6}_{3} \bar\chi^{m6}_{\frac{2}{3}} +  \chi^{m6}_{\frac{2}{5}} \bar\chi^{m6}_{\frac{1}{15}} +  \chi^{m6}_{\frac{7}{5}} \bar\chi^{m6}_{\frac{1}{15}} ,
 \nonumber \\ 
Z^{ b^2} &=  \chi^{m6}_{\frac{2}{3}} \bar\chi^{m6}_{0} +  \chi^{m6}_{\frac{2}{3}} \bar\chi^{m6}_{3} +  \chi^{m6}_{\frac{1}{15}} \bar\chi^{m6}_{\frac{2}{5}} +  \chi^{m6}_{\frac{1}{15}} \bar\chi^{m6}_{\frac{7}{5}} ,
 \nonumber \\ 
Z^{c} &= 0,
 \nonumber \\ 
Z^{ c^1} &= 0.
\end{align}
It satisfies the empirical rules (\ref{eqn:condensedbdy}) for the gapless boundary state with anyon condensed. 

\iffalse
There could be more than one way to arrange the vector of partition functions in the model after gauging. Let us discuss one. 

\begin{align}
\begin{pmatrix}
Z^{e} \\ Z^{e^2}
\end{pmatrix}=\begin{pmatrix}
1 & 1  \\ 
1 & -1  \\
\end{pmatrix}\begin{pmatrix}
Z^{([e], +)} \\ Z^{([e], -)}
\end{pmatrix}.
\end{align}
In particular, $Z^{([e],-)}=0$, since the $Z_2$ exchange symmetry has been broken. The gauged theory is a symmetry-breaking phase of the $Z_2$ exchange symmetry. 
\fi
\subsubsection{The electric-magnetic duality symmetry in $\ZZ_3$ topological order}\label{subsubsec:EMinZ3}
In the example in \ref{subsubsec:EMinZ2TO}, there is a prefactor $\frac{1}{\sqrt{2}}$ appearing in the $S$ matrix given in (\ref{eqn:EMcrossedST}), describing the topological order enriched by the $e-m$ exchange symmetry. It means that  on the boundary of this SET phase, the component in the twisted sector and that  in untwisted sector are not related by a one-to-one map under the $S$ transformation. Now let us show with a further example that the prefactor is tied to the number of types of invariant anyons under the symmetry. Since two symmetry defects fuse into those invariant anyons, this prefactor is also related to that the symmetry defect is non-Abelian -- its quantum dimension is greater than $1$.

We consider the EM exchange symmetry in the $\ZZ_3$ topological order. The symmetry defect now has quantum dimension $\sqrt{3}$. We will see that when applying the $S$ transformation on the boundary of the SET phase, the unitary $S$ matrix involves a prefactor $1/\sqrt{3}$.

We start with the boundary of the $\ZZ_3$ topological order. The anyons in the $\ZZ_3$ topological orders are $e^am^b, a,b=0,1,2$. 
We begin with the same vector of partition functions (\ref{eqn:Z3bdy}), composed of characters of the three-state Potts conformal field theory.

The $\ZZ_3$ topological order can be equivalently written as the $\ZZ_3^{(1)}\boxtimes \bar{\ZZ}_3^{(1)}$ topological order.\cite{barkeshli2019symmetry} That is a tensor product of two chiral topological orders. The $\ZZ_3^{(1)}$ TO has chiral central charge $c_-=2\mod 8$. It has anyons $a_+^i, i=0,1,2$, obeying the fusion rules $a_+^i\times a_+^j= a^{[i+j]_3}$, and all trivial $F$-symbols. The spin for $a^i_+$ is $\theta_{a_+^i}=e^{\ii \frac{2\pi}{3} i^2}$. The $\bar{\ZZ}_3^{(1)}$ TO is the time-reversal conjugate of $\ZZ_3^{(1)}$ TO. The charge anyon $e$ and the flux anyon $m$ are related to anyons $a_+\in Z_3^{(1)}$ and $a_-\in \bar{Z}_3^{(1)}$ via
\begin{align}
e= a_+a_-, ~~m=a_+a_-^2. 
\end{align}

It follows that the vector of partition functions can also be in the basis labeled by anyons in $\ZZ_3^{(1)}\boxtimes \bar{\ZZ}_3^{(1)}$ TO, as in the following. From now on, we omit the superscript of the character indicating the rational CFT that the character comes from. 
\begin{align}
\label{eqn:DZ3}
 Z^{\bf 1} &=  |\chi_0+\chi_3|^2 +  |\chi_\frac{2}{5}+\chi_\frac{7}{5}|^2,  
 \nonumber \\ 
 Z^{a_+} &=  \chi_0 \bar\chi_\frac{2}{3} +  \chi_3 \bar\chi_\frac{2}{3} +  \chi_\frac{2}{5} \bar\chi_\frac{1}{15} +  \chi_\frac{7}{5} \bar\chi_\frac{1}{15},\nonumber\\
Z^{a_+^2} &=  \chi_0 \bar\chi_\frac{2}{3} +  \chi_3 \bar\chi_\frac{2}{3} +  \chi_\frac{2}{5} \bar\chi_\frac{1}{15} +  \chi_\frac{7}{5} \bar\chi_\frac{1}{15} , 
 \nonumber \\ 
Z^{a_-} &=  \chi_\frac{2}{3} \bar\chi_0 +  \chi_\frac{2}{3} \bar\chi_3 +  \chi_\frac{1}{15} \bar\chi_\frac{2}{5} +  \chi_\frac{1}{15} \bar\chi_\frac{7}{5},  
 \nonumber \\ 
 Z^{a_+a_-} &=  |\chi_\frac{2}{3}|^2 +  |\chi_\frac{1}{15}|^2,  
 \nonumber \\ 
Z^{a_+^2a_-} &=  |\chi_\frac{2}{3}|^2 +  |\chi_\frac{1}{15}|^2, 
 \nonumber \\ 
 Z^{a_-^2} &=  \chi_\frac{2}{3} \bar\chi_0 +  \chi_\frac{2}{3} \bar\chi_3 +  \chi_\frac{1}{15} \bar\chi_\frac{2}{5} +  \chi_\frac{1}{15} \bar\chi_\frac{7}{5},  
 \nonumber \\ 
 Z^{a_+a_-^2} &=  |\chi_\frac{2}{3}|^2 +  |\chi_\frac{1}{15}|^2,  
 \nonumber \\ 
Z^{a_+^2a_-^2} &=  |\chi_\frac{2}{3}|^2 +  |\chi_\frac{1}{15}|^2.
\end{align}

Then the EM exchange symmetry on the $\ZZ_3$ topological order acts as the charge conjugation (CC) on the anyons in $\bar{\ZZ}_3^{(1)}$ topological order, $\ZZ_2^{EM}: a_+^la_-^m\rightarrow a_+^la_-^{-l}$. In particular, there is a single type of topological defect $\sigma$ with quantum dimension $\sqrt{3}$. Its fusion rules with other anyons in $\bar{\ZZ}_3^{(1)}$ are 
\begin{align}
\sigma \times a_-^l= a_-^l\times \sigma = \sigma,~~\sigma\times \sigma=1+a_-+a_-^2,
\end{align}
where $l=0,1,2$. Further solving the $F$ symbols, one finds two $\ZZ_2$-crossed fusion categories, identified with two types of $\ZZ_3$ Tambara-Yamagami fusion categories. 

More specifically, the boundary partition function in the symmetry twisted basis of this SET phase is the following,
\begin{align}
\label{eqn:Z3twist}
Z^\sigma_{-1,1}=&\left(\chi_\frac{1}{8}+\chi_\frac{13}{8}\right)\left(\bar{\chi}_0+\bar{\chi}_3\right)\nn\\
&+\left(\chi_\frac{1}{40}+\chi_\frac{21}{40}\right)\left(\bar{\chi}_\frac{2}{5}+\bar{\chi}_\frac{7}{5}\right),\nn\\
Z^{a_+\sigma}_{-1,1}=&\left(\chi_\frac{1}{8}+\chi_\frac{13}{8}\right)\bar\chi_\frac{2}{3}\nn\\
&+\left(\chi_\frac{1}{40}+\chi_\frac{21}{40}\right)\bar\chi_\frac{1}{15},\nonumber \\
Z^{a_+^2\sigma}_{-1,1}=&Z^{a_+\sigma}_{-1,1}.
\end{align}

These are obtained from performing a $S$ transformation on the following components in the untwisted sector and acted by the symmetry 
\begin{align}
\label{eqn:Z3EMaction}
Z^{{\bf 1}}_{1,-1}=&\left(\chi_0-\chi_3\right)\left(\bar{\chi}_0+\bar{\chi}_3\right)\nonumber \\
&+\left(-\chi_\frac{2}{5}+\chi_\frac{7}{5}\right)\left(\bar{\chi}_\frac{2}{5}+\bar{\chi}_\frac{7}{5}\right),\nn\\
Z^{a_+}_{1,-1} =& \left( \chi_0 \bar\chi_\frac{2}{3} -  \chi_3 \bar\chi_\frac{2}{3} -  \chi_\frac{2}{5} \bar\chi_\frac{1}{15} +  \chi_\frac{7}{5} \bar\chi_\frac{1}{15}\right),\nonumber\\
Z^{a_+^2}_{1,-1} =& Z^{a_+}_{1,-1},
\end{align}
and the components in the untwisted sector without symmetry action are those in (\ref{eqn:DZ3}), each $Z^a$ in (\ref{HeisCFT16}) is relabeled as $Z^a_{1,1}$ as a component on the SET boundary. 

The remaining components can be obtained by a $T$ transformation, up to an overall phase that will be absorbed in the modular $T$ matrix,
\begin{align}
\label{eqn:Z3twist}
Z^\sigma_{-1,-1}=&\left(\chi_\frac{1}{8}-\chi_\frac{13}{8}\right)\left(\bar{\chi}_0+\bar{\chi}_3\right)\nn\\
&-\left(\chi_\frac{1}{40}-\chi_\frac{21}{40}\right)\left(\bar{\chi}_\frac{2}{5}+\bar{\chi}_\frac{7}{5}\right),\nn\\
Z^{a_+\sigma}_{-1,-1}=&\left(\chi_\frac{1}{8}-\chi_\frac{13}{8}\right)\bar\chi_\frac{2}{3}\nn\\
&-\left(\chi_\frac{1}{40}-\chi_\frac{21}{40}\right)\bar\chi_\frac{1}{15},\nonumber \\
Z^{a_+^2\sigma}_{-1,-1}=&Z^{a_+\sigma}_{-1,-1}.
\end{align}

The $S$ matrix acting on the boundary of the $\ZZ_2^{EM}$ SET phase is the following,
\begin{align}
S=\begin{pmatrix}
S^{\calD (\ZZ_3)} & & & \\
& & \frac{1}{\sqrt{3}}M^{\ZZ_3} &  \\
& \frac{1}{\sqrt{3}}M^{\ZZ_3} & & \\
& & & -\frac{1}{\sqrt{3}}M^{\ZZ_3} \\
\end{pmatrix},
\end{align}
where 
\begin{align}
M^{\ZZ_3}=\begin{pmatrix}
1 & 1 & 1 \\
1 & \omega & \omega^2 \\ 
1 & \omega^2 & \omega
\end{pmatrix},
\end{align}
is the character table of $\ZZ_3$ group and $\omega=e^{\ii \frac{2\pi}{3}}$. And the $T$ matrix is 
\begin{align}
T=\begin{pmatrix}
T^{\calD (\ZZ_3)} & & & \\
& T^{\ZZ_3}& &  \\
&  & & \eta_{\ZZ_3} T^{\ZZ_3}\\
& & \eta_{\ZZ_3} T^{\ZZ_3}&  \\
\end{pmatrix},
\end{align}
where 
\begin{align}
T^{\ZZ_3}= \begin{pmatrix}
1 & 0 & 0 \\ 0 & \omega & 0 \\ 0 & 0 & \omega 
\end{pmatrix},
\end{align}
is the same as the $T$ matrix for $\ZZ_3^{(1)}$ topological order, and $\eta_{\ZZ_3}$ here equals $e^{\ii 2\pi/8}$. The $S$ matrix here is unitary. Indeed, it is a theorem that the $\calC_{G}^\times$ BFC is $G$-crossed modular if and only if $\calC_0$ is a UMTC. We say that  the $\calC_{G}^\times$ is $G$-crossed modular if its $S$-matrix is $G$-graded unitary.\cite{barkeshli2019symmetry}

Notice that the prefactor appearing in the $S$ matrix expanded by the component twisted by symmetries (along temporal or spacial direction) is $\frac{1}{d}$ with $d=\sqrt{3}$, the same as the quantum dimension of the $EM$ exchange defect in the $\ZZ_3$ topological order. We take the assumption that in the presence of a topological defect in the bulk, the whole system is still a Hilbert space of integral dimensions. Specifically, the coefficients in $Z_{-1,1}^a$ are required to be integral. 

The $Z_{-1,1}$ is consistent with the sector twisted by the non-invertible line in the $\ZZ_3$ Tambara-Yamagami category. Indeed, after we group the components exchanged by the symmetry into the $\ZZ_2$-orbits, the vector becomes of $15$ dimensions, with the following components, 
\begin{align}
&\left(Z_{1,1}^{{\bf 1}}, Z_{1,-1}^{{\bf 1}}, Z_{-1,1}^{{\sigma}},Z_{-1,-1}^{\sigma},\right. \nn\\
&Z_{1,1}^{a_+}, Z_{1,-1}^{a_+}, Z_{-1,1}^{a_+\sigma},Z_{-1,-1}^{a_+\sigma},\nn\\
&Z_{1,1}^{a_+^2}, Z_{1,-1}^{a_+^2}, Z_{-1,1}^{a_+^2\sigma},Z_{-1,-1}^{a_+^2\sigma} ,\nn\\
&Z^{a_-\oplus a_-^2},Z^{a_+a_-\oplus a_+a_-^2},Z^{a_+^2a_-\oplus a_+^2a_-^2})^T.
\end{align}
The bulk becomes the $Z_3^{(1)}\boxtimes \bar{SU(2)}_4$ topological order after the EM symmetry is gauged.\cite{barkeshli2019symmetry} For our purpose, we only describe the anyon types here. After gauging, $a_-$ and $a_-^2$ form a $\ZZ_2$ orbit, a single anyon with quantum dimension $2$. For both $\one$ and $\sigma$, the remaining stabilizer group is $\ZZ_2$, so each splits into two anyons, one with trivial representation and one with sign representation. We denote them as $\one, z, \sigma, \sigma^z$, respectively. Their quantum dimensions are $1,1, \sqrt{3}, \sqrt{3}$. The dictionary between these anyons and their spin under the $SU(2)$, as well as the topological spins in $SU(2)_4$ topological order is as follows.
\begin{align}
\renewcommand*{\arraystretch}{1.5}
\begin{tabular}{| c | c c c c c | }
\hline 
anyon &$ \one$ &$ \sigma$ & $[a_-]$ &$ \sigma^z$ & $z$ \\
\hline
$SU(2)$ spin & $0$ & $\frac{1}{2}$ & $1$ & $\frac{3}{2}$ & $2$ \\
\hline
topological spin & $1$ & $e^{\ii 2\pi\frac{1}{8}}$ & $e^{\ii 2\pi\frac{1}{3}}$ & $e^{\ii 2\pi\frac{5}{8}}$ & $1$ \\
\hline
\end{tabular}\nn
\end{align}

Up to certain prefactors, the vector above can be transformed to the vector labeled by those anyons in $\ZZ_3^{(1)}\boxtimes \bar{SU(2)}_4$. 
\begin{align}
\begin{pmatrix}
Z^{{a_+^l}}_{1,1} \\ Z^{{a_+^l}}_{1,-1} \\ Z^{{a_+^l\sigma}}_{-1,1} \\ Z^{{a_+^l\sigma}}_{-1,-1}
\end{pmatrix}=\begin{pmatrix}
1 & 1 & & \\
1 & -1 & & \\
 & & 1 & 1 \\
 & & 1 & -1 \\
\end{pmatrix}\begin{pmatrix}
Z^{(a_+^l,\one)} \\ Z^{(a_+^l,z)} \\ Z^{(a_+^l,\sigma)} \\ Z^{(a_+^l,\sigma^z)}
\end{pmatrix},
\end{align}
\begin{align}
Z^{a_+^la_-\oplus a_+^la_-^2} = Z^{[a_+^la_-]},
\end{align}
where $l=0,1,2$. However, unfortunately, the vector produced by this transformation does not describe a gapless boundary of $\ZZ_3^{(1)}\boxtimes \bar{SU(2)}_4$. Especially, the component $Z^{(a_+,\sigma)}=Z^{(a_+,\frac{1}{2})}$ has a conformal spin that does not match that of the $(a_+, \frac{1}{2})$ anyon, which is $\frac{5}{24}$.   In fact, in $\calM(6,5)$, there is no combination of left and right characters with a chiral conformal spin $\frac{5}{24}$.

\iffalse
The modular matrices for the $CC$ gauged $\bar{Z}_3^{(1)}$ are
\begin{align}
&S=\begin{pmatrix}
 S^{\bar{Z}_3^{(1)}} & & &  \\
  & 0 & 1 &0 \\
  & 1 & 0 & 0 \\
  & 0 & 0 &1 \\
\end{pmatrix},~T=\begin{pmatrix}
T^{\bar{Z}_3^{(1)}}  & & &  \\
 & 1 & 0 &0 \\
 & 0 & 0 & \ii \\
 & 0 & \ii & 0 \\
\end{pmatrix},
\end{align}
where
\begin{align}
&S^{\bar{Z}_3^{(1)}}=\frac{1}{\sqrt{3}}\begin{pmatrix}
1 & 1 & 1 \\
1 & \omega & \omega^2 \\
1 & \omega^2 & \omega \\
\end{pmatrix},~T^{\bar{Z}_3^{(1)}}=\begin{pmatrix}
1 & 0 & 0 \\
0 & \omega^2 & 0 \\
0 & 0 & \omega^2 \\
\end{pmatrix}.
\end{align}
\fi

Nevertheless, the boundary of $\ZZ_3^{(1)}\boxtimes \bar{SU(2)}_4$, where the spin-$2$ anyon in $\bar{SU(2)}_4$ is condensed, has the following vector of partition function,
\begin{align}
 Z^{\one,0} &=  |\chi_0+\chi_3|^2 +  |\chi_\frac{2}{5}+\chi_\frac{7}{5}|^2 ,
 \nonumber \\ 
Z^{\one,\frac{1}{2}} &= 0,
 \nonumber \\ 
Z^{\one,1} &=  \chi_\frac{2}{3} \left(\bar\chi_0 +  \bar\chi_3\right) +  \chi_\frac{1}{15} \left(\bar\chi_\frac{2}{5} +   \bar\chi_\frac{7}{5} \right),
 \nonumber \\ 
Z^{\one,\frac{3}{2}} &= 0,
 \nonumber \\ 
Z^{\one,2} &=  |\chi_0+\chi_3|^2 +  |\chi_\frac{2}{5}+\chi_\frac{7}{5}|^2 ,
 \nonumber \\ 
Z^{a_+,0} &= \left( \chi_0  +  \chi_3\right) \bar\chi_\frac{2}{3} +  \left(\chi_\frac{2}{5}+  \chi_\frac{7}{5}\right) \bar\chi_\frac{1}{15}  ,
 \nonumber \\ 
Z^{a_+,\frac{1}{2}} &=0 ,
 \nonumber \\ 
Z^{a_+,1} &=  |\chi_\frac{2}{3}|^2 +  |\chi_\frac{1}{15}|^2  ,
 \nonumber \\ 
Z^{a_+,\frac{3}{2}} &= 0,
 \nonumber \\ 
Z^{a_+,2} &= \left( \chi_0  +  \chi_3\right) \bar\chi_\frac{2}{3} +  \left(\chi_\frac{2}{5}+  \chi_\frac{7}{5}\right) \bar\chi_\frac{1}{15}  ,
 \nonumber \\ 
Z^{a_+^2,0} &= \left( \chi_0  +  \chi_3\right) \bar\chi_\frac{2}{3} +  \left(\chi_\frac{2}{5}+  \chi_\frac{7}{5}\right) \bar\chi_\frac{1}{15}  ,
 \nonumber \\ 
Z^{a_+^2,\frac{1}{2}} &= 0,
 \nonumber \\ 
Z^{a_+^2,1} &=  |\chi_\frac{2}{3}|^2 +  |\chi_\frac{1}{15}|^2 ,
 \nonumber \\ 
Z^{a_+^2,\frac{3}{2}} &= 0,
 \nonumber \\ 
Z^{a_+^2,2} &=  \left( \chi_0  +  \chi_3\right) \bar\chi_\frac{2}{3} +  \left(\chi_\frac{2}{5}+  \chi_\frac{7}{5}\right) \bar\chi_\frac{1}{15}  .
\end{align}

\section{Summary}

In this paper, we study anomalous symmetries via multi-component partition
function and their transformation properties under mapping class group
transformations of the space-time.  This allows us to treat symmetry, anomalous
symmetry, and gravitational anomaly at equal footing.  In other words, we can
treat symmetry and anomalous symmetry as a non-invertible gravitational
anomaly.  This gives us a very general and unified point of view about symmetry
and anomaly.  Since gravitational anomaly is just topological order in one
higher dimension,\cite{W1313} symmetry and anomalous symmetry can be viewed as topological
order in one higher dimension, which leads to a holographic point of view of
symmetry and anomalous symmetry.

~
We thank Shu-Heng Shao, Cenke Xu, Po-Shen Hsin, Dominic Else, Meng Cheng and Zhen Bi for helpful discussions.
This research is partially supported by NSF DMR-2022428 and by the Simons
Collaboration on Ultra-Quantum Matter, which is a grant from the Simons
Foundation (651440).

\appendix
\allowdisplaybreaks

\section{Group cohomology theory}
\label{gcoh}

\subsection{Homogeneous group cocycle}

In this section, we will briefly introduce group cohomology.  The group
cohomology class $\cH^d(G,\M)$ is an Abelian group constructed from a group $G$
and an Abelian group $\M$.   We will use ``+'' to represent the multiplication
of the Abelian groups.  Each elements of $G$ also induce a mapping $\M\to \M$,
which is denoted as
\begin{eqnarray}
g\cdot m = m', \ \ \ g\in G,\ m,m'\in \M.
\end{eqnarray}
The map $g\cdot$ is a group homomorphism:
\begin{eqnarray}
g\cdot (m_1+m_2)= g\cdot m_1 +g \cdot m_2.
\end{eqnarray}
The  Abelian group $\M$ with such a $G$-group homomorphism, is called a
$G$-module.

A homogeneous $d$-cochain
is a function $\nu_d: G^{d+1}\to \M$, that satisfies
\begin{align}
\nu_d(g_0,\cdots,g_d)
=g\cdot \nu_d(gg_0,\cdots,gg_d), \ \ \ \ g,g_i \in G.
\end{align}
We denote the set of $d$-cochains as $\cC^d(G,\M)$. Clearly $\cC^d(G,\M)$ is an
Abelian group.
homogeneous group cocycle

Let us define a mapping $\dd$ (group homomorphism) from
$\cC^d(G,\M)$ to $\cC^{d+1}(G,\M)$:
\begin{align}
  (\dd \nu_d)( g_0,\cdots, g_{d+1})=
  \sum_{i=0}^{d+1} (-)^i \nu_d( g_0,\cdots, \hat g_i
  ,\cdots,g_{d+1})
\end{align}
where $g_0,\cdots, \hat g_i ,\cdots,g_{d+1}$ is the sequence $g_0,\cdots, g_i
,\cdots,g_{d+1}$ with $g_i$ removed.
One can check that $\dd^2=0$.
The homogeneous $d$-cocycles are then
the homogeneous $d$-cochains that also satisfy the cocycle condition
\begin{eqnarray}
 \dd \nu_d =0.
\end{eqnarray}
We denote the set of $d$-cocycles as
$\cZ^d(G,\M)$. Clearly $\cZ^d(G,\M)$ is an Abelian subgroup of $\cC^d(G,\M)$.

Let us denote  $\cB^d(G,\M)$ as the image of the map $\dd: \cC^{d-1}(G,\M) \to
\cC^d(G,\M)$ and $\cB^0(G,\M)=\{0\}$.
The elements in $\cB^d(G,\M)$ are called $d$-coboundary.
Since $\dd^2=0$, $\cB^d(G,\M)$  is a subgroup of $\cZ^d(G,\M)$:
\begin{eqnarray}
\cB^d(G,\M) \subset \cZ^d(G,\M).
\end{eqnarray}
The group cohomology class $\cH^d(G,\M)$ is then defined as
\begin{eqnarray}
\cH^d(G,\M) =  \cZ^d(G,\M)/ \cB^d(G,\M) .
\end{eqnarray}
We note that the $\dd$ operator and the cochains $\cC^d(G,\M)$ (for all values
of $d$) form a so-called cochain complex,
\begin{align}
\cdots
\stackrel{\dd}{\to}
\cC^d(G,\M)
\stackrel{\dd}{\to}
\cC^{d+1}(G,\M)
\stackrel{\dd}{\to}
\cdots
\end{align}
which is denoted as $C(G,\M)$.  So we may also write the group cohomology
$\cH^d(G,\M)$ as the standard cohomology of the cochain complex $H^d[C(G,\M)]$.

\subsection{Nonhomogeneous group cocycle}
\label{nonhomo}

The above definition of group cohomology class can be rewritten in terms of
nonhomogeneous group cochains/cocycles.  An nonhomogeneous group $d$-cochain is
a function $\om_d: G^d \to M$. All $\om_d(g_1,\cdots,g_d)$ form $\cC^d(G,\M)$.
The nonhomogeneous group cochains and the homogeneous group cochains are
related as
\begin{eqnarray}
\nu_d(g_0,g_1,\cdots,g_d)=
\om_d( h_{01},\cdots, h_{d-1,d}),
\end{eqnarray}
with
\begin{align}
g_0=1,\ \
g_1=g_0 h_{01}, \ \
g_2=g_1 h_{12}, \ \ \cdots \ \
g_d=h_{d-1} h_{d-1,d}.
\end{align}
Now the $\dd$ map has a form on $\om_d$:
\begin{align}
&
(\dd\om_d)(h_{01},\cdots, h_{d,d+1})=
 h_{01}\cdot \om_d( h_{12},\cdots,h_{d,d+1})
\nonumber\\
& \ \ \
+\sum_{i=1}^d (-)^i \om_d(h_{01},\cdots, h_{i-1,i}h_{i,i+1},\cdots, h_{d,d+1})
\nonumber\\
& \ \ \
+(-)^{d+1}\om_d(h_{01},\cdots,h_{d-2,d-1})
\end{align}
This allows us to define the nonhomogeneous group $d$-cocycles which satisfy
$\dd \om_d=0$ and  the nonhomogeneous group $d$-coboundaries which have a form
$\om_d = \dd \mu_{d-1}$.  In the following, we are going to use nonhomogeneous
group cocycles to study group cohomology.  Geometrically, we may view $g_i$ as
living on the vertex $i$, while $h_{ij}$ as living on the edge connecting the
two vertices $i$ to $j$.

\subsection{``Normalized'' cocycles}
\label{norm}

We know that each element in $\cH^d(G,\R/\Z)$ can be represented by many
cocycles.  In the following, we are going to describe a way to simplify the
cocycles, so that the simplified cocycles can still represent all the elements
in $\cH^d(G,\R/\Z)$.

The simplification is obtained by considering ``normalized''
cochains,\cite{HS5310} which satisfy
\begin{align}
\om_d(g_1,\cdots, g_d)=0, \text{ if one of } g_i=1.
\end{align}
One can check that the $\dd$-operator maps a ``normalized'' cochain to a
``normalized'' cochain.  The group cohomology classes obtained from the
ordinary cochains is isomorphic to the group cohomology classes obtained from
the ``normalized'' cochains.  Let us use $\bar\cC^d(G,\M)$, $\bar\cZ^d(G,\M)$,
and $\bar\cB^d(G,\M)$ to denote the ``normalized'' cochains, cocycles, and
coboundaries.  We have $\cH^d(G,\M)=\bar\cZ^d(G,\M)/\bar\cB^d(G,\M)$.

\section{Topological path integral on a space-time}
\label{toppath}

\subsection{Space-time lattice and branching structure}

\begin{figure}[tb]
\begin{center}
\includegraphics[scale=0.6]{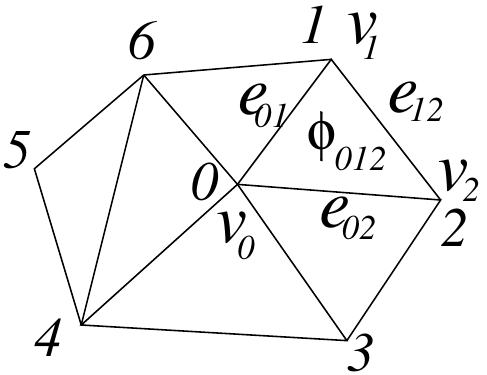} \end{center}
%Fig. 1
\caption{ 
A two-dimensional complex.  The vertices (0-simplices) are labeled by $i$.  The
edges (1-simplices) are labeled by $\<ij\>$.  The faces (2-simplices) are
labeled by $\<ijk\>$.  The degrees of freedom may live on the vertices
(labeled by $v_i$), on the edges (labeled by $e_{ij}$) and on the faces
(labeled by $\phi_{ijk}$).
}
\label{comp}
\end{figure}

To find the phase factors in \eqn{ZpropG}, we need to use extensively the
space-time path integral.  So we will first describe how to define a space-time
path integral. We first triangulate the $3$-dimensional space-time to obtain a
simplicial complex $\cM^3$ (see Fig. \ref{comp}).  Here we assume that all
simplicial complexes are of bounded geometry in the sense that the number of
edges that connect to one vertex is bounded by a fixed value.  Also, the number
of triangles that connect to one edge is bounded by a fixed value, \etc.

\begin{figure}[tb]
\begin{center}
\includegraphics[scale=0.6]{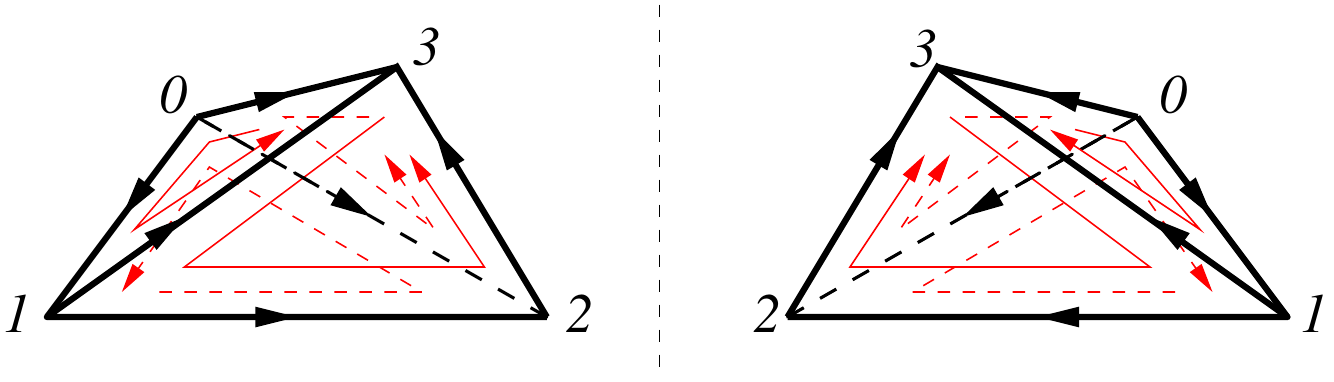} \end{center}
%Fig. 2
\caption{ (Color online) Two branched simplices with opposite orientations.
(a) A branched simplex with positive orientation and (b) a branched simplex
with negative orientation.  }
\label{mir}
\end{figure}

In order to define a generic lattice theory on the space-time complex $\cM^3$,
it is important to give the vertices of each simplex a local order.  A nice
local scheme to order  the vertices is given by a branching
structure.\cite{C0527,CGL1314,CGL1204} A branching structure is a choice of the orientation of each edge in the $n$-dimensional complex so that there is no
oriented loop on any triangle (see Fig. \ref{mir}).

The branching structure induces a \emph{local order} of the vertices on each
simplex.  The first vertex of a simplex is the vertex with no incoming edges,
and the second vertex is the vertex with only one incoming edge, \etc.  So the
simplex in  Fig. \ref{mir}a has the following vertex ordering: $0<1<2<3$.

The branching structure also gives the simplex (and its sub simplexes) an
orientation denoted by $s_{ij \cdots k}=1,*$.  Fig. \ref{mir} illustrates two
$3$-simplices with opposite orientations $s_{0123}=1$ and $s_{0123}=*$.  The
red arrows indicate the orientations of the $2$-simplices which are the
subsimplices of the $3$-simplices.  The black arrows on the edges indicate the
orientations of the $1$-simplices.

The degrees of freedom of our lattice model live on the vertices  (denoted by
$v_i$ where $i$ labels the vertices), on the edges (denoted by $e_{ij}$ where
$\<ij\>$ labels the edges), and on other high dimensional simplicies of the
space-time complex (see Fig. \ref{comp}).


\begin{figure}[t]
\begin{center}
\includegraphics[scale=0.6]{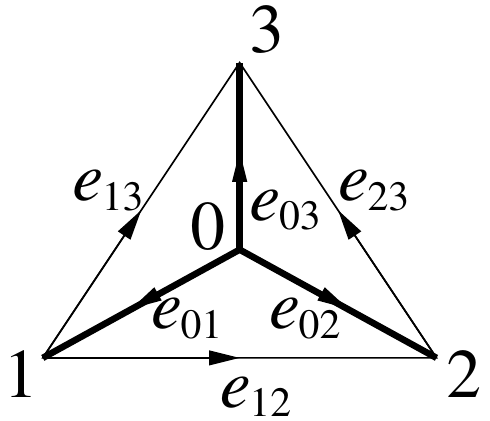}
%Fig. 5
\end{center}
\caption{
The tensor $\tC{C}0123$ is associated with a tetrahedron, which has a branching
structure.  If the vertex-0 is above the triangle-123, then the tetrahedron
will have an orientation $s_{0123}=*$.  If the vertex-0 is below the
triangle-123, the tetrahedron will have an orientation $s_{0123}=1$. The
branching structure gives the vertices a local order: the $i^{th}$ vertex has
$i$ incoming edges.  
}
\label{tetr}
\end{figure}

\subsection{Discrete path integral}

In this paper, we will only consider a type of 2+1D path integral that can be
constructed from a tensor set $T$ of two real and one complex tensor:
$T=(w_{v_0}, \tAw{d}01,\tC{C}0123)$.  The complex tensor $\tC{C}0123$ can be
associated with a tetrahedron, which has a branching structure (see Fig.
\ref{tetr}).  A branching structure is a choice of the orientation of each edge in
the complex so that there is no oriented loop on any triangle (see Fig.
\ref{tetr}).  Here the $v_0$ index is associated with the vertex-0, the
$e_{01}$ index is associated with the edge-$01$, and the $\phi_{012}$ index is
associated with the triangle-$012$.  They represent the degrees of freedom on
the vertices, edges, and triangles.

Using the tensors, we can define the path integral on any 3-complex
that has no boundary:
\begin{align}
\label{Z3d}
 Z(\cM^3)&=\sum_{ v_0,\cdots; e_{01},\cdots; \phi_{012},\cdots}
\prod_\text{vertex} w_{v_{0}} 
\prod_\text{edge} \tAw{d}01\times
\\
&\ \ \ \ \ \ \ \ \ \ 
\prod_\text{tetra} [\tC{C}0123 ]^{s_{0123}}
\nonumber 
\end{align}
where $\sum_{v_0,\cdots; e_{01},\cdots; \phi_{012},\cdots}$ sums over all the
vertex indices, the edge indices, and the face indices, $s_{0123}=1$ or $*$
depending on the orientation of tetrahedron (see Fig.  \ref{tetr}).  We believe
such type of path integral can realize any 2+1D topological order.

\section{SPT invariant and properties of the boundary partition function}
\label{SPTinv}

In this section, we will give a brief description of SPT
invariant\cite{W1447,HW1339,K1459} and its relation to the properties of
boundary partition function. This allows us to gain a more general
understanding of 't Hooft anomaly and its consequence on the anomalous boundary
partition function.

\subsection{Topological partition function as topological invariant}

A very general way to characterize a topologically ordered phase is via its
partition function $Z(M^D)$ on closed spactime $M^D$ with all possible
topologies. A detailed discussion on how to define the partition function via
tensor network is given in \Ref{KW1458} and Appendix \ref{toppath}, from which,
we see that the partition function also depends on the branched triangulation of
the space-time (see Appendix \ref{toppath}), as well as the tensor associated
with each simplex.  We collectively denote the triangulation,the branching
structure, and the tensors as $\cT$.  Thus the partition function should be
more precisely denoted as $ Z_\text{TN}(M^D,\cT)$.  

However, $Z_\text{TN}(M^D,\cT)$ is not a
topological invariant since it contains a so-called volume term
$\ee^{-\int_{M^D} \eps\, \dd^D x }$ where $\eps$ is the energy density. After
removing the volume term, we can obtain a topological partition function
$Z_\text{TN}^\text{top}(M^D)$ which is a topological
invariant:\cite{KW1458,WW180109938}
\begin{align}
 Z_\text{TN}(M^D,\cT) = \ee^{-\int_{M^D} \eps\, \dd^D x } Z_\text{TN}^\text{top}(M^D,\cT) .
\end{align}
Appendix \ref{toppath} describes the way to make volume term vanish (\ie
$\eps=0$), in this case, the path integral directly produces the
topological partition function.  Such a topological invariant may completely
characterize the topological order.

In the above, we have ignored the symmetry.  To characterize a topological order
with symmetry $G$, we need to include the symmetry twists described by the flat
connection $A$ on the space-time $M^D$, ann use the corresponding topological
partition function $Z_\text{TN}^\text{top}(M^D,A,\cT)$ (after removing the
volume term).  Since topological order with symmetry includes trivial
topological order with symmetry, therefore $Z_\text{TN}^\text{top}(M^D,A,\cT)$
can also be used to characterize trivial topological order with symmetry, which
is nothing but the SPT order.  Thus $Z_\text{TN}^\text{top}(M^D,A,\cT)$ include
the SPT invariant mentioned above.  

\subsection{Properties of the boundary partition function}

In this paper, we will concentrate on global anomaly, and we will assume that
there is no perturbative anomaly.  In this case, the global anomaly is
characterized by the bulk topological invariant
$Z_\text{TN}^\text{top}(M^D,A,\cT)$, which  is described by the topological
path integral described in Appendix \ref{toppath}.

One way to systematically generate some of the bulk topological invariant
$Z_\text{field}^\text{top}(M^D,A,\cT)$ is to consider the following topological
partition function $Z^\text{top}(B^d \gext_\vphi S^1,A)$.  Here where $d=D-1$
and $B^d \gext_\vphi S^1$ is the mapping torus obtained from $I\times B^d$ by
gluing its two boundaries via map $\vphi$: $B^d\to B^d$ in the mapping class
group of $B^d$.  In other words, $B^d \gext_\vphi S^1$ is a fiber bundle with
fiber $B^d$ and base space $S^1$.  We see that we can obtain a topological
invariant from each element of mapping class group of $B^d$, provided that
there is no perturbative anomaly.

To link such a topological invariant for invertible orders with symmetry,
$Z_\text{TN}^\text{top}(M^D,A,\cT)$, to the partition function on the boundary
$B^d$, we note that the gapped state on $M^D$ can have a boundary $B^d=\prt
M^D$.  The  boundary partition function is given by
\begin{align}
 Z(B^d,A,\cT_B) =  Z_\text{TN}^\text{top}(M^D,A,\cT).
\end{align}
We may obtain a more general boundary by attaching a $d$-dimensional system
described by a $d$-dimensional tensor network, $Z_\text{TN}(B^d,A,\cT_B)$, to
the boundary. The resulting boundary partition function has a form
\begin{align}
 Z(B^d,A,\cT_B) =  Z_\text{TN}(B^d,A,\cT_B) Z_\text{TN}^\text{top}(M^D,A,\cT)
\end{align}
We see that the boundary partition function $Z(B^d,A,\cT_B)$ is not purely
given by a tensor network on the boundary $B^d$, which gives rise to a
partition function $Z_\text{TN}(B^d,A,\cT_B)$.  The boundary partition function
also contain a bulk topological term $Z_\text{TN}^\text{top}(M^D,A,\cT)$. This
makes the boundary quantum system defined by $Z(B^d,A,\cT_B)$ to be potentially
anomalous.  If the boundary partition function is given purely by a tensor
network $Z_\text{TN}(B^d,A,\cT_B)$ on the boundary (\ie when
$Z_\text{TN}^\text{top}(M^D,A,\cT)=1$), such a quantum system will be
anomaly-free.

\begin{figure}[t]
\begin{center}
\includegraphics[scale=0.6]{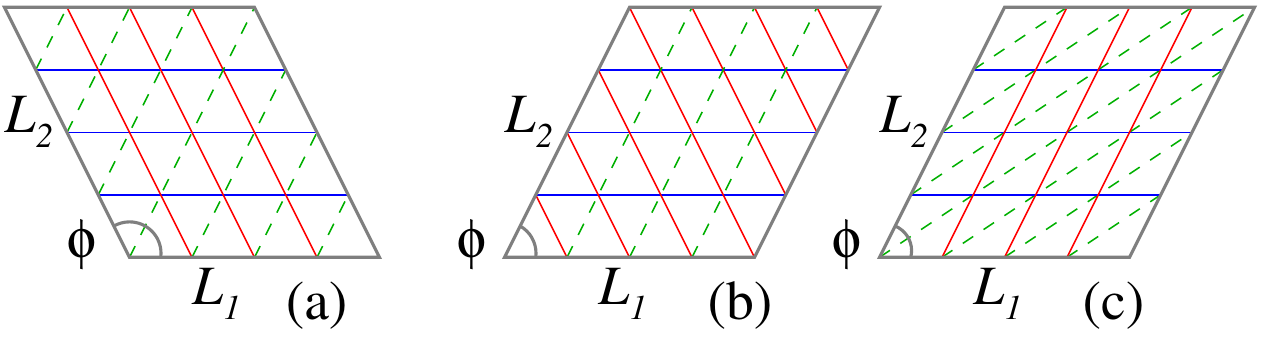} \end{center}
%6
\caption{(Color online) 
(a) A torus with a shape $\tau$.
(b,c) two tori with a shape $\tau+1$.
(a,b) The two tori have the same triangulation $\cT$.
(b) The torus has another triangulation $\cT'$.
(a) and (b) actually describe the same thing and are equivalent:
$(\tau,\cT) \sim (\tau+1,\cT')$
}
\label{triMT}
\end{figure}

To see the anomaly from the  boundary partition function, as an example, we
assume $d=2$ and $B^d=S^1\times S^1\equiv T^2$.  The partition function on
$T^2$ will depend on the shape of the torus, \ie the sizes $L_1,L_2$ and the
angle $\phi$ (see Fig. \ref{triMT}).  The partition function also depends on
the symmetry twist characterized by a flat connection $A$, as well as
the triangulation $\cT$ on $T^2$.  So we will denote that partition function as
$Z(T^2,A,\cT)$.  In this paper, we choose the energy zero such that the energy
density $\eps=0$.  In this case $Z(T^2,A,\cT)$ does not depend on the size of
the spacetime $L_1,L_2$. $Z(T^2,A,\cT)$ only depends on the shape of the
space-time characterized by $L_2/L_1$ and $\phi$.  We will introduce a complex
number 
\begin{align}
 \tau \equiv \frac{L_2}{L_1}\ee^{\ii \phi}
\end{align}
to characterize the shape of the torus. Thus partition function cab be written
as
\begin{align}
 Z(\tau,A,\cT)
\end{align}

For example, Fig. \ref{triMT}a and Fig. \ref{triMT}c describe torus with
different shapes but with the same triangulation $\cT$ Fig.  \ref{triMT}b has a
different triangulation $\cT'$.  Although  Fig. \ref{triMT}a and Fig.
\ref{triMT}b have different shapes described by $\tau$ and $\tau+1$ and
different triangulations described by $\cT$ and $\cT'$, they actually describe
the same thing (\ie the same simplicial complex).  Thus
\begin{align}
 Z(\tau,0,\cT)= Z(\tau+1,0,\cT').
\end{align}
(For the time being, we set the symmetry twist $A=0$.) However, the partition
function for Fig. \ref{triMT}b, $Z(\tau+1,0,\cT')$, may not equal to the
partition function for Fig. \ref{triMT}c, $Z(\tau+1,0,\cT)$.
$Z(\tau+1,0,\cT')$ and $Z(\tau+1,0,\cT)$ may differ by a phase.  Such a
triangulation dependent phase factor represent a gravitational anomaly.  This
resembles the non-invariance under diffeomorphism giving rise to gravitational
anomaly.

The phase difference from retriangulation $\cT' \to \cT$ is given by the
partition function of the bulk topological order
\begin{align}
 Z(\tau,0,\cT)&=
 Z(\tau+1,0,\cT')
\\
&= Z^\text{top}(T^2\times I,A=0) Z(\tau+1,0,\cT)
\nonumber 
\end{align}
Here the bulk $T^2\times I$ has a particular tianglation $\cT_\text{bulk}$,
such that one boundary of $T^2\times I$, together with the triangulation $\cT'$
on the boundary, is given by Fig. \ref{triMT}b, while the other boundary of
$T^2\times I$, together with the triangulation $\cT$ on the boundary, is given
by Fig.  \ref{triMT}c.  Since Fig. \ref{triMT}b and Fig. \ref{triMT}a describe
the same thing and Fig. \ref{triMT}a differ from Fig. \ref{triMT}c by a modular
transformation $T: \tau \to \tau+1$, the factor $Z^\text{top}(T^2\times I,A=0)$
is a topological invariant discussed above
\begin{align}
 Z^\text{top}(T^2\times I,A=0) = Z^\text{top}(T^2\gext_{\vphi_T} S^1,A=0) ,
\end{align}
where $\vphi_T$ is an element of the mapping class group of the torus
$SL(2,\Z)$.

The main point of the above discussion is that, in absence of perturbative
anomaly,  the change of the boundary partition function under the mapping class
group transformation is given by the bulk topological invariant $
Z^\text{top}(T^2\gext_{\vphi} S^1,A=0)$ that characterizes the bulk invertible
topological order.  Thus the non-invariance of the boundary partition function
under the mapping class group transformation is a sign of
anomaly.\cite{RZ1232,L13017355}  In other words, an anomaly-free system has an
invariant partition function under the mapping class group transformation.

However, in two-dimensional spacetime, in absence of perturbative gravitational
anomaly, there is no other gravitational anomaly. In other words, without
symmetry there is no other invertible topological orders in 2+1D, except the
ones generated by abelian topological order characterized by the $K$-matrix
$K_{E_8}$.  Therefore, without symmetry twist, we always have
$Z^\text{top}(T^2\gext_{\vphi} S^1,A=0)=1$.  As a result, the boundary partition
function $Z^\text{top}(T^2\times I,A=0)$ is always modular invariant without
symmetry twist, provided that the chiral central charge $c-\bar c=0$.  So the
non-trivial results only appear when we have symmetry and symmetry twist $A\neq
0$.

\begin{figure}[t]
\begin{center}
\includegraphics[scale=0.6]{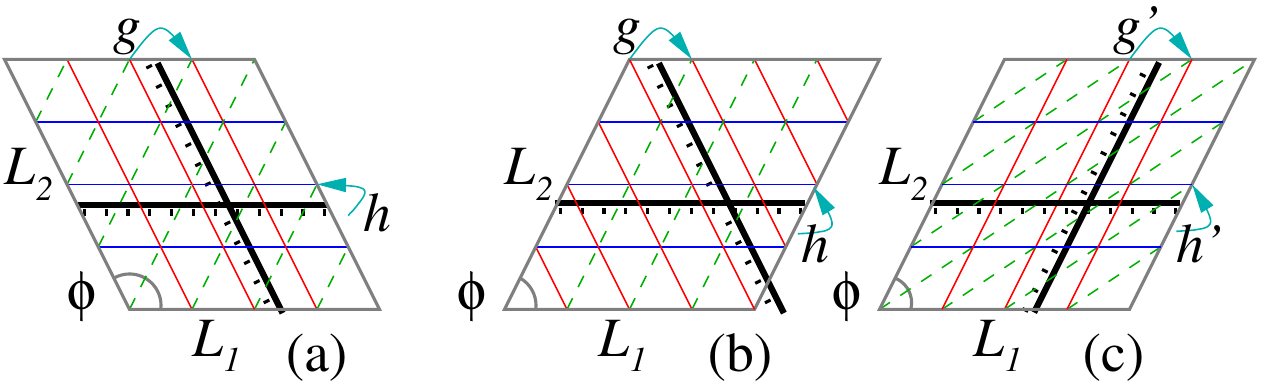} \end{center}
%6
\caption{
A torus with symmetry twist.
(a,b,c) all have the same symmetry twist.
(a,b) The symmetry twist is described by the same cuts $(g,h)$.
(c) The symmetry twist is described by different cuts $(g',h')=(gh,h)$.
}
\label{triMTA}
\end{figure}

In the following, we will consider partition functions with symmetry twist
$A=0$, $Z(\tau,A,\cT)$. We note that on a torus the symmetry twist $A$ can be
described by symmetry transformation across two cuts $g,h$ (see Fig.
\ref{triMTA}a), where $g,h$ satisfy
\begin{align}
 gh=hg,\ \ \ \ \ g,h \in G .
\end{align}
So we can denote the partition function as $Z(\tau,g,h,\cT)$.  Since Fig.
\ref{triMTA}a and  Fig.  \ref{triMTA}b are equivalent, we have
\begin{align}
 Z(\tau,g,h,\cT)=
 Z(\tau+1,g,h,\cT')
\end{align}
But 
$ Z(\tau,g,h,\cT)$ and
$ Z(\tau+1,g',h',\cT)$ differ by a re-triangulation, and may differ by a phase
\begin{align}
\label{ZtopT}
 Z(\tau,g,h,\cT)
&= Z^\text{top}(T^2\times I,A_{g,h;g',h'}) Z(\tau+1,g',h',\cT),
\nonumber\\
(g',h') &= (gh,h).
\end{align}
Here the bulk $T^2\times I$ has a particular triangulation $\cT_\text{bulk}$
and symmetry twist $A_{g,h;g'h'}$, such that one boundary of $T^2\times I$,
together with the triangulation $\cT'$ and symmetry twist on the boundary, is
given by Fig.  \ref{triMTA}b, while the other boundary of $T^2\times I$,
together with the triangulation $\cT$ and the symmetry twist on the boundary,
is given by Fig.  \ref{triMTA}c.  Similarly, we also have
\begin{align}
\label{ZtopS}
 Z(\tau,g,h,\cT)
&= Z^\text{top}(T^2\times I,A_{g,h;g',h'}) Z(-\tau^{-1},g',h',\cT),
\nonumber\\
(g',h') &= (h^{-1},g).
\end{align}
Even if we do not change the shappe of the torus, but just change
the symmetry twist, we may still have a phase factor
\begin{align}
\label{ZtopU}
 Z(\tau,g,h,\cT)
&= Z^\text{top}(T^2\times I,A_{g,h;g',h'}) Z(\tau,g',h',\cT),
\nonumber\\
(g',h') &= (ugu^{-1},uhu^{-1}).
\end{align}

\subsection{Properties of the boundary partition function for SPT states}

\begin{figure}[t]
\begin{center}
\includegraphics[scale=0.6]{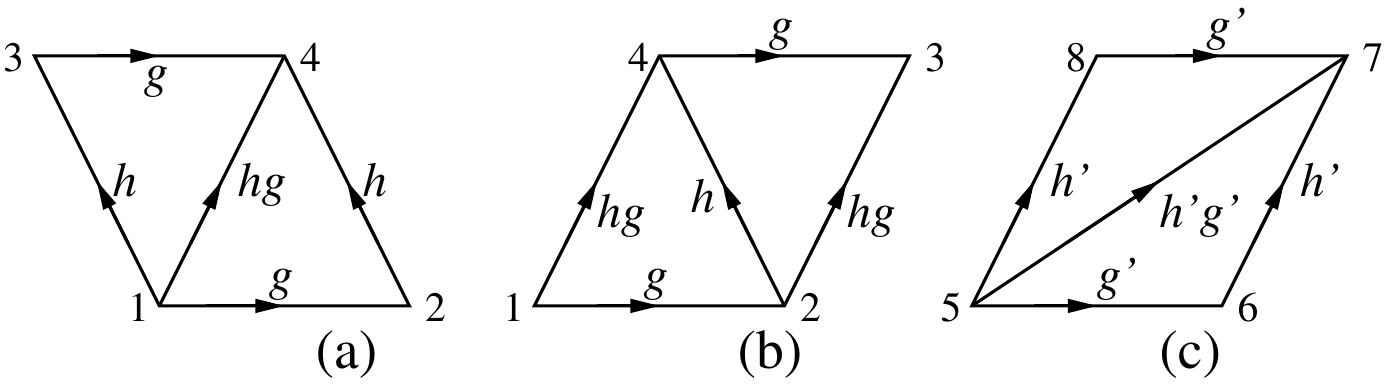} \end{center}
%6
\caption{
A simplification of Fig. \ref{triMTA} to have the simplest triangulation. 
}
\label{cocT}
\end{figure}

\Eqn{ZtopT}, \eqn{ZtopS} and \eqn{ZtopU} are one of the main results of this
paper.  They link the properties of the boundary partition function to the bulk
topological invariance.  Now let us calculate the phase factors
$Z^\text{top}(T^2\times I,A_{g,h;g'h'})$ in \eqn{ZtopT}, \eqn{ZtopS},  and
\eqn{ZtopU}, for boundary of SPT state.  We know that the SPT invariant of the
bulk SPT state with symmetry $G$ can be expressed in terms of group cocycle
$\om_3 \in \cH^3(G;\R/\Z)$.  Thus $Z^\text{top}(T^2\times I,A_{g,h;g'h'})$ can
be computed directly from the group cocycle that characterize the SPT
order.\cite{HW1339} We would like to remark that $Z^\text{top}(T^2\times
I,A_{g,h;g'h'})$ has boundary.  As a result, $Z^\text{top}(T^2\times
I,A_{g,h;g'h'})$ may depend on the choices of coboundary for the group cocycle.
Only a certain choice of the coboundary gives rise to the correct properties
of the boundary partition function, which respects the positivity of the
partition function.

\begin{figure}[t]
\begin{center}
\includegraphics[scale=0.8]{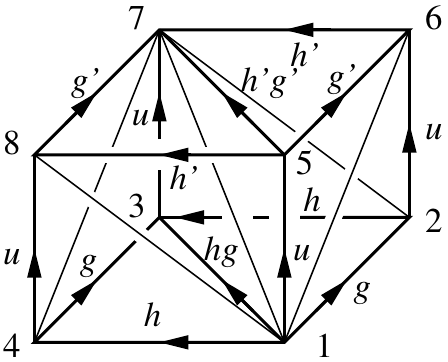} \end{center}
%6
\caption{
A simple triangulation of $T^2\times I$ with the symmetry twist
$A_{g,h;ugu^{-1},uhu^{-1}}$.  The two squares 1265 and 4378 are identified.
The two squares 2376 and 1485 are also identified.  Here $g'=ugu^{-1}$ and
$h'=uhu^{-1}$.
}
\label{cocUev}
\end{figure}

\begin{figure}[t]
\begin{center}
\includegraphics[scale=0.7]{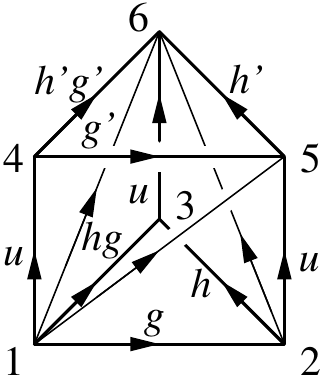} \end{center}
%6
\caption{
A wedge with its canonical triangulation is formed by three tetrahedrons 1236, 1456, 1256.
Here $g'=ugu^{-1}$ and $h'=uhu^{-1}$.
}
\label{wedge}
\end{figure}

First, let us calculate the phase factor induced by changing the symmetry twist
with a symmetry transformation: $(g,h) \to (g',h')=(ugu^{-1},uhu^{-1})$,
without changing the shape of the torus (see \eqn{ZtopU}).
We may triangulate $T^2\times I$ as in Fig. \ref{cocUev} (see Appendix
\ref{toppath})).  The group elements on the links describe the symmetry
twist.\cite{HW1339} The 3-cocycle $\om_3$ on $T^2\times I$ is a function of
those group elements on the links. We note that $T^2\times I$ is formed by two
wedges 123567, 143587,  with their canonical triangulation. Each wedge is given
by Fig.  \ref{wedge}:
\begin{align}
&\ \ \ \
 w(g,h,u) \equiv \ee^{\ii 2\pi \int_{123456} \om_3}
\\
&=
 \ee^{\ii 2\pi \om_3(1236)}
 \ee^{-\ii 2\pi \om_3(1256)}
 \ee^{\ii 2\pi \om_3(1456)}
\nonumber\\
&=
 \ee^{\ii 2\pi \om_3(g,h,u)}
 \ee^{-\ii 2\pi \om_3(g,u,uhu^{-1})}
 \ee^{\ii 2\pi \om_3(u,ugu^{-1},uhu^{-1})}
\nonumber 
\end{align}
Thus $Z^\text{top}(T^2\times I,A_{g,h;ugu^{-1},uhu^{-1}})$ in \eqn{ZtopU}
can be expressed in terms 
of group cocycle $\om_3(g_1,g_2,g_3)$:
\begin{align}
&\ \ \ \
 Z^\text{top}(T^2\times I,A_{g,h;ugu^{-1},uhu^{-1}}) = 
\frac{w(g,h,u)}{w(h,g,u)} . 
\end{align}
Note that the wedge 143587 has a ``$-$'' orientation, and hence
it contributes a phase factor $w^{-1}(h,g,u)$.

\begin{figure}[t]
\begin{center}
\includegraphics[scale=0.8]{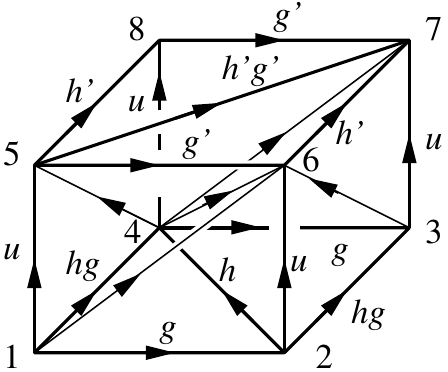} \end{center}
%6
\caption{
A simple triangulation of $T^2\times I$ with the symmetry twist $A_{g,h;gh,h}$,
which contains two wedges 458367, 145236.  The wedge 458367 has the canonical
triangulation as in Fig. \ref{wedge}, but the wedge 145236 does not has the
canonical triangulation, and is formed by three tetrahedrons 1456, 1246, 2436.
The two squares 1265 and 4378 are identified.  The two squares 2376 and 1485
are also identified.  The bottom boundary matches Fig. \ref{cocT}b  and the top
boundary matches Fig.  \ref{cocT}c.  Here  $u=1$, $g'=g$ and $h'=hg$.
}
\label{cocTev}
\end{figure}

Next, let us try to compute the phase factor $Z^\text{top}(T^2\times
I,A_{g,h;gh,h})$ for transformation $\tau \to \tau+1$ in \eqn{ZtopT}.  We first
simplify the triangulation of Fig. \ref{triMTA} to Fig. \ref{cocT}.  The group
elements on the links describe the symmetry twist.\cite{HW1339} We then use the
triangulation in  Fig. \ref{cocTev} to triangulate $T^2\times I$ in
$Z^\text{top}(T^2\times I,A_{g,h;gh,h})$. 
 The phase factor
$Z^\text{top}(T^2\times I,A_{g,h;gh,h})$ is given by $\om_3$ via
\begin{align}
 Z^\text{top}(T^2\times I,A_{g,h;gh,h}) = 
\ee^{\ii 2\pi \int_{T^2\times I} \om_3}
\end{align}
In Fig. \ref{cocTev} we divided $T^2\times I$ into one wedge 458367
with the canonical triangulation, plus three tetrahedrons 1456, 1246, 2436.
Thus
\begin{align}
\label{ZtopTw}
&\ \ \ \
 Z^\text{top}(T^2\times I,A_{g,h;gh,h}) = 
\ee^{\ii 2\pi \int_{T^2\times I} \om_3}
\\
&=
\ee^{\ii 2\pi \int_{1456} \om_3} 
\ee^{\ii 2\pi \int_{1246} \om_3}  
\ee^{-\ii 2\pi \int_{2436} \om_3} 
\ee^{-\ii 2\pi \int_{458367} \om_3} 
\nonumber\\
&= 
\frac{
\ee^{\ii 2\pi \om_3(hg,g^{-1}h^{-1},g)}
\ee^{\ii 2\pi \om_3(g,h,h^{-1})}  
\ee^{-\ii 2\pi \om_3(h,g,g^{-1}h^{-1})}
}{
w(g^{-1}h^{-1},hg,g)
}
,
\nonumber 
\end{align}
or
\begin{align}
\label{ZtopTom}
&\ \ \ \
 Z(\tau,g,h,\cT)
\nonumber\\
&=  
\ee^{\ii 2\pi \om_3(hg,g^{-1}h^{-1},g)}
\ee^{\ii 2\pi \om_3(g,h,h^{-1})}  
\ee^{-\ii 2\pi \om_3(h,g,g^{-1}h^{-1})}
\nonumber\\
&\ \ \ \
w^{-1}(g^{-1}h^{-1},hg,g)
Z(\tau+1,g',h',\cT),
\nonumber\\
& \ \ \ \ \ (g',h') = (gh,h).
\end{align}
This way, we express $Z^\text{top}(T^2\times I,A_{g,h;gh,h})$ in \eqn{ZtopT} in
terms of group cocycle $\om_3(g_1,g_2,g_3)$.

\begin{figure}[t]
\begin{center}
\includegraphics[scale=0.6]{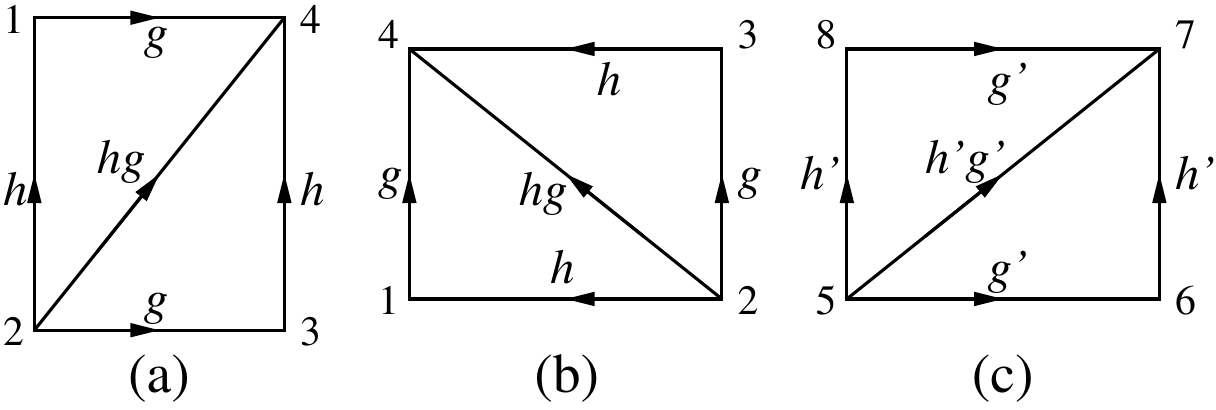} \end{center}
%6
\caption{
A torus with symmetry twist.
(a,b,c) all have the same symmetry twist.
(a) has a shape $\tau$, and
(b,c)  has a shape $-1/\tau$.
(a,c) has the same trianglation $\cT$
and (b) has another trianglation $\cT'$.
(a) and (b) are equivalent.
Here $g'=h^{-1}$ and $h'=g$.
}
\label{cocS}
\end{figure}

\begin{figure}[t]
\begin{center}
\includegraphics[scale=0.8]{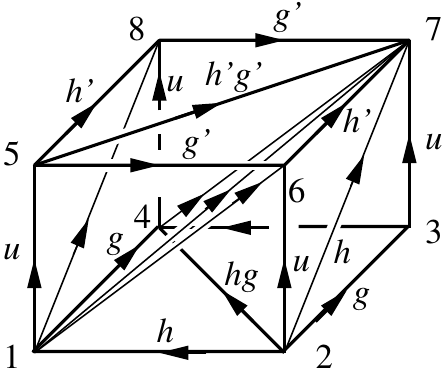} \end{center}
%6
\caption{
A simple triangulation of $T^2\times I$ with the symmetry twist
$A_{g,h;h^{-1},g}$, which contains wedges 216347, 156487, with the canonical
triangulation.  The two squares 1265 and 4378 are identified.  The two squares
2376 and 1485 are also identified.  The bottom boundary gives rise to Fig.
\ref{cocS}b and the top boundary gives rise to Fig. \ref{cocS}c.  Here $u=1$,
$g'=h^{-1}$ and $h'=g$.
}
\label{cocSev}
\end{figure}

Similarly, to calculate $Z^\text{top}(T^2\times I,A_{g,h;h^{-1},g})$ in
\eqn{ZtopS}, we need to consider Fig. \ref{cocS}. We see that the phase factor
$Z^\text{top}(T^2\times I,A_{g,h;h^{-1},g})$ is induced by the retriangulation:
Fig. \ref{cocS}b to Fig. \ref{cocS}c.  We may triangulate $T^2\times I$ as in
Fig. \ref{cocSev}, which allows us to express $Z^\text{top}(T^2\times
I,A_{g,h;h^{-1},g})$ in \eqn{ZtopS} in terms of group cocycle
$\om_3(g_1,g_2,g_3)$:
\begin{align}
&\ \ \ \
 Z^\text{top}(T^2\times I,A_{g,h;h^{-1},g}) = 
\ee^{\ii 2\pi \int_{T^2\times I} \om_3}
\nonumber\\
&=
\ee^{\ii 2\pi \int_{216347} \om_3} 
\ee^{\ii 2\pi \int_{156487} \om_3} 
\nonumber\\
&= 
w(h,h^{-1},g)w(1,h^{-1},g)
,
\end{align}
In other words
\begin{align}
\label{ZtopSom}
&\ \ \ \
 Z(\tau,g,h,\cT)
\nonumber\\
&=  
w(h,h^{-1},g)w(1,h^{-1},g)
Z(-1/\tau,g',h',\cT),
\nonumber\\
&\ \ \ \
(g',h') = (h^{-1},g).
\end{align}

To summarize, the partition function for a boundary of a 2+1D SPT state
characterized by $\om_3\in \cH^3(G,\R/\Z)$ has the following properties
\begin{align}
\label{Zprop}
&
Z(\hat T(\tau);\hat T(g,h))
=  
\ee^{-\ii 2\pi \om_3(hg,g^{-1}h^{-1},g)}
\ee^{-\ii 2\pi \om_3(g,h,h^{-1})}  
\nonumber\\
&\ \ \ \ \ \ \ \ \ \ 
\ee^{\ii 2\pi \om_3(h,g,g^{-1}h^{-1})}
w(g^{-1}h^{-1},hg,g)
Z(\tau;g,h) ,
\nonumber\\
&
Z(\hat S(\tau);\hat S(g,h))
=  
w^{-1}(h,h^{-1},g)w^{-1}(1,h^{-1},g) 
Z(\tau;g,h) ,
\nonumber\\
&
Z(\hat R_u(\tau);\hat R_u(g,h))
= \frac{w(h,g,u)}{w(g,h,u)} 
Z(\tau;g,h) ,
\nonumber\\
& \ \ \ \hat T(\tau) =\tau+1, \ \ \ \ 
\hat T(g,h) =gh,h,  
\nonumber\\
& \ \ \ \hat S(\tau) =-1/\tau, \ \ \ \ 
\hat S(g,h) =h^{-1},g,
\nonumber\\
& \ \ \ \hat R_u(\tau) =\tau, \ \ \ \
\hat R_u(g,h)=ugu^{-1},uhu^{-1}.
\end{align}
We remark that the transformation $\hat R_u$ is in general non-trivial for the
non-Abelian symmetry group and needs to be included.  For Abelian symmetry
group, we can ignore $\hat R_u$ as in \Ref{TR171004730}.  We also would like to remark
that in \eqn{Zprop}, the partition function is labeled by a pair of group
elements $(g,h)$ in one-to-one fashion.  In \Ref{TR171004730}, the  partition
functions are labeled by a larger set of indices than a pair of group elements
$(g,h)$.  So there are many partition functions for the same symmetry twist
$(g,h)$.  The difference of the partition functions that correspond to the same
symmetry twist is viewed as the gauge non-invariance in \Ref{TR171004730} that
characterize the anomaly .  The different conventions lead to different
expressions for the transformation, and our 1-to-1 labeling of the partition
functions in terms of  a pair of group elements $(g,h)$ is more convenient for
non-Abelian symmetry.

From the above expression, we see that, starting from
a symmetry twist $(g,h)$, the three transformations
\begin{align}
\hat T(g,h) &=(g',h')=(gh,h),  
\nonumber\\
\hat S(g,h) &=(g',h')=(h^{-1},g),
\nonumber\\
\hat R_u(g,h)&=(g',h')=(ugu^{-1},uhu^{-1}). 
\end{align}
generate a ``walk'' in the space of possible symmetry twists.  
For some
combinations of the three transformations, the generated ``walk'' may form a
loop.  In this case, we obtain a property of the boundary partition function
(\ie the partition function for an anomalous system)
\begin{align}
Z(\tau;g,h)=\ee^{\ii \th} Z(\tau',g,h). 
\end{align}
where the phase factor $\ee^{\ii \th}$ is a cobimation of the phase factors in
\eqn{Zprop}.  Such a combination correspond to a topological invariant
$\ee^{\ii \th}=Z^\text{top}(T^2 \gext_\vphi S^1,A)$ for the 2+1D bulk SPT state
and is indenpendent of the choice of the coboundaries in the 3-cocycle $\om_3$.

In the example of $Z_2$ symmetry on to 1+1D partition
function with a $Z_2$ 't Hooft anomaly, the $Z_2$-anomaly is characterized by a 3-cocycle
$\om_3\in \cH^3(Z_2;\R/\Z)$, whose ``normalized'' form (see Appendix
\ref{gcoh})  is given by
\begin{align}
\om_3(g_1,g_2,g_3)=\frac{ s(g_1)s(g_3)s(g_3)}{2},\ \ \
s(g)= \frac{1-g}{2} 
.
\end{align}
For an Abelian group, the action of $\hat R_u$ is trivial and indeed
$\frac{w(g,h,u)}{w(h,g,u)}=1$. The cocycles appear in the actions of $T,S$ in the following way,
\begin{align}
\label{ZpropZ2}
Z(\hat T(\tau);\hat T(g,h ))
&=  
(-)^{s(g)s(h)} 
Z(\tau;g,h ) ,
\nonumber\\
Z(\hat S(\tau);\hat S(g,h ))
&=  
(-)^{s(g)s(h)} 
Z(\tau;g,h ) ,
\nonumber\\
\hat T(\tau) &=\tau+1, \ \ \ \ 
\hat T(g,h ) =gh ,h ,  
\nonumber\\
\hat S(\tau) &=-1/\tau, \ \ \ \ 
\hat S(g,h ) =h^{-1} ,g.
\end{align}

\iffalse
The four partition function for the four symmetry twists can be labeled as
\begin{align}
&Z_\one=Z_{1,1}=\zb,
&
& Z_2=Z_{1,-1}=\zbh,\nn\\
&Z_3=Z_{-1,1}= \zbv, 
&
& Z_4=Z_{-1,-1}= \zbx,
\end{align}
where $Z_{g,h}(\tau)\equiv Z(\tau;g,h)$.
We introduce the $S,T$ matrices as
\begin{align}
\label{ZST}
 Z_a(\tau+1)=T_{ab}Z_b(\tau),\ \ \
 Z_a(-1/\tau)=S_{ab}Z_b(\tau) .
\end{align}
In such a basis, the $S$ and $T$ matrix for $Z_2$ anomalous theory is
\begin{align}
\label{Z2STmatA}
S=\begin{pmatrix}
1 & 0 & 0 & 0 \\
0 & 0 & 1 & 0 \\
0 & 1 & 0 & 0 \\
0 & 0 & 0 & -1
\end{pmatrix},\quad 
T=\begin{pmatrix}
1 & 0 & 0 & 0 \\
0 & 1 & 0 & 0 \\
0 & 0 & 0 & -1 \\
0 & 0 & 1 & 0
\end{pmatrix}
\end{align}
and for anomaly-free theory is
\begin{align}
\label{Z2STmat}
S=\begin{pmatrix}
1 & 0 & 0 & 0 \\
0 & 0 & 1 & 0 \\
0 & 1 & 0 & 0 \\
0 & 0 & 0 & 1
\end{pmatrix},\quad 
T=\begin{pmatrix}
1 & 0 & 0 & 0 \\
0 & 1 & 0 & 0 \\
0 & 0 & 0 & 1 \\
0 & 0 & 1 & 0
\end{pmatrix} .
\end{align}
\fi

\iffalse
\section{$1+1D$ partition functions with $\ZZ_N$ 't Hooft anomaly}\label{TNZN}

For $a,g,h\in G$, we introduce the auxiliary quantities, 
\begin{align}
\beta_a (h,g)=\frac{\omega (a,h,g)\omega (h,g,(hg)^{-1}ahg)}{\omega (h,h^{-1}ah,g)}
\end{align}

In particular, when $G=\ZZ_N$, the following quantities are simplified
\begin{align}
\beta_a (a,b)= \omega (a, b , a)
\end{align}
\begin{align}
T^N Z_{a,1}=&\ee^{\ii \theta_T} Z_{a,1}
\end{align}
\begin{align}
\ee^{\ii\theta_T}=\prod_{k=0}^{N-1}\beta_a (a,a^k)=\prod_{k=0}^{N-1}\omega (a,a^k,a)
\end{align}
\fi

\section{Proof of $S, T$ as a permutation in symmetry twist basis with phases given by $3$-cocycle}\label{basistrans}

%Taking the partition functions as a function of the finite group $Z_{a,b}$. The basis transformation is the generalized Fourier transform on the finite group.

Here, we prove that for CT-twisted quantum double, the S, T matrices quasiparticle basis (\ref{Sqp}) are the same as $S, T$ matrices (\ref{Ssym}) in the symmetry twist basis, after the basis transformation (\ref{qptosym}). Let us begin with the matrix element of $S$. 

\begin{align}
& \langle a,\mu^a |S |b,\lambda^b\rangle =  \frac{1}{|G|}\sum_{\substack{a'\in [a] \\ g\in Z_{a'}}}\sum_{\substack{b'\in [b] \\ h\in Z_{b'}}} \tilde{\chi}_\mu^{a'} (g) \tilde{\chi}_\mu^{b'} (h)^* \langle a',g |S|b',h\rangle\nn\\
=& \frac{1}{|G|}\sum_{\substack{a'\in [a] \\ g\in Z_{a'}}}\sum_{\substack{b'\in [b] \\ h\in Z_{b'}}} \tilde{\chi}_\mu^{a'} (g) \tilde{\chi}_\lambda^{b'} (h)^* \langle a',g |h,{b'}^{-1}\rangle\beta_h^*(b',{b'}^{-1})\nn\\
=&\frac{1}{|G|}\sum_{\substack{a'\in [a],\, b'\in [b] \\ a'b'=b'a'}}\tilde{\chi}_\mu^{a'} ({b'}^{-1}) \tilde{\chi}_\lambda^{b'} (a')^* \beta_{a'}^*(b',{b'}^{-1})\nn\\
=&\frac{1}{|G|}\sum_{\substack{a'\in [a],\, b'\in [b] \\ a'b'=b'a'}}\tilde{\chi}_\mu^{a'} (b')^* \tilde{\chi}_\lambda^{b'} (a')^* 
\end{align}
where we use the relation $\tilde \chi_\mu^{a'}(b')^*=\tilde \chi_\mu^{a'}({b'}^{-1})\beta_{a'}^*(b',{b'}^{-1})$.This is obtained by taking the trace of the last equation of the following,
\begin{align}
\begin{split}
&\tilde{\rho}^{a'}_\mu (b') \tilde{\rho}^{a'}_\mu ({b'}^{-1})=\beta_{a'}(b',{b'}^{-1})\mathbf{1}\,,\\
& \tilde{\rho}^{a'}_\mu (b')^\dagger =\beta_{a'}^*(b',{b'}^{-1})\tilde{\rho}^{a'}_\mu ({b'}^{-1})\,.
 \end{split}
\end{align}

\begin{align}
T|a,\mu^a\rangle=&\frac{1}{\sqrt{|G|}}\sum_{\substack{a'\in [a] \\ g\in Z_{a'}}}\tilde{\chi}^{a'}_\mu (g)^* T|a',g\rangle\nn\\
=&\frac{1}{\sqrt{|G|}}\sum_{\substack{a'\in [a] \\ g\in Z_{a'}}}\tilde{\chi}^{a'}_\mu (g)^* \beta_{a'}(a',g)|a',a'g\rangle\nn\\
=&\frac{1}{\sqrt{|G|}}\sum_{\substack{a'\in [a] \\ g'\in Z_{a'}}}\tilde{\chi}^{a'}_\mu ({a'}^{-1}g')^* \beta_{a'}(a',{a'}^{-1}g')|a',g'\rangle\nn\\
=&\frac{1}{\sqrt{|G|}}\sum_{\substack{a'\in [a] \\ g'\in Z_{a'}}}\frac{\tilde{\chi}^{a'}_\mu (a')}{\dim \mu}\tilde{\chi}^{a'}_\mu (g')^*|a',g'\rangle\nn\\
=&\frac{\tilde{\chi}^{a'}_\mu (a')}{\dim \mu}\,|a,\mu^a\rangle
\end{align}
where we have used $\tilde{\rho}^a(a)=\frac{\tilde \chi_\mu^a (a)}{\dim \mu}\mathbf{1}$ from the Schur's lemma, and $\tilde{\chi}^{a'}_\mu ({a'}^{-1}g')^* \beta_{a'}(a',{a'}^{-1}g')=\frac{\tilde{\chi}^{a'}_\mu (a')}{\dim \mu}\tilde{\chi}^{a'}_\mu (g')^*$, coming from the trace of $\tilde \rho^{a'}(a')\tilde \rho^{a'}({a'}^{-1}g')=\beta_{a'}(a',{a'}^{-1}g')\tilde \rho^{a'}(g')$. In particular, the projective characters are related to the characters through $1$-cochain of the centralizer group $\epsilon_a: \mathcal{Z}_a\rightarrow U(1)$, 
\begin{align}
\tilde{\chi}_\mu^a(g)= \epsilon_a(g) \chi_\mu^a(g).
\end{align}

To digest, let us look at some examples. First, we consider the untwisted topological order with a non-Abelian group $G$. 

\begin{itemize}
\item $G=D_4=\langle r,s| r^4=s^2=1, (rs)^2=1\rangle= [1]\cup [r^2=-1]\cup [r]\cup [s]\cup [rs]=\{ 1\} \cup \{ -1\} \cup \{r,r^3\}\cup \{s, r^2s\} \cup \{rs, r^3s\}$. The normal subgroups are $N_1= D_4, N_{-1}=D_4, N_{r}=\langle r\rangle =\ZZ_4, N_s=\{1, s, r^2s, -1\}=\ZZ_2\times \ZZ_2$ There are $22$ anyons. The basis transformation $M_{\text{sym},\text{any}}$ is block-diagonalized, with the diagonal blocks as the following. 
\begin{align}
&M^{C(1)=D_4}=\begin{bmatrix}
1 & 1 & 1 & 1 & 2 \\
1 & 1 & 1 & 1 & -2 \\
1 & -1 & -1 & 1 & 0 \\
1 & 1 & -1 & -1 & 0 \\
1 & -1 & 1 & -1 & 0
\end{bmatrix},\nn\\
&M^{C(s^2)=D_4}=\begin{bmatrix}
1 & 1 & 1 & 1 & 2 \\
1 & 1 & 1 & 1 & -2 \\
1 & -1 & -1 & 1 & 0 \\
1 & 1 & -1 & -1 & 0 \\
1 & -1 & 1 & -1 & 0
\end{bmatrix},\nn\\
& M^{C(s)=\ZZ_4}=\begin{bmatrix}
1 & 1 & 1 & 1 \\ 1 & \ii & -1 & -\ii \\ 1 & -1 & 1 & -1 \\ 1 & -\ii & -1 & \ii
\end{bmatrix},\nn\\
&M^{C(r)}=\Z_2\times \Z_2=M^{C(sr)}=\begin{bmatrix}
1 & 1 & 1 & 1 \\
1 & 1 & -1 & -1 \\
1 & -1 & 1 & -1 \\
1 & -1 & -1 & 1 \\
\end{bmatrix},\nn\\
\end{align}

\item $G=Q_8=\langle \bar e, X^1,X^2,X^3| \bar e^2=e, (X^1)^2=(X^2)^2=(X^3)^2=X^1X^2X^3=\bar e\rangle= [e]\cup [\bar e] \cup [X^1] \cup [X^2]\cup [X^3]  $, in which $e$ is the identity element. There are $22$ anyons as follows,
\begin{align}
V_{\text{anyon}}\,: ~~ &1=(e,1)\, ~J_a=(e,J_a)\,,~ \chi=(e,\chi) \,,~~a=1,2,3\nn\\
&\bar 1=(\bar e,1)\,~ \bar J_a=(\bar e, \bar J_a)\, ~ \bar \chi=(\bar e, \bar \chi)\,,\nn\\
&X^a_i=(X^a, \Gamma_i)\,, ~~i=0,1,2,3\,, ~a= 1,2,3.
\end{align}
The basis transformation $M_{\text{sym},\text{any}}$ is block-diagonalized, with the diagonal blocks as the following. 
\begin{align}
&M^{C(1)=Q_8}=M^{C(s^2)=Q_8}=\begin{bmatrix}
1 & 1 & 1 & 1 & 2 \\
1 & 1 & 1 & 1 & -2 \\
1 & -1 & -1 & 1 & 0 \\
1 & 1 & -1 & -1 & 0 \\
1 & -1 & 1 & -1 & 0
\end{bmatrix},\nn\\
& M^{C(s)=\ZZ_4}=\begin{bmatrix}
1 & 1 & 1 & 1 \\ 1 & \ii & -1 & -\ii \\ 1 & -1 & 1 & -1 \\ 1 & -\ii & -1 & \ii
\end{bmatrix}.
\end{align}
\end{itemize}

Next, we consider the twisted topological order with group $\ZZ_3$. The character table and the 1-cochain table are as follows,
\begin{align}
\begin{tabular}{| c | c c c |}
\hline
$\ZZ_3$ & $[1]$ & $[a]$ & $[a^2]$ \\
\hline
$\lambda_1$ & $1$ & $1$ & $1$ \\
$\lambda_\omega$ & $1$ & $\omega$ & $\omega^2$ \\
$\lambda_{\omega^2}$ & $1$ & $\omega^2$ & $\omega$ \\
\hline
\end{tabular}~,~~~
\begin{tabular}{ |c | c c c |}
\hline
$\ZZ_3$ & $[1]$ & $[a]$ & $[a^2]$ \\
\hline
$\epsilon_1$ & $1$ & $1$ & $1$ \\
$\epsilon_a$ & $1$ & $s$ & $s^2$ \\
$\epsilon_{a^2}$ & $1$ & $s^2$ & $s^4$ \\
\hline
\end{tabular}~.
\end{align}
Here, $\omega=\ee^{\ii \frac{2\pi}{N}}$ and $s=\ee^{\ii \frac{2\pi}{N^2}}$ with $N=3$. Explicitly, the basis transformation is
\begin{align}
\begin{bmatrix}
(1,1) \\ (1,a) \\ (1,a^2)
\end{bmatrix}=&M_0\begin{bmatrix}
(1,\tlambda_1 ) \\ (1, \tlambda_\omega) \\ (1, \tlambda_{\bar\omega})
\end{bmatrix},\\
\begin{bmatrix}
(a,1) \\ (a,a) \\ (a,a^2)
\end{bmatrix}=&\begin{bmatrix}
1 & 0 & 0 \\ 0 & s^* & 0 \\ 0 & 0 & (s^2)^*
\end{bmatrix}M_0\begin{bmatrix}
(a,\tlambda_1 ) \\ (a, \tlambda_\omega) \\ (a, \tlambda_{\bar\omega})
\end{bmatrix},\\
\begin{bmatrix}
(a^2,1) \\ (a^2,a) \\ (a^2,a^2)
\end{bmatrix}=&\begin{bmatrix}
1 & 0 & 0 \\ 0 & (s^2)^* & 0 \\ 0 & 0 & (s^4)^*
\end{bmatrix}M_0\begin{bmatrix}
(a^2,\tlambda_1 ) \\ (a^2, \tlambda_\omega) \\ (a^2, \tlambda_{\bar\omega})
\end{bmatrix},
\end{align}
where 
\begin{align}
M_0=&\frac{1}{\sqrt{3}}\begin{bmatrix}
1 & 1 & 1 \\ 1 & \omega & \bar\omega \\ 1 & \bar \omega & \omega
\end{bmatrix},
\end{align}
is the Hermitian conjugate matrix of $\ZZ_3$ character table normalized to have determinate $1$. The basis transformation tells us immediately the spin selection rules. The CFT with $\ZZ_N$ (generated by $a$) 't Hooft anomaly labeled by the $3$-cocycle $\omega_q$, the conformal spin of the primary fields in the $a$-twisted sector satisfies
\begin{align}
h_L-h_R\in \frac{q}{N^2}+\frac{\ZZ}{N}.
\end{align}

%{\bf [In general, how to determine 1-cochain $\epsilon_a (a)$ from the
%3-cocycle $\omega_q$?]}

\section{Gauge the charge conjugation symmetry of a topological order}\label{app:CC}
In broad strokes, a \emph{topological symmetry}\cite{barkeshli2019symmetry} on a topological order $\calC$ is characterized by a map: $[\rho]: G\rightarrow \operatorname{Aut} (\calC)$ from the group $G$ to the automorphisms of $\calC$. Given such a topological symmetry $\rho$, there is a unique element $[\mathfrak{O}]\in H^3_{[\rho]}(G, \calA)$  telling us if the symmetry can be fractionalized. If and only if $[\mathfrak{O}]=[0]$, the Abelian anyons $\calA$ in $\calC$, invariant under the map $[\rho]$, can carry projective representations of the symmetry $G$. This symmetry fractionalization is classified by the $H^2 (G, \calA)$ torsor.

Let us consider the charge conjugation symmetry ${\bf C}$ on the fusion space $V_c^{ab}$, representing the $c$ superselection sector with $a$ and $b$ anyon that split from a $c$-anyon. On the local state $|a,b;c, \mu\rangle\in V_c^{ab}$, the symmetry acts as, 
\begin{align}
\rho_{{\bf C}} (g)|a,b;c,\mu\rangle =\sum_\nu \left[U_{{\bf C}} (\bar a, \bar b; \bar c)\right](g)_{\mu\nu} |\bar{a},\bar{b};\bar{c},\nu\rangle.
\end{align}

The local symmetry action operators $U_{{\bf C}}$ can admit $\ZZ_2$ symmetry fractionalization. When there is no obstruction $[\mathfrak{O}]=[0]$, the symmetry fractionalization is given by \cite{barkeshli2019symmetry}
\begin{align}
H^2_{[\rho]}(\ZZ_2,\calA)=\calA^{{\bf C}},
\end{align}
where $\calA^{{\bf C}}=\{a\in \calA |a=\bar{a}\}$ is the group defined by the set of self-dual Abelian anyons, with group multiplication given by the fusion rules. For example, for $\ZZ_{3}^{(-1)}$ we consider, $\calA^{{\bf C}}=\{1\}$. There is only one trivial symmetry fractionalization. The operators $U_{{\bf C}}$ on each space $V^{ab}_c$ will be a linear representation of the $\ZZ_2$ group.

\section{Anomaly for the group beyond $\ZZ_N$}
\subsection{Anomaly of an Abelian group}

Another way to detect the anomaly is via the deformation and fusion of the
lines of the symmetry twist.  In particular, the symmetry twist is anomalous
when the following $F$ move corresponds to a nontrivial cocycle
\begin{align}
F^{gg^{-1}g}_{g}=\omega (g,g^{-1},g)
\end{align}
where $g$ runs through the generators of the group $G$. 

We are now ready to determine the abelian group $G$ anomaly through $S$ and $T$ matrices on the boundary. We denote the generators in $G$ as $g_1,\cdots, g_r$, and $r$ is the rank of the group. Any symmetry action is represented by $a=g_1^{n_1}\cdots g_r^{n_r}$. Given the action of the generators $g_j$ on the representations, as a matrix $(g_j)_{\mu\lambda}$. And $[g_j,g_{j'}]=0$. The $g_j$ action twisting the spatial boundary condition is 
\begin{align}
G_j =S^\dagger g_j S
\end{align}
where we have used that $S$ is an unitary matrix $S^{-1}=S^\dagger$.
It follows that $[G_j,G_{j'}]=0$. 

Consider the sector, twisted in the spatial direction by $a_x=g^{n_1}_1\cdots g^{n_r}_r$, and in the temporal direction by $a_t=g^{m_1}_1\cdots g^{m_r}$ is captured in the matrix
\begin{align}
&Z_{\{n_j\},\{m_j\}}=\sum_{\mu \lambda} \bar \chi_{\mu} \calN [ \{ n_j\}, \{m_j\}]_{\mu\lambda}\chi_\lambda\nn\\
&\calN[{\{ n_j\}, \{m_j\}}]=\prod_{j=1}^r G_j^{n_j}\cdot\prod_{j=1}^rg_j^{m_j}
\end{align}

\begin{align}
\ee^{\ii \theta[{\{ n_j\}, \{m_j\}}]}=\omega_3 (a_x,a_t,a_x)
\end{align}
where $a_x=g^{n_1}_1\cdots g^{n_r}_r$, $a_t=g^{m_1}_1\cdots g^{m_r}$.

\iffalse
\begin{figure}[h]
\begin{tikzpicture}[baseline={(current bounding box.center)}]
\draw (0,0) rectangle (3,3);\node at (1.5,0.5) {$ghk^{-1}$};
\draw (1,3) rectangle (0,2);\node at (2.5,1.2) {$k$};
\draw (1,2) -- (2,2)--(2,0);\node at (1.5,1.7) {$gh$};
\draw (2,1)--(3,1);\node at (0.5,1.7) {$g$};\node at (1.2,2.5) {$h$};
\draw [decorate,decoration={brace,amplitude=2pt},xshift=-2pt,yshift=0pt]
(0,0) -- (0,0.95) node [black,midway,xshift=-0.6cm] 
{\footnotesize $\calH_{ghk^{-1}}$};
\draw [decorate,decoration={brace,amplitude=2pt},xshift=-2pt,yshift=0pt]
(0,1.05) -- (0,1.95) node [black,midway,xshift=-0.6cm] 
{\footnotesize $\calH_{gh}$};
\draw [decorate,decoration={brace,amplitude=2pt},xshift=-2pt,yshift=0pt]
(0,2.05) -- (0,2.95) node [black,midway,xshift=-0.6cm] 
{\footnotesize $\calH_{h}$};
\end{tikzpicture}
\end{figure}
\fi

\begin{figure}[h]
\begin{pspicture}(0,-2.97)(10.96,2.97)
\psframe[linewidth=0.04,dimen=outer](3.18,2.95)(0.0,-0.23)
\psline[linewidth=0.04cm](0.92,2.95)(0.94,2.01)
\psline[linewidth=0.04cm](0.0,2.03)(1.92,2.03)
\psline[linewidth=0.04cm](1.92,2.03)(1.92,1.03)
\psline[linewidth=0.04cm](3.16,1.01)(1.9,1.03)
\psline[linewidth=0.04cm](1.92,1.03)(1.92,-0.25)
\psline[linewidth=0.04cm](10.92,-2.95)(10.94,-2.95)
\end{pspicture} 
\end{figure}

\iffalse
\begin{table}
\setlength\extrarowheight{4.5pt}
\setlength{\tabcolsep}{6pt}
\begin{tabular}{  |c | c| c|c|}
\hline 
Symbol & $(r,s)$ & $h$ & under $\ZZ_3$ \\
\hline
$1$ & $(1,1)$ or $(4,5)$ & $0$ & $1$ \\
$\sigma$ & $(3,3)$ or $(2,3)$ & $\frac{1}{15}$ & $\ee^{\ii\frac{2\pi}{3}}\sigma$ \\
 $\sigma^\dagger$ & $(3,3)$ or $(2,3)$ & $\frac{1}{15}$ & $\ee^{-\ii\frac{2\pi}{3}}\sigma^\dagger$ \\
  $\psi$ & $(4,3)$ or $(1,3)$ & $\frac{2}{3}$ & $\ee^{\ii\frac{2\pi}{3}}\psi$ \\
   $\psi^\dagger$ & $(4,3)$ or $(1,3)$ & $\frac{2}{3}$ & $\ee^{-\ii\frac{2\pi}{3}}\psi^\dagger$ \\
    $\epsilon$ & $(2,1)$ or $(3,5)$ & $\frac{2}{5}$ & $\epsilon$ \\
     $X$ & $(3,1)$ or $(2,5)$ & $\frac{7}{5}$ & $X$ \\
      $W$ & $(4,1)$ or $(1,5)$ & $3$ & $W$ \\
\hline
\end{tabular}
\caption{Primary fields in $\ZZ_3$ parafermion CFT.}\label{z3}
\end{table}

\begin{table}[h]
\setlength\extrarowheight{4.5pt}
\setlength{\tabcolsep}{6pt}
\begin{tabular}{|  c| c| c |}
\hline
$\{l,m\}$ & $h\mod 1$ & Symbol \\
\hline
$\{0,0\}$ & $0$ & $1$ \\
$\{0,2\}$ & $\frac{2}{3}$ & $\psi,\psi^\dagger$ \\
$\{0,-2\}$ & $\frac{2}{3}$ & $\psi,\psi^\dagger$ \\
$\{1,3\}$ & $\frac{2}{5}$ & $\epsilon$ \\
$\{1,1\}$ & $\frac{1}{15}$ & $\sigma,\sigma^\dagger$ \\
$\{1,-1\}$ & $\frac{1}{15}$ & $\sigma,\sigma^\dagger$ \\
\hline
\end{tabular}
\caption{$\ZZ_3$ Parafermion fields in coset construction.}
\end{table}
\fi

\subsection{Anomaly of a non-abelian group}

We can also detect the anomaly of non-abelian subgroup by symmetry twisted partition functions. The setup is as follows. Consider the non-abelian subgroup $G$. For each element $g\in G$, we can construct a twisted Hilbert space $\calH_g$, depending only on the conjugacy class $[g]$. In each sector, we can project onto the states invariant under the stabilizer group $N_g$, withe the projector
\begin{align}
P^g=\frac{1}{|N_g|}\sum_{h\in N_g} h
\end{align}
and $|N_g|=|G|/|[g]|$.

\subsubsection{$S_3$ symmetry }

The 't Hooft anomaly of $S_3$ is classified by $p\in H^3(S_3,U(1))=\ZZ_6$.
Denote the group element $g=r^As^a$, with $A=0,1,\ a=0,1,2$ for $S_3$. The
group multiplicity is $(A,a)(B,b)= (\langle A+B\rangle_2,
\langle(-1)^Ba+b\rangle_3)$, where $\langle x\rangle_n= x\mod n$. Then the
$3$-cocycles are given by \cite{deWildPropitius:1995cf}
\begin{align}
&\omega_q\left( (A,a),(B,b),(C,c)\right)\nn\\
=&\exp \left\{-\frac{\ii 2\pi}{9}p\left[ (-1)^{B+C}a\left[ (-1)^C b+c-\langle (-1)^Cb+c\rangle_3\right]\right.\right.\nn\\
&\left.\left.+\frac{9}{2}ABC\right]\right\}
\end{align}

We find the anomaly indexed by $p\in\ZZ_6$ can be detected by measuring two gauge invariant 3-cocycle, 
\begin{align}
\omega_p(s,s^{-1},s)=\ee^{-\ii \frac{2\pi}{3}p},~~\omega_p(r,r,r)=\ee^{\ii \pi p}.
\end{align}

We examine the $D_N$ anomaly in $SU(2)_k$ CFT. For this purpose, we write the primary fields of $SU(2)_k$ CFT in terms of parafermion and $U(1)$ primary fields.
\begin{align}
\phi^l_m(z)=f^l_m(z) \ee^{\ii \frac{m}{\sqrt{2k}}\varphi(z)}
\end{align}
with $l=0,1,\cdots k$, $-l\leq m\leq l,m+l=0\mod 2$.

\subsubsection{$su(2)_1$}
\begin{align}
s \left(\ee^{\ii \frac{m}{2}\varphi (z)}\right)=&\ee^{-\ii \frac{m}{2}\varphi (z)}\nn\\
r \left(\ee^{\ii \frac{m}{2}\varphi (z)}\right)=&\ee^{\ii 2\pi\frac{m}{N}}\ee^{-\ii \frac{m}{2}\varphi (z)}
\end{align}
since $m=0,1$, defined $\mod 2$. On the character, the action of $s$ on the characters is trivial.
\begin{align}
s=\operatorname{Id}_2,\quad r=\begin{pmatrix}
1 & 0 \\ 0 & \ee^{\ii \frac{2\pi}{N}}
\end{pmatrix}
\end{align}

\begin{align}
R=&S^\dagger rS=\frac{1}{2}\begin{pmatrix}
1+\omega & 1-\omega \\ 1-\omega & 1+\omega
\end{pmatrix} , \nn\\
R^{-1}=&\frac{1}{2}\begin{pmatrix}
1+\omega^{-1} & 1-\omega^{-1} \\ 1-\omega^{-1} & 1+\omega^{-1}
\end{pmatrix}
\end{align}
where $\omega=\ee^{\ii \frac{2\pi}{N}}$ and $R^N=1$. 
\begin{align}
T=\begin{pmatrix}
1 & 0 \\ 0 & \ee^{\ii \frac{\pi}{2}}
\end{pmatrix}
\end{align}

\begin{align}
T^N R (T^\dagger)^N=\frac{1}{2}\begin{pmatrix}
1+\omega & \ee^{-\ii \frac{N\pi}{2}}(1-\omega) \\ \ee^{-\ii \frac{N\pi}{2}}(1-\omega) & 1+\omega
\end{pmatrix}
\end{align}

There is no anomaly if $N\in 4\ZZ$.

\bibliography{../../bib/all,../../bib/allnew,../../bib/publst,./local}

\end{document}